\documentclass[aps,prb]{revtex4}
\usepackage{amsmath,amssymb}
\usepackage{graphicx}
\usepackage{dcolumn}
\usepackage{bm}
\usepackage{color}
\usepackage{relsize}
\newcommand{\mbb}{\mathbb}
\newcommand{\mc}{\mathcal}

\newcommand{\tet}{\texttt}

\begin{document}
\title{Suppressed plasmon excitations, enhanced damping and static screening \\
in Kek-Y strained $\alpha-\mc{T}_3$ model}
\author{
Jean Marseille$^{1}$,
Teresa Lee$^{2}$,
Andrii Iurov$^{1}$\footnote{E-mail contact: aiurov@mec.cuny.edu, theorist.physics@gmail.com},
Liubov Zhemchuzhna$^{1,2,3}$\footnote{E-mail contact: lzhemchuzhna@mec.cuny.edu, lzhemchuzhna@fordham.edu},
Godfrey Gumbs$^{3,4}$, 
and
Danhong Huang$^{5}$}

\affiliation{
$^{1}$Department of Physics and Computer Science, Medgar Evers College of City University of New York, Brooklyn, NY 11225, USA\\ 
$^{2}$Department of Physics and Astronomy, Hunter College of the City University of New York, 695 Park Avenue, New York, New York 10065, USA\\ 
$^{3}$Department of Physics \& Engineering Physics, Fordham University, Bronx, NY 10458, USA\\
$^{4}$Donostia International Physics Center (DIPC), P de Manuel Lardizabal, 4, 20018 San Sebastian, Basque Country, Spain\\ 
$^{5}$Space Vehicles Directorate, US Air Force Research Laboratory, Kirtland Air Force Base, New Mexico 87117, USA
}

\date{\today}

\begin{abstract}
We performed a rigorous theoretical and numerical investigation into the polarization function, plasmon excitations, and plasmon damping in the Kek-$\alpha$ model, a two-dimensional material combining the key features of the $\alpha-\mc{T}_3$ lattice and Kekule-distorted graphene. Unlike conventional Kek-Y graphene, the Kekule modulation in the Kek-$\alpha$ model affects only one of the two sublattices, giving rise to a fundamentally new model with unusual electronic properties. The low-energy spectrum consists of two degenerate flat bands and two inequivalent Dirac cones with different Fermi velocities, referred to as the fast and slow cones. The particle-hole continuum responsible for Landau damping exhibits two distinct branches associated with transitions involving these Dirac cones. An additional particle-hole mode originates from electron transitions associated with the fast Dirac cone, appearing above the main diagonal. As the parameter $\alpha$ increases, the contribution from the fast Dirac cone becomes dominant. The additional transitions involving the flat bands and the fast Dirac cone substantially reduce the region where undamped plasmons can exist, similarly to the conventional $\alpha-\mc{T}_3$. Consequently, stable plasmons are observed only for relatively small values of $\alpha$ or at very small wave vectors. These unusual electronic and collective properties make the Kek-$\alpha$ model a promising platform for future plasmonic and nanoscale electronic applications.
\end{abstract}

\maketitle

\section{Introduction} 
\label{sec1}

Since the discovery of graphene \,\cite{geim2009graphene,geim2007rise}, all two-dimensional Dirac cone materials received tremendous attention by the researchers. Graphene has demonstrated a wealth of unique electronic, mechanical, thermal, and optical properties,\,\cite{castro2009electronic,ando2009electronic,gumbs2014revealing} making it one of the most extensively studied two-dimensional materials. Shortly after its discovery, several other Dirac materials were identified. Unlike graphene, many of these materials possess a finite band gap,\,\cite{wehling2014dirac} which enables the confinement of charge carriers within a given region and makes them more suitable for electronic and optoelectronic device applications.\,\cite{pedersen2009optical,pereira2008supercritical}

\par 
Importantly, the band gap can be further tuned by applying resonant circularly or elliptically polarized light, giving rise to a substantially modified electronic band structure through a mechanism known as Floquet engineering. \,\cite{kibis2010metal,ibarra2019dynamical,iurov2022floquet,oka2019floquet,iurov2024floquet,bukov2015universal} The Floquet engineering is applied to modify its transport properties \,\cite{iurov2020quantum,islam2018driven,iurov2022finite,aitouni2026laser,kristinsson2016control,iurov2017exploring} in a specific way need for the applications. 

\par 
Dirac materials with flat band \,\cite{liu2014exotic,roy2019unconventional,milicevic2019type} have attracted significant interest because their nearly dispersionless energy bands strongly suppress the kinetic energy of charge carriers and enhance the role of electron-electron interactions. This gives rise to a variety of correlated quantum phenomena.Flat bands naturally occur in several lattices, such as the Lieb lattice  \,\cite{slot2017experimental,mukherjee2015observation} and the Kagome lattice,\,\cite{jo2012ultracold,xue2019acoustic} and are also realized in the dice $\alpha-\mc{T}_3$ lattice through destructive quantum interference \cite{Leykam2018}. A major experimental breakthrough was the observation of correlated insulating states and unconventional superconductivity in magic-angle twisted bilayer graphene\,\cite{Cao2018a,Cao2018b,Ortiz2019,Yin2022}. 

\par 
The $\alpha-\mc{T}_3$ model is a promising two-dimensional material characterized by a honeycomb lattice with an additional hub site at the center of each hexagon. Its electronic properties are governed by the parameter $\alpha=\tan\phi$, which determines the relative hopping strength between the hub and rim sites. A unique feature of the model is the continuous tunability of $\alpha$ from 0 to 1, interpolating between graphene $\alpha \rightarrow  0$ and pseudospin-1 dice-lattice materials $\alpha \rightarrow 1$.\,\cite{gorbar2019electron} The general electronic,\,\cite{illes2017properties,roldan2011theory, weekes2021generalized} magnetic \,\cite{raoux2014dia, islam2023role, islam2023effect}  collective and transport properties \,\cite{malcolm2016frequency,oriekho2023quantum,oriekhov2020rkky,huang2019interplay} including the  Klein tunneling \,\cite{illes2017klein,iurov2020klein,roslyak2010unimpeded} of the $\alpha-\mc{T}_3$ have been also investigated and developed. An important question of research became how those properties are different from graphene. The Floquet theory for the irradiated at3 materials \,\cite{tamang2023probing,iurov2019peculiar,dey2018photoinduced} has been also thoroughly developed.

\medskip 
Kekule-strained graphene is a bond-modulated phase of graphene characterized by a periodic reconstruction of the honeycomb lattice, which triples the unit cell and couples the two inequivalent Dirac valleys. \,\cite{Hou2007,Jackiw2007,herrera2020dynamic} Two main Kekule patterns are the Kek-O phase, which opens a gap at the Dirac point by generating a Dirac mass, and the Kek-Y phase, which preserves gapless Dirac cones but produces valley-momentum locking.\,\cite{Naumis2017,OlivaLeyva2013,Pereira2009} Theoretical studies \,\cite{mojarro2020dynamical,herrera2020electronic,iurov2023application} have demonstrated that strain strongly modifies the Kekule electronic structure by tuning hopping amplitudes, Dirac velocities, valley coupling, and topological properties.\,\cite{deJuan2012,Andrade2019Nanoribbons,RuizTijerina2019,andrade2019valley} Experimental observations using scanning tunneling microscopy have confirmed Kekule ordering and strain-induced distortions.\,\cite{andrade2025topical,Eom2020} Kek-Y strained $\alpha-\mc{T}_3$ model was  derived and analyzed in Ref.~[\onlinecite{sanchez2025band}] which is in part the subject of the present paper. 

\medskip 
Plasmons, the self-sustained quantum oscillations of electrons in a conducting material, have demonstrated
considerable potential for various technological applications and have become an important research subject for any
newly discovered low-dimensional structure, The nature of the plasmon dispersion is directly
related to the low-energy electron dispersion in a given material.  Dynamical polarization function, dielectric function, screening, and plasmons have been investigated for graphene, \,\cite{politano2014plasmon,yan2012infrared,hwang2007dielectric,polini2008plasmons,jablan2013plasmons,zhang2012surface,wunsch2006dynamical,constant2016all} bilayer \,\cite{gamayun2011dynamical} and gapped graphene \,\cite{pyatkovskiy2008dynamical}, silicene and other buckled honeycomb lattices, \,\cite{tabert2014dynamical} tilted and anisotropic Dirac materials \,\cite{torbatian2021hyperbolic,hayn2021plasmons,gomes2021tilted,ross2025dynamical,badalyan2009anisotropic,trescher2015quantum} transition metal dichalcogenides, \,\cite{sriram2020hybridizing,andersen2013plasmons,scholz2013plasmons} Dirac semimetals, \,\cite{sadhukhan2020novel} noncentrosymmetric systems, \,\cite{dutta2023intrinsic,dey2022dynamical,shitrit2013spin} including the intrinsic plasmons at a finite temperature. \,\cite{sarma2013intrinsic,iurov2022finite}

\par 

Plasmon excitations have been also investigated in graphene nanoribbons, \,\cite{karimi2017plasmons,brey2007elementary,fei2015edge,wang2005plasmon}
flat band materials,\,\cite{kajiwara2016observation} including $\alpha-\mc{T}_3$ nanoribbons \,\cite{iurov2021tailoring} graphene-based heterostructures and  multi-component systems \,\cite{gumbs2015nonlocal,yao2018broadband,li2017first,iurov2017controlling,sarma1981collective,yerin2023dielectric} and buckyballs (fullerenes), \, \cite{ju1993excitation,gumbs2014strongly} surface plasmon polaritons \,\cite{berini2012surface,barnes2006surface} Collective excitations show very unusual properties in the presence of magnetic field \,\cite{balassis2020magnetoplasmons,mawrie2014magnetotransport,tamang2023orbital,dutta2022collective}
Separately, plasmon instabilities \,\cite{gumbs2015tunable,koseki2016giant,petrov2017amplified} have been also investigated. Optical conductivity closely related to polarization function and collective properties \,\cite{mojarro2026topical,xiong2023optical,mojarro2021optical,stauber2013optical,tan2021anisotropic,wareham2023optical,iurov2023optical,wild2023optical,oriekhov2022optical,malik2026probing} also represent and important direction in low-dimensional condensed matter physics. 

\medskip 
\par 

The remaining part of the present publication is organized as follows: in Section \ref{sec2}, we present and explain the detailed formalism of the low-energy Hamiltonian, the corresponding energy dispersions given by the slow and fast Dirac cones and two degenerate flat bands, as well as the associated wave functions (eigenstates) for the Kekule-modulated $\alpha-\mc{T}_3$ model. We then calculate the polarization function and provide detailed analytical derivations of the required wave-function overlap factors in Section \ref{sec3}. Section \ref{sec4} is devoted to our calculation and analysis of the dielectric function, plasmon dispersions obtained as the zeros of this dielectric function, the corresponding particle-hole excitation modes, and plasmon damping. We also investigate static screening in Sec.~\ref{sec5} and compare our results with the non-strained $\alpha-\mc{T}_3$ model for different values of the relative hopping parameter $\alpha$, including the limiting cases of graphene $\alpha \longrightarrow 0$ and a dice lattice $\alpha \longrightarrow 0$. Our concluding remarks are presented and discussed in Section \ref{sec6}. In Appendix \ref{apa}, we present and outline an alternative derivation of the polarization function based on Green's functions and projector operators. We derive several key projector operators and demonstrate that this approach yields results identical to those obtained from the approach based on the wave-function overlaps. In Appendices \ref{apa} and \ref{apa}, we demonstrate the graphene $\alpha \longrightarrow 0$ and dice lattice $\alpha \longrightarrow 1$ limits of our model, including the corresponding wave functions and wave-function overlap factors.

\section{Electronic states in Kek-Y strained $\alpha-\mc{T}_3$ model: general formalism} 
\label{sec2}

We begin with a general description discussion of the low energy Hamiltonian, energy dispersion, the corresponding wave functions for Kek-Y strained $\alpha-\mc{T}_3$ model. Its effective Hamiltonian, obtained in Ref.~[\onlinecite{sanchez2025band}] resulted in the following $6 \times 6$ matrix

\begin{equation}
\hat{\mathcal{H}}_\alpha (\bf k)
=
\begin{bmatrix}
\hbar v_F\,\mathbf{k}\cdot\mathbf{S} & \hat{Q}\\
\hat{Q}^{\dagger} & \hbar v_F\,\mathbf{k}\cdot\mathbf{S}^{*}
\end{bmatrix} \, , 
\end{equation}
where

\begin{equation}
\hat{Q}
=
\alpha\hbar v_F
\begin{pmatrix}
0 & 0 & 0\\
0 & 0 & ke^{i\theta_k}\\
0 & ke^{i\theta_k} & 0
\end{pmatrix} \, .
\end{equation}
Here, $k=|{\bf{k}}|$ is the absolute value of the wave vector ${\bf{k}}$, and $\Theta_{\bf{k}}= \tan^{-1} \left(\frac{k_x}{k_y}\right)$ is the angle associated with the wave vector. The Fermi velocity $v_F = 10^6\,m/s$ is the same as in graphene to ensure the proper limiting behavior for $\alpha \rightarrow 0$. 

\medskip 
The two $3 \times 3$ pseudospin-1 matrices $\mathbf{\hat{\Sigma}}^{(3)} = \left(\hat{\Sigma}_x^{(3)},\hat{\Sigma}_x^{(3)}\right)$ are defined as 

\begin{equation}
\label{sx01}
\hat{\Sigma}_x^{(3)} =
\begin{pmatrix}
0 & 1 & 0\\
1 & 0 & \alpha\\
0 & \alpha & 0
\end{pmatrix}
\end{equation}
and 

\begin{equation}
\label{sy01}
\hat{\Sigma}_x^{(3)} =
\begin{pmatrix}
0 & -i & 0\\
i & 0 & -\alpha i\\
0 & \alpha i & 0
\end{pmatrix} \,  .
\end{equation}
We verify that matrices \eqref{sx01} and \eqref{sy01} are reduced to the regular spin-1/2 Pauli matrices for $\alpha \longrightarrow 0$, even though they do not exactly match the $\phi$-dependent spin-1 Pauli matrices used to define the Hamiltonian for the regular $\alpha-\mc{T}_3$ model, \,\cite{weekes2021generalized} since the approximation only works well for small $\alpha \ll 1$. However, a formal limit of $\alpha \longrightarrow 1$ of a dice lattice also exists. Therefore, it could be considered a smooth interpolation between graphene and a dice lattice. However, the precision of this model is much higher for a small 
$\alpha$, close to graphene.

\par

Therefore, we arrive at the following Hamiltonian 

\begin{equation}
\label{MainHamalpha}
\hat{\mc{H}}_\alpha ({\bf k}) = \left\{
\begin{array}{ccc|ccc}
0 & e^{-i\theta}k & 0 & 0 & 0 & 0 \\[6pt]
e^{i\theta}k & 0 & e^{-i\theta}k\alpha & 0 & 0 & e^{i\theta}k\alpha \\[6pt]
0 & e^{i\theta}k\alpha & 0 & 0 & e^{i\theta}k\alpha & 0 \\[6pt]
\hline
0 & 0 & 0 & 0 & e^{i\theta}k & 0 \\[6pt]
0 & 0 & e^{-i\theta}k\alpha & e^{-i\theta}k & 0 & e^{i\theta}k\alpha \\[6pt]
0 & e^{-i\theta}k\alpha & 0 & 0 & e^{-i\theta}k\alpha & 0
\end{array}
\right\} 
\end{equation}
or, alternatively, introducing a notation $k_\pm = k_x \pm i k_y$, we can write

\begin{equation}
\hat{\mc{H}}_\alpha ({\bf k}) = \left\{
\begin{array}{ccc|ccc}
0 & k_- & 0 & 0 & 0 & 0 \\[6pt]
k_+ & 0 & \alpha k_- & 0 & 0 & \alpha k_+ \\[6pt]
0 & \alpha k_+ & 0 & 0 & \alpha k_+ & 0 \\[6pt]
\hline
0 & 0 & 0 & 0 & k_+ & 0 \\[6pt]
0 & 0 & \alpha k_- & k_- & 0 & \alpha k_+ \\[6pt]
0 & \alpha k_- & 0 & 0 & \alpha k_- & 0
\end{array}
\right\} \, . 
\end{equation}

\begin{figure} 
\centering
\includegraphics[width=0.75\textwidth]{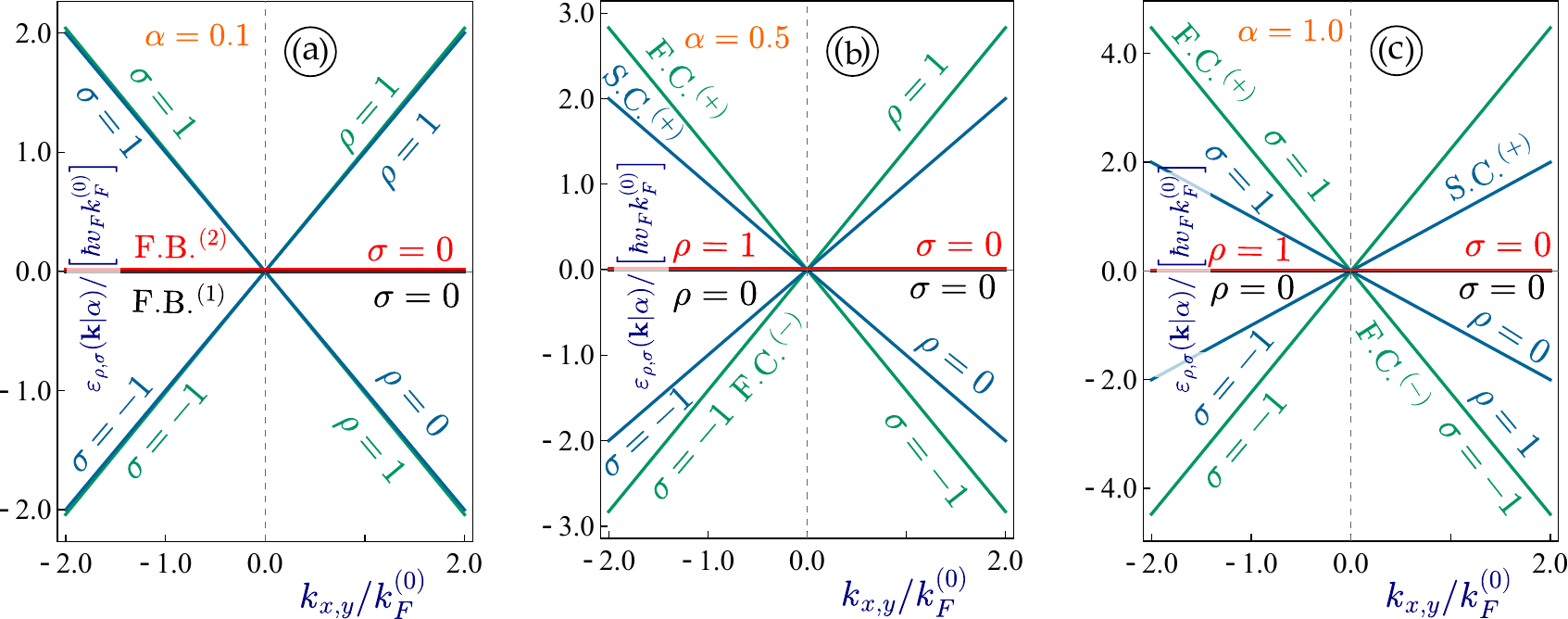}
\caption{(Color online) Schematics for the low-energy bandstructure of the Kek-$\alpha$ model obtained from Hamiltonian
\eqref{MainHamalpha}. The obtained dispersions represent two Dirac cone with differnt Fermi velocities $v_F$ and 
$\sqrt{1+4 \alpha^2}v_F$, and two degenerate flat bands at the zero energy.
}
\label{FIG:01}
\end{figure}
\medskip
The eigenvalue equation corresponding to Hamiltonian \eqref{MainHamalpha}

\begin{equation}
\hat{\mc{H}}_\alpha ({\bf k}) \, \Psi_\lambda (\alpha | {\bf k}) 
=
\varepsilon_\lambda (\alpha \vert {\bf k}) \, \Psi_\lambda (\alpha | {\bf k})  = 
\varepsilon_\lambda (\alpha | {\bf k}) \,\,
\begin{bmatrix}
\Psi_{\lambda,K} (\alpha | {\bf k}) \\
\Psi_{\lambda,K'} (\alpha | {\bf k})
\end{bmatrix} \, ,
\end{equation}
is utilized to obtain the energy dispersions and the eigenstates for the considered Kek-$\alpha$ model.
The low-energy band structure of Kek-$\alpha$ model is obtained as 

\begin{equation}
\label{mainE}
\varepsilon_{\sigma}^{\rho}(\mathbf{k})
= \pm \hbar v_F \, \sqrt{1 + 2 \alpha^2 \pm 2 \alpha^2} \, k
= \sigma \, r_{\alpha}^\rho \, (\hbar v_F k)
\end{equation}
where $r_{\alpha}=\sqrt{4\alpha^2+1}$ specifies the two Dirac cones with the non-equivalent Dirac velocities $v_F$ and $r_{[\alpha]}\,v_F$ specified by the index $\rho$, $\sigma = 0, \pm 1$ is the band index -- $\sigma=1$ for the conduction band,
$\sigma=-1$ for the valence band, and $\sigma=0$ for the flat band. 

\medskip 
The general schematic for the energy dispersions for the Kek-Y strained $\alpha-\mc{T}_3$ model (which we will also call the Kek-$\alpha$ model) is presented in Fig.~\ref{FIG:01}. We observe that the band structure depends significantly on our only parameter $\alpha$. In general, the spectrum consists of two non-equivalent Dirac cones, known as the fast ($F.C^{(\pm)}$) and slow ($S.C^{(\pm)}$) cones, and two degenerate flat bands. The latter must be present because the Hamiltonian must have 6 solutions in total, yielding two flat bands with identical zero-energy dispersions but distinct wave functions. 

\par
We emphasize one more time that the energy band spectrum directly depends on the relative hopping parameter $\alpha$, which is not possible for the regular $\alpha-\mc{T}_3$ model. Also, we see that only one cone depends on $\alpha$, and the other one remains $\alpha$-independent. The difference between the two Dirac cones stemming from their non-equivalent Fermi velocities depends strongly on the value of $\alpha$. This difference vanishes for $\alpha=0$, corresponding to graphene, shown in panel $(a)$, and becomes most pronounced for $\alpha=1$, which is the case of a dice lattice. For most intermediate values of $\alpha$, the energy separation between the two cones remains substantial. Importantly, electronic transitions between the two non-equivalent Dirac cones are also allowed, making the Kek-modulated $\alpha-\mc{T}_3$ model qualitatively different from the previously investigated $\alpha-\mc{T}_3$ model and other Dirac-cone materials. 

\medskip 
\par 

Now we look at the corresponding wave functions 

\begin{equation}
\label{psm1}
\Big| \Psi_1 ({\bf k}) \Big \rangle = \Big| \Psi_{\rho = 0}^{\sigma = 0} ({\bf k}) \Big \rangle =
\frac{1}{\sqrt2}
\begin{pmatrix}
0\\0\\-e^{i\Theta_{{\bf k}}}\\0\\0\\e^{-i\Theta_{{\bf k}}}
\end{pmatrix} \, ,
\end{equation}

\begin{equation}
\label{psm2}
\Big| \Psi_2 ({\bf k})  \Big \rangle  = \Big| \Psi_{\rho = 1}^{\sigma = 0} ({\bf k})  \Big \rangle =
\frac{1}{\sqrt{1+2\alpha^2}}
\begin{pmatrix}
\alpha \tet{e}^{-i\Theta_{\bf k}}\\0\\- \tet{e}^{i\Theta_{{\bf k}}}\\\alpha \tet{e}^{i\Theta_{\bf k}}\\0\\0
\end{pmatrix}
\end{equation}

and

\begin{equation}
\label{psm3}
 \Big| \Psi_{3,4}\Big \rangle = \Big| \Psi_{\rho = 0}^{\sigma = \pm 1} ({\bf k})  \Big \rangle  =
\frac{1}{2}
\begin{pmatrix}
\mp \tet{e}^{-i\Theta_{{\bf k}}}\\-1\\0\\ \pm \tet{e}^{i\Theta_{{\bf k}}}\\1\\0
\end{pmatrix} = \frac{1}{2}
\begin{pmatrix}
-\sigma \tet{e}^{-i\Theta_{{\bf k}}}\\-1\\0\\ \sigma \tet{e}^{i\Theta_{{\bf k}}}\\1\\0
\end{pmatrix}  \,  , 
\end{equation}
as well as 

\begin{equation}
\label{psm6}
\Big| \Psi_{5,6}\Big \rangle  =  \Big| \Psi_{\rho = 1}^{\sigma = \pm 1} ({\bf k}) \Big \rangle = \frac{1}{2 \sqrt{1 + 4 \alpha^2}}
\begin{pmatrix}
1\\[2mm]
\pm \tet{e}^{i\Theta_{{\bf k}}}\sqrt{1+4\alpha^2}\\[2mm]
2\alpha\,\tet{e}^{2 i \Theta_{{\bf k}}}\\[2mm]
\tet{e}^{2i\theta}\\[2mm]
\pm \tet{e}^{i \Theta_{{\bf k}}}\sqrt{1+4\alpha^2}\\[2mm]
2\alpha
\end{pmatrix}  = 
\frac{1}{2 r_{[\alpha]}}
\begin{pmatrix}
1\\[2mm]
\sigma \tet{e}^{i\Theta_{{\bf k}}} \, r_{[\alpha]}\\[2mm]
2\alpha\,\tet{e}^{2 i \Theta_{{\bf k}}}\\[2mm]
\tet{e}^{2i\theta}\\[2mm]
\sigma \tet{e}^{i \Theta_{{\bf k}}} \, r_{[\alpha]}\\[2mm]
2\alpha
\end{pmatrix}
\, .
\end{equation}
Once again, we observe that only a subset of the eigenstates depends on the relative hopping parameter $\alpha$. These eigenstates correspond to the energy bands whose dispersions are also $\alpha$-dependent. This behavior is in sharp contrast to the conventional $\alpha-\mc{T}_3$ model.

\section{Polarization function} 
\label{sec3}

The dynamical polarization function, also known as the density-density response function,\,\cite{hwang2007dielectric,wunsch2006dynamical,iurov2022finite,pyatkovskiy2008dynamical,scholz2013plasmons} is one of the fundamental quantities in many-body physics, cooperative phenomena and the response theory. It is defined in the following way  

\begin{equation}
\Pi^{0}({\mathbf q},\omega\,\vert \, \alpha)
=
g_s 
\sum\limits_{\lambda,\lambda'=1}^{6}
\int \frac{d^2 {\bf k}}{(2\pi)^2} \, \mbb{O}_{\lambda,\lambda'}(\mathbf k,\mathbf q)  \, 
\frac{
f[\varepsilon_\lambda(\alpha |\mathbf k)]-f[\varepsilon_{\lambda'}(\alpha | \mathbf k+\mathbf q)]
}{
\hbar\omega  + i 0^ + \varepsilon_\lambda(\alpha | \mathbf k)-\varepsilon_{\lambda'}(\alpha | \mathbf k+\mathbf q)
}  \, , 
\end{equation}
where $\varepsilon_\lambda(\alpha | \mathbf k)$ are the energy dispersions obtained in Eq.~\eqref{mainE}; $f[\varepsilon_\lambda(\alpha |\mathbf k)]$ are the 
Fermi-Dirac distribution functions which become equivalent to the Heaviside function $\Theta(x)$ at zero temperature. 

For six-band $\alpha$-$T_3$ materials, the band indices run over $\lambda,\lambda'=1,\ldots,6$

\begin{equation}
\sum\limits_{\lambda=1}^{6} \Longrightarrow \sum\limits_{\rho=(0,1)} \,\,\,\,\,\,\, \sum\limits_{\sigma=(0, \pm 1)} \ldots \, , 
\end{equation}
equivalently, 

\begin{equation}
\sum\limits_{\lambda,\lambda'=1}^{6} \Longrightarrow \sum\limits_{\rho,\rho'=0,1} \,\,\,\,\,\,\, \sum\limits_{\sigma,\sigma'=(0, \pm 1)} \ldots \, . 
\end{equation}
Since each of the six energy subbands $\varepsilon_\lambda(\alpha | \mathbf k)$ is labeled by two quantum numbers, $\rho = 0,1$ and $\sigma = 0,\pm1$, the summation over all electronic states is performed by summing over all possible combinations of these two indices, which are summarized in the following table:

\par 

\begin{table}[ht]
\centering
\label{tab1}
\begin{tabular}{c| cc}
\toprule
$\lambda$ & $\sigma$ & $\rho$ \\
\hline
1 & 0 & 0 \\
2 & 0 & 1 \\
3 & -1 & 0 \\
4 & 1 & 0 \\
5 & -1 & 1 \\
6 & 1 & 1 \\
\hline
\end{tabular}
\caption{Correspondence of the indices $\lambda=1,\ldots,6$ and $\rho = 0,1$ and $\sigma = 0,\pm1$, the summation over all electronic states.}
\end{table}

\medskip 

There are thirty-six overlap factors,  

\begin{equation}
\mbb{O}_{\lambda,\lambda'}(\mathbf k,\mathbf q)
=
\Big|
\Big \langle
\Psi_\lambda(\mathbf k)
\Big|
\psi_{\lambda'}(\mathbf k+\mathbf q)
\Big \rangle
\Big|^2,
\qquad \lambda,\lambda'=1,\ldots,6 \, , 
\end{equation}

which we now need to calculate.

\par 
Each of these overlap functions is defined simply a dot product of two wave functions $\Big| \Psi_{\lambda} (\mathbf{k}) \Big \rangle  = \Big| \Psi_{\rho}^{\sigma} (\mathbf{k}) \Big \rangle$ and $\Big| \Psi_{\lambda'} (\mathbf{k}) \Big \rangle  = \Big| \Psi_{\rho'}^{\sigma'} (\mathbf{k}+\mathbf{q}) \Big \rangle$
with the wave vectors $\mathbf{k}$ and $\mathbf{k}'=\mathbf{k}+\mathbf{q}$ as well as the band indices $\lambda$ and $\lambda'$,  also referred to the initial and scattered states. The magnitude of vector $\mathbf{k}' = \mathbf{k}+\mathbf{q}$ is $\vert \mathbf{k}+\mathbf{q} \vert =\sqrt{k^2+q^2+2kq\cos\delta_{\mathbf{k},\mathbf{k}'}}$, and we also define an angle $\delta_{\mathbf{k},\mathbf{k}'} =\delta_{\mathbf{k},\mathbf{k}+\mathbf{q}} = \Theta_{\mathbf{k}'}-\Theta_{\mathbf{k}}$ as the angle between $\mathbf{k}$ and $\mathbf{k}'$.

\medskip 

First, we obtain the overlap elements associated with the flat bands 

\begin{equation}
\label{O00001}
\mbb{O}_{[1,1]}({\bf k},{\bf q}) = \mbb{O}_{\rho=0,\sigma=0}^{\rho'=0,\sigma'=0}
=\cos^2 (\delta_{{\bf k},{\bf k}+{\bf q}}) \, 
\end{equation}
The other intraband overlap factors are derived as 

\begin{equation}
\label{O00002}
\mbb{O}_{[2,2]}(\mathbf k,\mathbf q) = \mbb{O}_{\rho=1,\sigma=0}^{\rho'=1,\sigma'=0} = 
\frac{1+4\alpha^2(1+\alpha^2)\cos(\delta_{{\bf k},{\bf k}+{\bf q}})}
{(1+2\alpha^2)^2}.
\end{equation}
since

\begin{eqnarray}
\label{O00003}
\Big \langle\Psi_0^1(\mathbf{k}) \Big| \Psi_0^1(\mathbf{k}+\mathbf{q})\Big \rangle &=&
\Big \langle\Psi_1(\mathbf{k}) \Big| \Psi_1(\mathbf{k}+\mathbf{q})\Big \rangle = 
\frac{1}{1+2\alpha^2}
\Big[
\alpha^2 \tet{e}^{-i \delta_{{\bf k},{\bf k}+{\bf q}}}
+\tet{e}^{i \delta_{{\bf k},{\bf k}+{\bf q}}}
+\alpha^2 \tet{e}^{i\delta_{{\bf k},{\bf k}+{\bf q}}}
\Big] \, = \\
\nonumber 
&& = \, \frac{(1+\alpha^2) \,\tet{e}^{i \delta_{{\bf k},{\bf k}+{\bf q}}}
+\alpha^2 \tet{e}^{-i  \delta_{{\bf k},{\bf k}+{\bf q}}}}
{1+2\alpha^2} \, .
\end{eqnarray}
We now see that for the wave functions $\Big| \Psi_{\rho = 0}^{\sigma = \pm 1} ({\bf k})  \Big \rangle$, the overlap factors exactly resemble those for the Dirac cone states of a dice lattice

\begin{equation}
\label{O00004}
\mbb{O}_{[(3,4),(3,4)]}(\mathbf k,\mathbf q) = \mbb{O}_{\rho=0,\sigma=\pm}^{\rho'=0,\sigma'=\pm 1} = 
\frac{1}{4}\,
\Big[
1+\sigma \sigma' \cos (\delta_{{\bf k},{\bf k}+{\bf q}}) 
\Big]^2 \, .
\end{equation}
Next, we obtain the overlap factors for the most complicated wave functions $\Big| \Psi_{\rho = 1}^{\sigma = \pm 1} ({\bf k})  \Big \rangle$, which must also depend on parameter $\alpha$ of the $\alpha-\mc{T}_3$ lattice

\begin{eqnarray}
\label{O00005}
&& \mbb{O}_{[(5,6),(5,6)]}(\mathbf k,\mathbf q) = \mbb{O}_{\rho=1,\sigma=\pm1}^{\rho'=1,\sigma'=\pm1}
= \frac{\alpha^4}{4\left(1+4\alpha^2\right)^2} \, \times \\
\nonumber 
&& \, \times \, \Big\{
4(1+4\alpha^2)^2 \cos^2(\delta_{{\bf k},{\bf k}+{\bf q}})\left[\sigma_1 \sigma_2 + \cos(\delta_{{\bf k},{\bf k}+{\bf q}})\right]^2
+2\left[3+16\alpha^2+34\alpha^4+\cos(2 \delta_{{\bf k},{\bf k}+{\bf q}})\right]\sin^2(\delta_{{\bf k},{\bf k}+{\bf q}})
\Big\} \, = \\
&& = 2\alpha^4 \left\{
2\Big[\sigma_1 \sigma_2+\cos(\delta_{{\bf k},{\bf k}+{\bf q}})\Big]^2 \, \cos^2(\delta_{{\bf k},{\bf k}+{\bf q}})
+\Big[3+16\alpha^2+34\alpha^4+\cos(2\delta_{{\bf k},{\bf k}+{\bf q}})\Big] \frac{\sin^2(\delta_{{\bf k},{\bf k}+{\bf q}})}{\left(1+4\alpha^2\right)^2}
\right\}
\, .
\end{eqnarray}

\medskip 

The only overlap functions left are those related to interband transitions involving a flat band. One of the simplest results is 

\begin{equation}
\label{O00006}
\mbb{O}_{[1,2]}({\bf k},{\bf q}) = \mbb{O}_{\rho=0,\sigma=0}^{\rho'=1,\sigma'=0}
=\frac{1}{2(1+2\alpha^2)} \, ,
\end{equation}
since it has only one non-zero element in its inner product and $\vert \tet{e}^{- i \Delta_{{\bf k},{\bf k}+{\bf q}}} \vert^2 = 1$. It is obviously expected that such element which doesn't depend on angle delta or angle phi associated with the vector q will not contribute to the polarization function since the corresponding term will vanish under angular integration.

\par 
Now, immediately discern that $ \Big \langle \Psi_{\rho = 0}^{\sigma = 0} ({\bf k}) \Big| \Psi_{\rho = 0}^{\sigma = \pm 1} ({\bf k}) \Big \rangle  = 0$ and 

\begin{equation}
\label{O00007}
\mbb{O}_{[1,(3,4)]}({\bf k},{\bf q}) = \mbb{O}_{\rho=0,\sigma=0}^{\rho'=0,\sigma'=\pm1} = 0 \,  
\end{equation}
and

\begin{equation}
\label{O00008}
\mbb{O}_{[1,(5,6)]}({\bf k},{\bf q}) = \mbb{O}_{\rho=0,\sigma=0}^{\rho'=1,\sigma'=\pm 1} = \frac{\alpha^2}{1+ 4 \alpha^2} \, \big[ 1 - \cos(2 \Delta_{{\bf k},{\bf k}+{\bf q}})
\big] = \frac{2 \alpha^2}{1+ 4 \alpha^2} \,  \sin^2(\Delta_{{\bf k},{\bf k}+{\bf q}})
 \, . 
\end{equation}
Next, we calculate the inter-band overlap functions for the second  degenerate flat band

\begin{equation}
\label{O00009}
\mbb{O}_{[2,(3,4)]}({\bf k},{\bf q}) = \mbb{O}_{\rho=1,\sigma=0}^{\rho'=0,\sigma'=\pm 1} = \frac{\alpha^2}{1+2\alpha^2}
\sin^2(\delta_{{\bf k},{\bf k}+{\bf q}})
 \, . 
\end{equation}

Finally, the only remaining element is 

\begin{equation}
\label{O000011}
\mbb{O}_{[(3,4),(5,6)]}({\bf k},{\bf q}) = \mbb{O}_{\rho=0,\sigma=\pm1}^{\rho'=1,\sigma'=\pm1} = \frac{\sin^2(\delta_{{\bf k},{\bf k}+{\bf q}})}{4(1+ 4 \alpha^2)}
 \, . 
\end{equation}

\begin{equation}
\Big \langle\Psi_2(\mathbf{k}) \Big| \Psi_{5,6} (\mathbf{k}+\mathbf{q})\Big \rangle= \Big \langle\Psi_1^0(\mathbf{k}) \Big| \Psi_1^{\pm 1}1(\mathbf{k}+\mathbf{q})\Big \rangle
\frac{\alpha e^{i\Theta_\mathbf{k}}}
{2\sqrt{(1+2\alpha^2)(1+4\alpha^2)}}
\left(1-\tet{e}^{2 i \delta_{{\bf k},{\bf k}+{\bf q}}} \right).
\end{equation}

Using the identity $\left\vert 1-\tet{e}^{2 i \delta_{{\bf k},{\bf k}+{\bf q}}}\right \vert^2=4\sin^2(\delta_{{\bf k},{\bf k}+{\bf q}})$, we obtain the overlap factor
\begin{equation}
\label{O000012}
 \mbb{O}_{[2,(5,6)]}({\bf k},{\bf q}) = \mbb{O}_{1,\sigma = 0}^{\rho'=1,\sigma' = \pm1} =
\frac{\alpha^2}
{(1+2\alpha^2)(1+4\alpha^2)}
 \sin^2(\delta_{{\bf k},{\bf k}+{\bf q}}) \, . 
\end{equation}

Finally, we derive

\begin{equation}
\label{O000013}
\mbb{O}_{[(3,4),(5,6)]}({\bf k},{\bf q}) = \mbb{O}_{\rho = 0,\sigma = \pm1}^{\rho = 1,\sigma' = \pm1} = \frac{1}{4 \left(1+ 4 \alpha^2\right)} \, \sin^2(\delta_{{\bf k},{\bf k}+{\bf q}})
 \, , 
\end{equation}
which completes our calculation for all possible 36 wavefunction overlap factors, corresponding to all possible electron transitions between the valence, conduction, and flat bands for a Keck-alpha model.

\medskip

An important question is whether all band indices should be included in the summation of the polarization function. At first glance, the situation in Kek-Y graphene \,\cite{herrera2020dynamic} appears in contradiction with the previous results for silicene, \,\cite{tabert2014dynamical} where only a single band index is included and transitions between the two conduction or valence subbands are absent. As a result, the polarization function of silicene can be expressed as the sum of two independent polarization functions corresponding to gapped graphene with different band gaps. 

\par 
 To clarify this issue and establish the general form of the polarization function, we derive it directly from the single-particle Green's function $G(\mathbf{k},i\omega_n)$:

\begin{equation}
\label{Pi0GdefN9}
\Pi(\mathbf{q},i\Omega_m)
=
-gT
\sum_n
\int
\frac{d^2k}{(2\pi)^2}
\,
\mathrm{Tr}
\left[
G(\mathbf{k},i\omega_n)
G(\mathbf{k+q},i\omega_n+i\Omega_m)
\right] \, ,
\end{equation}
 Expressing the Green's function in terms of the corresponding band projector operators provides a rigorous framework for determining which interband and intraband transitions contribute to the polarization function. Also, we use $\mc{P}_\lambda = |\Psi_\lambda \rangle \langle \Psi_\lambda|$, $\lambda =1,\ldots,6$.
We multiply the projectors $\mc{P}_\lambda$ and $\mc{P}_{\lambda'} $to obtain

\begin{eqnarray}
\mc{P}_\lambda \mc{P}_{\lambda'} && = \Big( |\Psi_\lambda\rangle \langle \Psi_\lambda| \Big)
\times \Big( |\Psi_\lambda\rangle \langle \Psi_{\lambda'} | \Big) = |\Psi_\lambda\rangle \,
\langle \Psi_\lambda|\Psi_{\lambda'}\rangle \, 
\langle \Psi_{\lambda'}| \, .
\end{eqnarray}
Using a well-known expression $\text{Tr}
\left(
|a\rangle\langle b|
\right)
=
\langle b|a\rangle$, we finally derive $\mc{P}_\lambda \mc{P}_{\lambda'} = \langle \Psi_\lambda |\Psi_{\lambda'}\rangle \langle \Psi_{\lambda'}|\Psi_\lambda\rangle = \Big| \langle \Psi_\lambda|\Psi_{\lambda'}\rangle \Big|^2$.

\par 
Therefore, all 36 overlap functions $\mbb{O}_{\lambda,\lambda'}(\mathbf k,\mathbf q)$ must be included in the calculation of the polarization function. In particular, interband transitions between the two Dirac cones are allowed and provide a finite contribution to the electronic response. Nevertheless, the calculation is considerably simplified because the Hamiltonian has a block-diagonal structure, causing many of the corresponding projector matrix elements to vanish. 

\medskip 
Several representative diagonal and off-diagonal projector operators for the Kek-$\alpha$ model are derived in Appendix \ref{apa}. We also demonstrate that the trace of the product of two projector operators is exactly equal to the corresponding wave-function overlap factors. Now let's see how this calculation plays out for graphene Hamiltonian, Green's functions, and projector operators. For the wave functions given by equations , the projector operators are obtained by evaluating the outer product directly. Thus, for $s=\pm1$ we obtain 

\begin{equation}
\mc{P}_s (\vec{\mathbf{k}}) = \vert \Psi_{s} (\vec{\mathbf{k}}) \rangle  \otimes \langle  \Psi_{s} (\vec{\mathbf{k}}) \vert
= \frac{1}{2}
\begin{bmatrix}
1 & s\,\tet{e}^{-i\Theta_{\vec{\mathbf{k}}}} \\
s\,\tet{e}^{-i\Theta_{\vec{\mathbf{k}}}} & 1
\end{bmatrix} = \frac{1}{2}
\left[
1+s\,\vec{\Sigma}^{(2)} \cdot \vec{\mathbf{k}}
\right] \, , 
\end{equation}
since 

\begin{equation}
\vec{\Sigma}^{(2)} \cdot \vec{\mathbf{k}} = \begin{bmatrix}
0 & \tet{e}^{-i\Theta_{\vec{\mathbf{k}}}} \\
\tet{e}^{-i\Theta_{\vec{\mathbf{k}}}} & 0
\end{bmatrix} \, . 
\end{equation}

Similarly, 

\begin{equation}
\mc{P}_{s'} (\vec{\mathbf{k}}+\vec{\mathbf{q}})  = 
\begin{pmatrix}
1 & s'\tet{e}^{-i\phi_{\,\vec{\mathbf{k}}+\vec{\mathbf{q}}}} \\
s'\tet{e}^{i\phi_{\vec{\,\mathbf{k}}+\vec{\mathbf{q}}}} & 1
\end{pmatrix}
\end{equation}

The product of and is evaluated as 

\begin{equation}
\label{PPProdNC9}
\mc{P}_s (\vec{\mathbf{k}}) \, \mc{P}_{s'} (\vec{\mathbf{k}}+\vec{\mathbf{q}}) 
= \frac{1}{4} \, 
\begin{pmatrix}
1+s\,s' \tet{e}^{-i(\Theta_{\vec{\mathbf{k}}}-\Theta_{\vec{\,\mathbf{k}}+\vec{\mathbf{q}}})}
&
s\,\tet{e}^{-i \Theta_{\vec{\mathbf{k}}} }+s' \tet{e}^{-i \Theta_{\vec{\,\mathbf{k}}+\vec{\mathbf{q}}}}
\\[1.5mm]
s\,\tet{e}^{i \Theta_{\vec{\mathbf{k}}}} +s'\tet{e}^{i \Theta_{\vec{\,\mathbf{k}}+\vec{\mathbf{q}}}}
&
1+s\,s' \tet{e}^{i(\Theta_{\vec{\mathbf{k}}}-\Theta_{\vec{\,\mathbf{k}}+\vec{\mathbf{q}}})}
\end{pmatrix}.
\end{equation}

Trace of matrix \eqref{PPProdNC9} is immediately obtained as $\text{Tr} \left[ \mc{P}_s (\vec{\mathbf{k}}) \, \mc{P}_{s'} (\vec{\mathbf{k}}+\vec{\mathbf{q}})  \right] = \frac{1}{2} \Big| 1+s\,s' \tet{e}^{i(\Theta_{\vec{\mathbf{k}}}-\Theta_{\vec{\,\mathbf{k}}+\vec{\mathbf{q}}})} \Big|^2 $ $=  \langle  \Psi_{s} (\vec{\mathbf{k}})  \vert \Psi_{s} (\vec{\mathbf{k}}+\vec{\mathbf{q}})  \rangle $. 

The Green function becomes

\begin{equation}
G(\vec{\mathbf{k}},z)
=
\sum_{s=\pm 1}
\frac{
\mc{P}_s (\vec{\mathbf{k}}) 
}{
z-s v_F k
} = \frac{
\mc{P}_+ (\vec{\mathbf{k}}) 
}{
z- v_F k
} + \frac{
\mc{P}_- (\vec{\mathbf{k}}) 
}{
z+ v_F k
} =  \frac{1}{z^{2}-(v_F k)^2}
\begin{bmatrix}
z & v_F k e^{-i\phi} \\
v_F k e^{i\phi} & z
\end{bmatrix} \, , 
\end{equation}
which is in accordance with the well-known apporach to calculating the polarization funciton through the wave function overlaps.

\begin{figure} 
\centering
\includegraphics[width=0.75\textwidth]{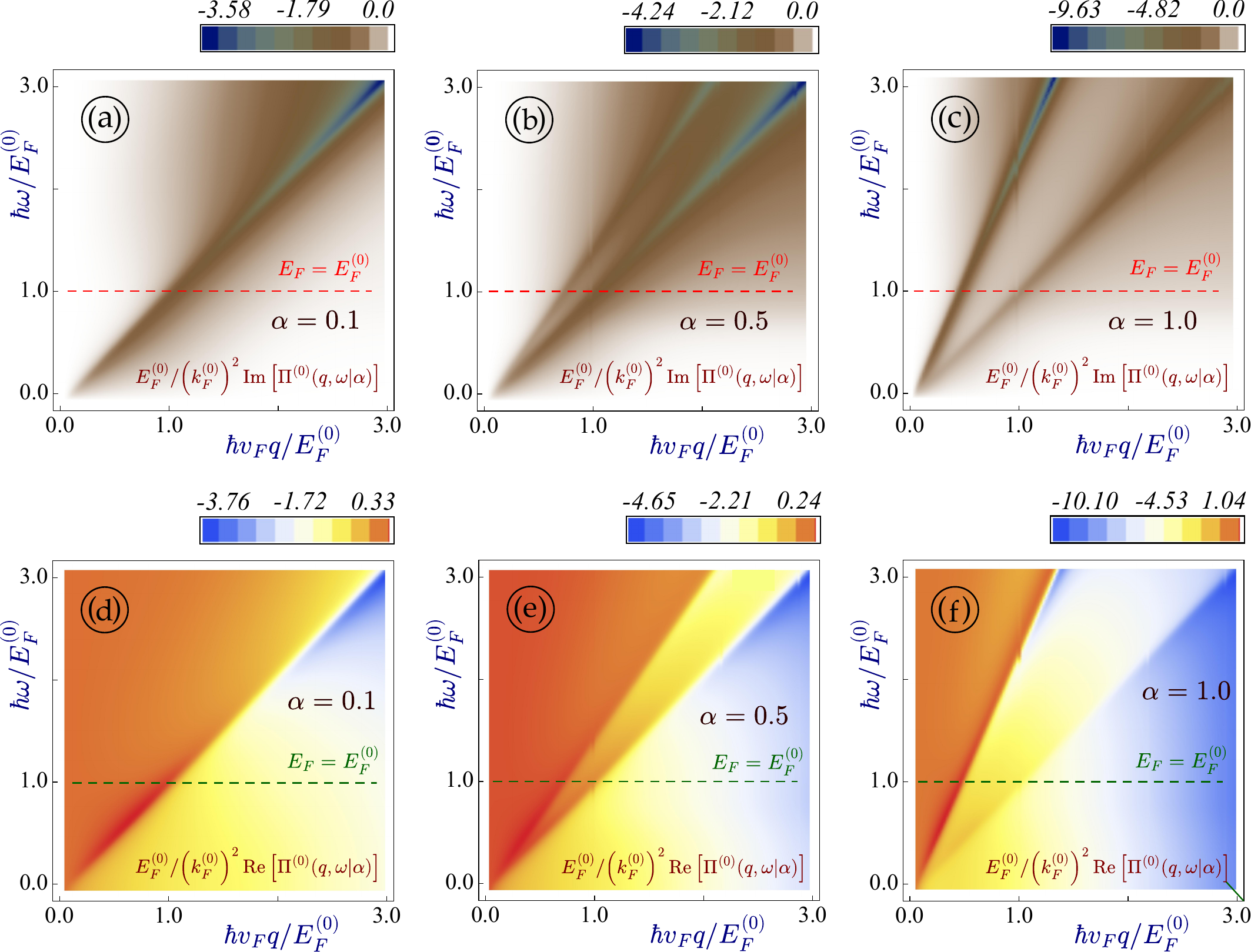}
\caption{(Color online) Numerically calculated dynamical polarization function $\Pi^{(0)}(q,\omega \, \vert \, \alpha)$ for a Kek-$\alpha$ strained lattice. Here, each of upper panels  $(a)$, $(b)$ and $(c)$ displays the imaginary part of $\Pi^{(0)}(q,\omega \, \vert \, \alpha)$ and the particle-hole modes (or the single-particle excitation regions), corresponding to a finite imaginary part of the polarization function.  The three lower plots represent the real part of the polarization function of the Kek-$\alpha$ model which determines the shape of a plasmon branch. Left panels $(a)$ and $(d)$, middle panels $(b)$ and $(e)$, and right ones $(c)$ and $(f)$ correspond to the different value of parameter $\alpha$: $\alpha =0.1$ (nearly graphene), $0.5$, and $1$ (a dice lattice).
}
\label{FIG:1}
\end{figure}
\medskip

\medskip 
Now we are in a position to discuss our numerical results for the polarization functions and plasmon excitations. The real and imaginary parts of the polarization function for the  Kek-$\alpha$ model are presented in Fig.~\ref{FIG:1}. The imaginary part of the polarization function $\Pi^{(0)}(q,\omega \vert \alpha)$ and, specifically, the regions where it is non-zero, determine the so-called particle-hole continuum, or the region in which plasmons decay into single-particle excitations. We observe that, for different values of $\alpha$, this region occupies a significant portion of the considered frequency and wave vector ranges. First, transitions to and from the flat bands are allowed, giving rise to particle-hole modes for most frequencies above the Fermi energy $\hbar \omega = E_F^{(0)}$. The only two regions free of particle-hole excitations are the triangular region above the $\omega = v_F \, \sqrt{1+4 \alpha^2} \, q$-line located higher than the main diagonal ($\omega =  v_F \, q$) and below the Fermi energy,  for larger values of $q$. 

\par 
However, plasmons can exist only above the diagonal and for small wave vectors. Consequently, the regions in which plasmons may exist are substantially reduced and become even smaller as $\alpha$ increases. We further observe two distinct lines $\omega = v_F\, \left(1+4 \alpha^2\right)^{\rho/2} \, q$, $\rho = (0,1)$, at which both the real and imaginary parts of the polarization function reach their maximum values. An interesting phenomenon is observed as the relative hopping parameter $\alpha$ varies. For small values of $\alpha$ ($\alpha <0.5$), the main diagonal $(\hbar v_F)  \, q$ dominates and exhibits the largest magnitude of the imaginary part of the polarization function. In contrast, for larger values of $\alpha$, approaching the dice-lattice limit, the higher off-diagonal line, corresponding to the fast Dirac cone, becomes dominant and exceeds the main diagonal in magnitude of the $\Pi^{(0)}(q,\omega \vert \alpha)$. To the best of our knowledge, such behavior has not been reported for any previously studied Dirac material.

\par 
The real part of the polarization function, presented in the lower panels $(d)$, $(e)$, and $(f)$, reveals the shape of the plasmon branch for each $\alpha$. The overall behavior is rather standard, with large positive values appearing above the diagonal, which allows for the existence of plasmons (a zero of the dielectric function). However, we observe a sudden drop in the real part between the main diagonal $(\hbar v_F) \, q$ and the line associated with the larger Fermi velocity $(\hbar v_F) \, \left(1+4 \alpha^2\right)^{\rho/2} \, q$ of the fast Dirac cone, which, in principle, could significantly modify the plasmon dispersion. Nevertheless, in this region the imaginary part of the polarization function is large, meaning that the Landau damping is strong, and the plasmon mode is generally suppressed in this area.

\begin{figure} 
\centering
\includegraphics[width=0.55\textwidth]{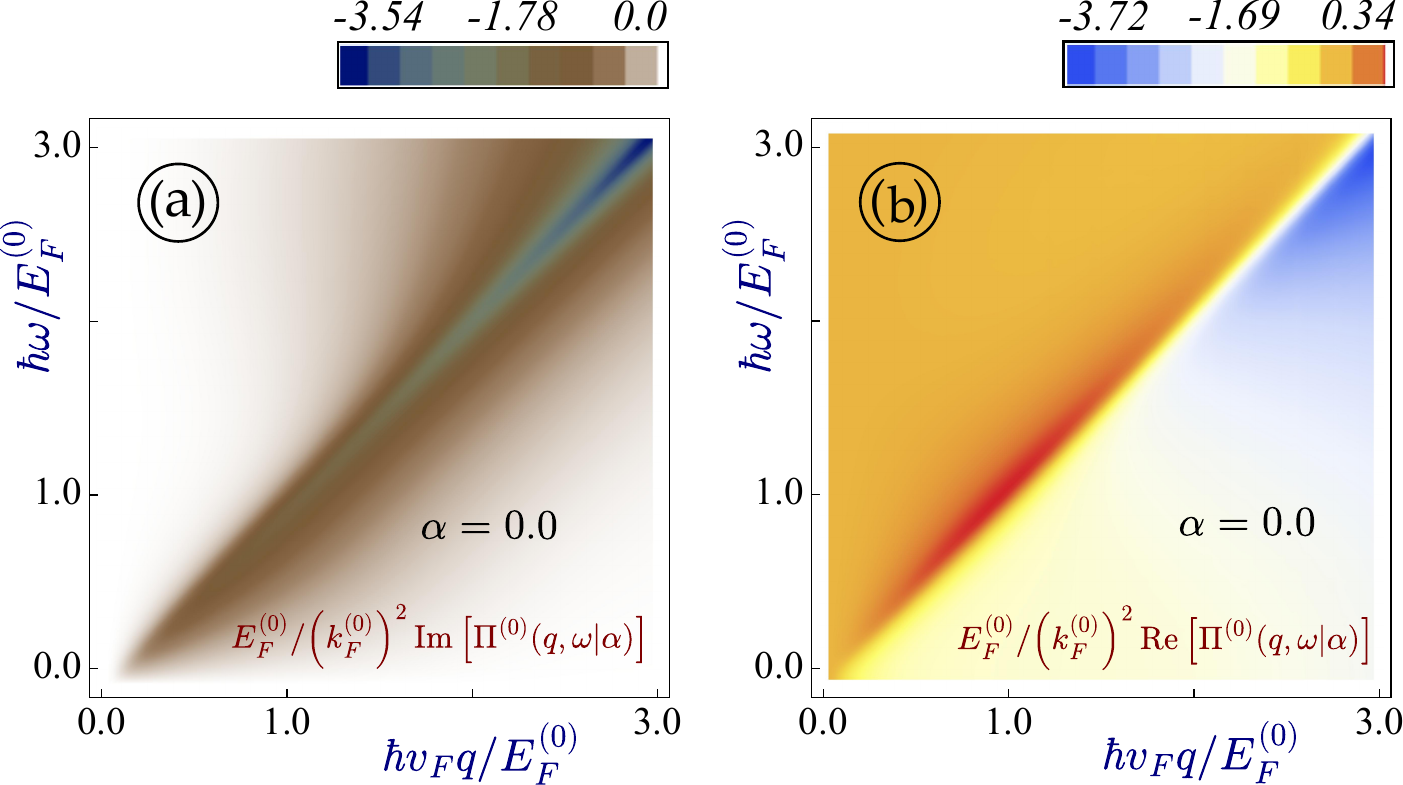}
\caption{(Color online) (Color online) Numerically calculated dynamical polarization function $\Pi^{(0)}(q,\omega \, \vert \, \alpha=0)$ for graphene, corresponding to the $\alpha \longrightarrow 0$ limit of the Kek-$\alpha$ strained lattice. Here, each of upper panels  $(a)$, $(b)$ and $(c)$ displays the imaginary part of $\Pi^{(0)}(q,\omega)$ and the particle-hole modes (or the single-particle excitation regions), corresponding to a finite imaginary part of the polarization function. The three lower plots represent the real part of the polarization function of the Kek-$\alpha$ model which determines the shape of a plasmon branch $\Pi^{(0)}(q,\omega)$.}
\label{FIG:100}
\end{figure}

Next, we consider the graphene limit $\alpha=0$, and present the corresponding real and imaginary parts of the dynamical polarization function $\Pi^{(0)}(q,\omega \, \vert \, \alpha=0)$ in Fig.~\ref{FIG:100}. It is important to emphasize that this limit is not trivial: the wave functions and overlap factors obtained as the eigenfunctions of the Hamiltonian \eqref{grlim0012} for $\alpha \longrightarrow 0$ are not identical to those obtained by simply taking the $\alpha \longrightarrow 0$-limit of the general expressions \eqref{psm1}-\eqref{psm6} derived for arbitrary $\alpha$. Therefore, this limiting case requires a separate verification.

\par 
In both approaches, we recover the well-known dynamical polarization function of graphene. The imaginary part exhibits the standard particle-hole continuum located below the main diagonal  $ v_F q$, together with the characteristic boundary at $\hbar \omega=2\mu-\hbar v_F q$. The real part displays a strong peak along the main diagonal, with large positive values above it and smaller values below, which once again result in a solution of the dielectric function equal to zero and the existence of a plasmon. Therefore, complete agreement for $\alpha=0$ with the known graphene has been confirmed. Our results for the wave functions and overlap function for the graphene ($\alpha = 0$) and dice lattice ($\alpha = 1$) limits of Kek-$\alpha$ model are presented in Appendices \ref{apb} and \ref{apc}.

\section{Dielectric function, plasmon dispersions and the loss function} 
\label{sec4}

As already discussed, the real part of the dynamical polarization function $\Pi^{(0)}(q,\omega \vert \alpha)$ plays the leading role
in determining the shape and exact location of the plasmon branches determined as the zeros of the dielectric
function $\epsilon(q,\omega\mid \phi,\lambda_0)$ which is in turn defined as
\begin{equation}
\label{eps011}
\epsilon(q,\omega \vert \alpha) =
1-v_C(q)\Pi^{(0)}(q,\omega \vert \alpha)\, ,
\end{equation}
where
\begin{equation}
v_C(q)=\frac{e^2}{2\epsilon_0\epsilon_r q} = \frac{\beta_r e^2}{q}
\end{equation}
is the two-dimensional Coulomb potential with $\epsilon_r$ as the
host-material dielectric constant. Here, we also introduced a relative dielectric constant $\beta_r = 1/(2 \epsilon_0\epsilon_r)$ which in fact plays the role of the inverse dielectric constant: a large $\epsilon_r$ corresponds to the smaller, reduced value of $\beta_r$.

\medskip 
In the random-phase approximation (RPA), the dielectric function in
equation is utilized to obtain Coulomb-affected polarization
function $\Pi_{\mathrm{RPA}}(q,\omega \vert \alpha)$, i.e.,
\begin{equation}
\Pi_{\text{RPA}}(q,\omega \vert \alpha) =
\frac{\Pi^{(0)}((q,\omega \vert \alpha))}
{\epsilon(q,\omega \vert \alpha)},
\end{equation}
where the plasmon dispersion, determined by
$\text{Re}\,[\epsilon(q,\omega\vert \alpha)]=0$, is shown as a peak in the density plot of
$\Pi_{\text{RPA}}(q,\omega \vert \alpha)$ as functions of both wave vector $q$ and plasmon frequency/energy $\omega$.

\begin{figure} 
\centering
\includegraphics[width=0.75\textwidth]{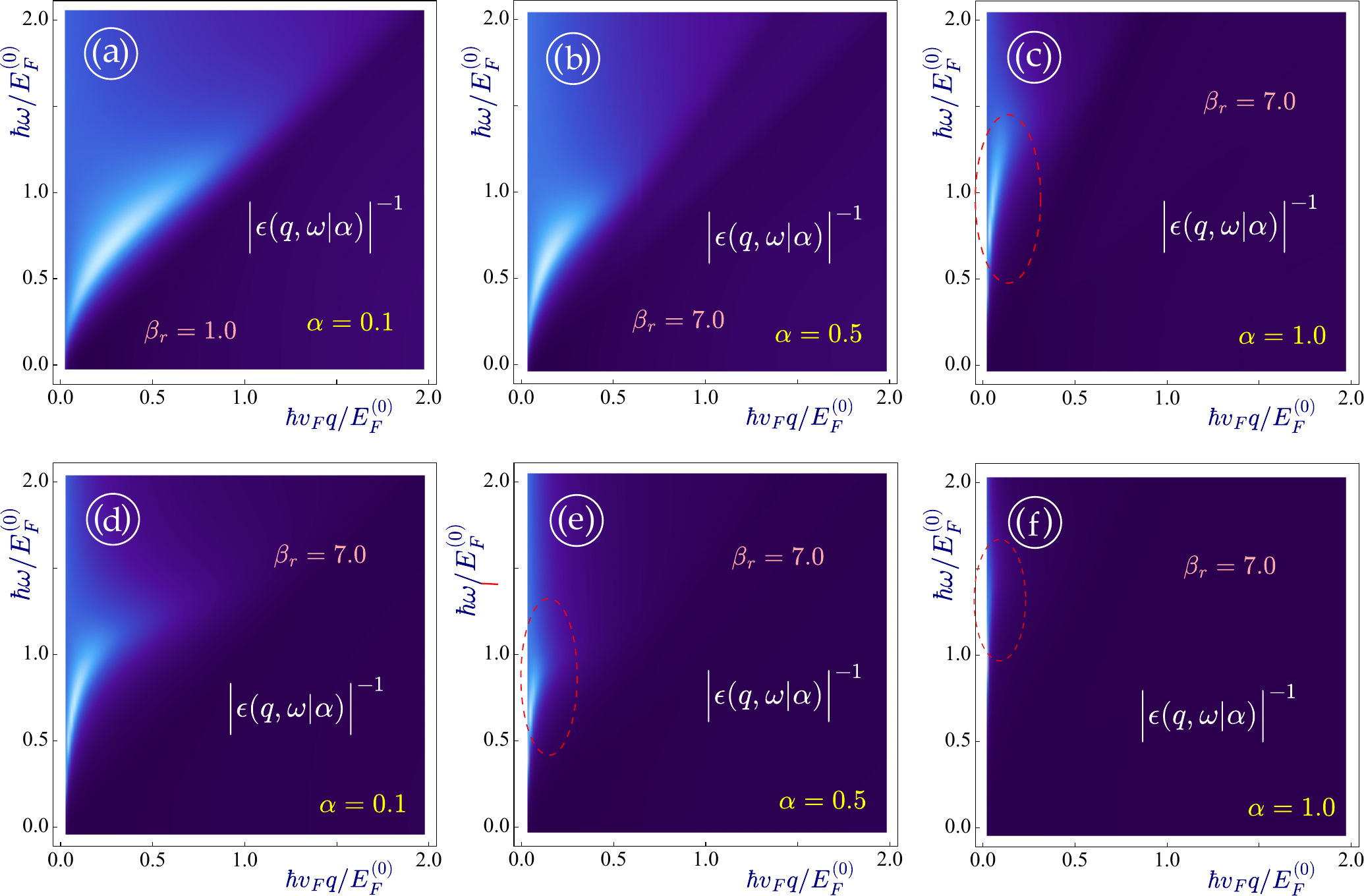}
\caption{
(Color online) Numerically calculated inverse dielectric function $\Big| \epsilon(q,\omega \vert \alpha)\Big|^{-1}$ for a for a Kek-$\alpha$ strained lattice so that the plasmon dispersion (for an undamped or low-dapmed plasmon mode) corresponds to a peak in the inverse dielectric function (ideally, a vanishing real and imaginary parts of the dielectric function $\text{Re}\left[ \epsilon(q,\omega \vert \alpha) \right] \longrightarrow 0$ and $\text{Im}\left[ \epsilon(q,\omega \vert \alpha) \right] \longrightarrow 0$. The left panels $(a)$ and $(d)$, middle panels $(b)$ and $(e)$, and right ones $(c)$ and $(f)$ correspond to the different value of parameter $\alpha$: $\alpha =0.1$ (nearly graphene), $0.5$, and $1$ (a dice lattice). The upper panels represent the situation for a relative dielectric constant $\beta_r =1.0$, and the lower ones - for $\beta_r =7.0$.
}
\label{FIG:2}
\end{figure}
\medskip

The plasmon dispersions for Kekule-modulated materials with different parameters $\alpha$ are presented in Fig.~\ref{FIG:2}. For $\alpha=0.1$ (nearly graphene), the plasmon behavior is expectedly very similar to the earlier discussed case of graphene, and we observe the standard undamped plasmon mode located above the main diagonal $\omega = (\hbar v_F) \,q$. In this case, the separation between the two Dirac cones is small, and the electronic transitions between them do not significantly affect the plasmon dispersion. As $\alpha$ increases, the separation between the Dirac cones becomes more noticeable, and the undamped plasmon region is restricted to the area above the fast Dirac cone dispersion $\omega \geq v_F \, q$. Consequently, the available phase space for weakly damped plasmons is substantially reduced. The plasmon mode becomes increasingly suppressed and can be observed only for small wave vectors $q  \ll k_F^{(0)}$. Finally, for $\alpha=1$, corresponding to the dice-lattice limit, only a very narrow range of undamped plasmon excitations survives (see panels $(c)$ and $(f)$ of Fig.~\ref{FIG:2}).

\par
The lower panels are related to a larger relative dielectric constant $\beta_r$, which shifts the plasmon dispersion to higher frequencies. However, for sufficiently large values of $\alpha$, the plasmon mode becomes strongly limited in the $q-\omega$-space and could barely be observed. 

\begin{figure} 
\centering
\includegraphics[width=0.6\textwidth]{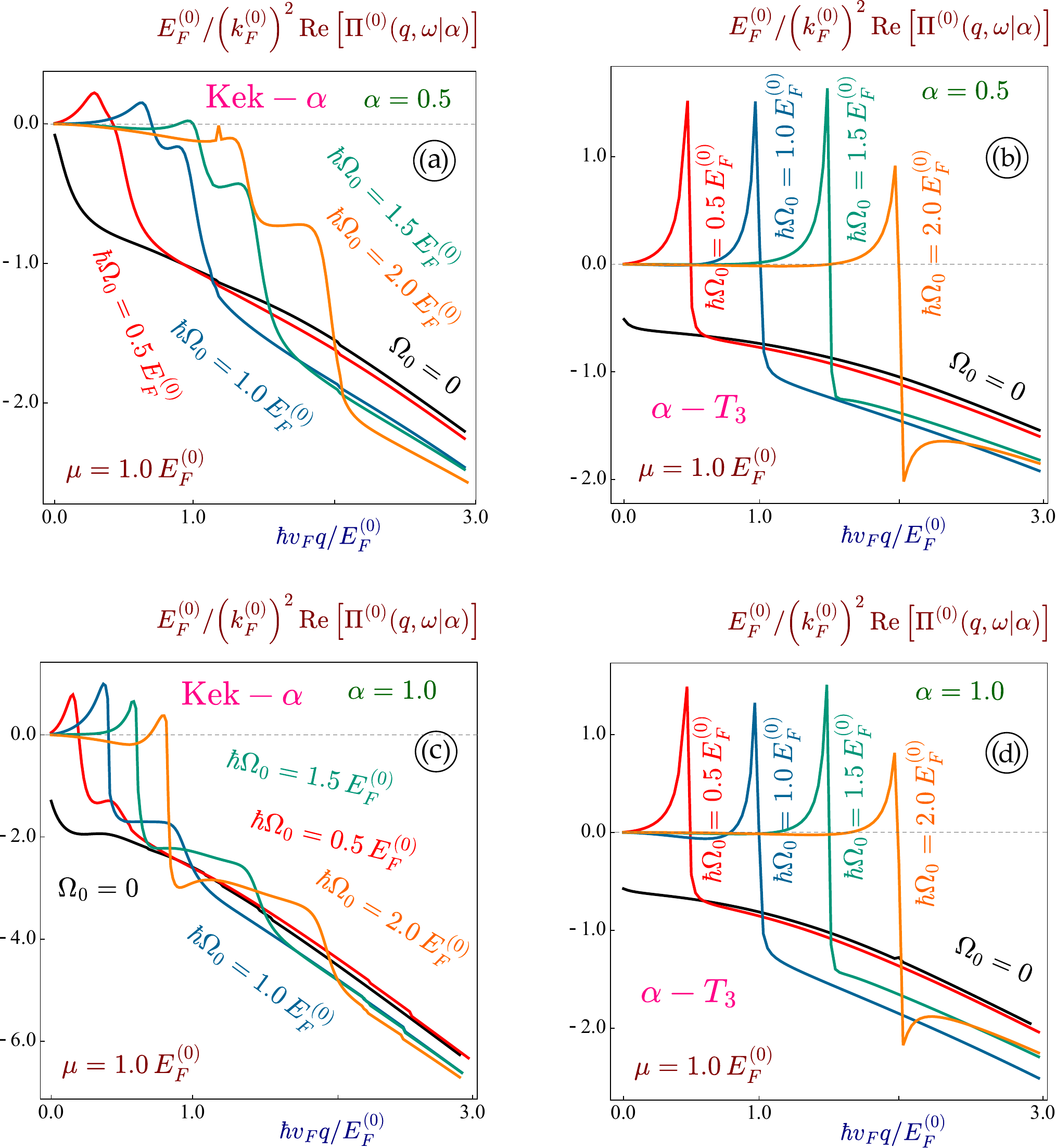}
\caption{
(Color online) Constant-frequency cuts to the real part of the polarization function $\Pi^{(0)}(q,\omega \, \vert \, \alpha=0)$ for a 
Kek-$\alpha$ model (left panels $(a)$ and $(c)$) and unstrained $\alpha-\mc{T}_3$ model as a function of wave vector $q$. The two upper plots  $(a)$ and $(b)$ describe the situation for $\alpha = 0.5$, and the two lower ones  $(c)$ and $(d)$ - for a dice lattice with $\alpha = 1.0$. 
Each curve is related to a specific fixed frequency $\omega = 0.5\,E_F^{(0)}/\hbar$, $1.0\,E_F^{(0)}/\hbar$, $1.5\,E_F^{(0)}/\hbar$, $2.0\,E_F^{(0)}/\hbar$ and $2.5\,E_F^{(0)}/\hbar$ for all panels, as labeled.
}
\label{FIG:4}
\end{figure}

We also discuss the behavior of the constant frequency-cuts of the real part of the polarization function $\text{Re}\left[\Pi^{(0)}(q,\omega \vert \alpha)\right]$for the Kekule-modulated $\alpha-\mc{T}_3$ and the conventional unstrained $\alpha-\mc{T}_3$ model, shown in Fig.~\ref{FIG:4}. First, we note that both models exhibit positive peaks of the real part of the polarization function. However, for the Kek-$\alpha$ model, these peaks are less pronounced, and we observe two closely located peaks corresponding to the two different Fermi velocities of the fast and slow Dirac cones. As the frequency increases, they evolve into a step-like structure. The zero-frequency (static) limit of the polarization function is qualitatively similar in all cases, exhibiting a monotonic decrease with increasing wave vector $q$.

\par 
For $\hbar \omega=2E_F^{(0)}$, the Kek-$\alpha$ model displays two distinct positive peaks, whereas the conventional unstrained $\alpha-\mc{T}_3$ exhibits one positive and one negative peak. In general, the Kekule-modulated materials demonstrate more chaotic and less similar frequency dependence, with qualitatively different behavior for different frequencies. The peaks are located at substantially smaller values of the wave vector than in the conventional unstrained model. This two-peak structure is a direct consequence of the existence of two non-equivalent Dirac cones with different Fermi velocities. At the same time, for a regular $\alpha-\mc{T}_3$ lattice, all the peaks are located at the main diagonal $\omega = v_F q$.

\begin{figure} 
\centering
\includegraphics[width=0.6\textwidth]{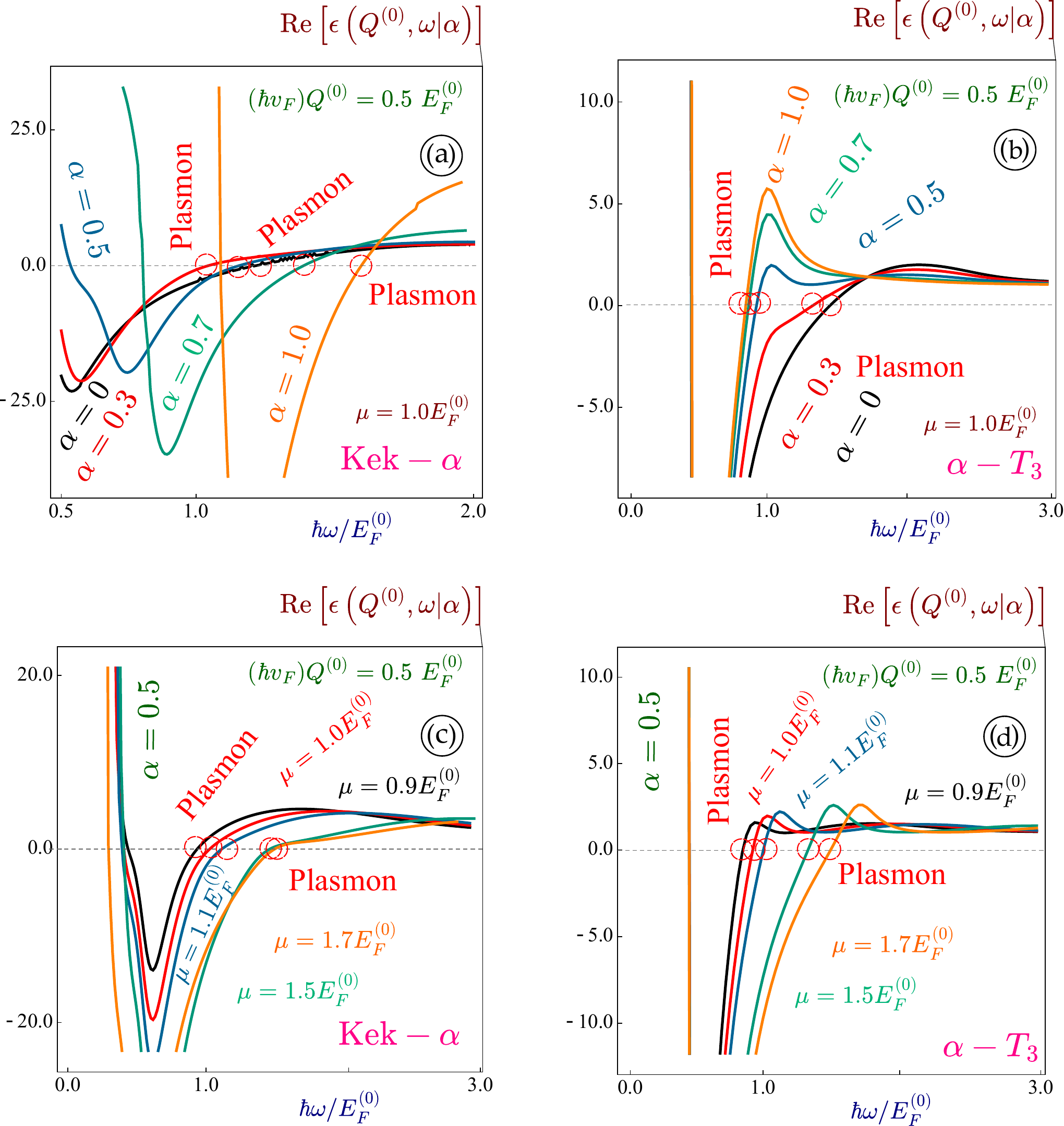}
\caption{
(Color online) Dielectric function, $\text{Re}\left[ \epsilon(q,\omega \vert \alpha) \right]$, and the corresponding plasmon frequencies (points where each curve crosses zero line) for the Kek-modulated $\alpha$-$\mathcal{T}_3$ model (left panels $(a)$ and $(c)$) and the regular unstrained $\alpha$-$\mathcal{T}_3$ model (right panels $(b)$ and $(d)$). In the upper panels $(a)$ and $(b)$ the chemical potential (Fermi energgy) is fixed at $\mu=1.0 E_F^{(0)}$, corresponding to the location of the flat band, while the relative hopping parameter $\alpha$ is varied as $\alpha=0$ (graphene), $0.3$, $0.5$, $0.7$, and $1.0$ (a dice lattice). In the lower panels $(c)$ and $(d)$, the hopping parameter is fixed at $\alpha=0.5$, and the curves correspond to different values of the chemical potential, $\mu=0.9 E_F^{(0)} $, $1.0E_F^{(0)}$ (flat-band location), $1.5 E_F^{(0)}$, $1.7 E_F^{(0)}$, and $2.0 E_F^{(0)}$. In all calculations, the wave vector is fixed at $Q^{(0)}=0.5E_F^{(0)}/(\hbar v_F)$.
}
\label{FIG:5}
\end{figure}
\medskip

As a next step, we investigate how the plasmon frequency and the location of the plasmon branches depend on the parameter $\alpha$ for both the Kek-$\alpha$ model and the unstrained $\alpha-\mc{T}_3$ model. To this end, we plot the real part of the dielectric function $\text{Re}\left[ \epsilon(q,\omega \vert \alpha) \right]$ in Fig.~\ref{FIG:5} and identify the plasmon frequency for a fixed wave vector $Q^{(0)}$ as the point at which the dielectric function crosses zero.
\par
We observe a tremendous difference between the Kek-$\alpha$ and unstrained $\alpha-\mc{T}_3$ models. First, for the regular $\alpha-\mc{T}_3$ materials, the plasmon frequency decreases monotonically with increasing $\alpha$,  implying that each plasmon branch shifts to lower frequencies as $\alpha$ increases. Consequently, the lowest plasmon branch corresponds to the dice lattice $\alpha=1$, while the highest one is obtained for graphene $\alpha=0$. In contrast, for the Kek-$\alpha$ model, this dependence is non-monotonic. The plasmon frequency initially increases with $\alpha$ and subsequently exhibits a significant decrease for larger values of $\alpha$.

\par
Another crucial feature of the regular $\alpha-\mc{T}_3$ model is that all curves shown in panel $(b)$ intersect at a single point, which is analogous to the well-known pinching effect, where different plasmon branches meet at the same point. This behavior is clearly present in the unstrained $\alpha-\mc{T}_3$ model but is absent in the Kek-$\alpha$ model.

\par 
Next, we examine the dependence of the plasmon branch frequency on the chemical potential for a fixed value of $\alpha$. In both models, we observe a monotonic increase of the plasmon frequency with increasing chemical potential. However, for the Kek-$\alpha$ model, this dependence is considerably more nonlinear, exhibiting a weaker variation at small chemical potentials, followed by a more rapid increase at intermediate values and, finally, a reduced rate at larger chemical potentials.

\begin{figure} 
\centering
\includegraphics[width=0.6\textwidth]{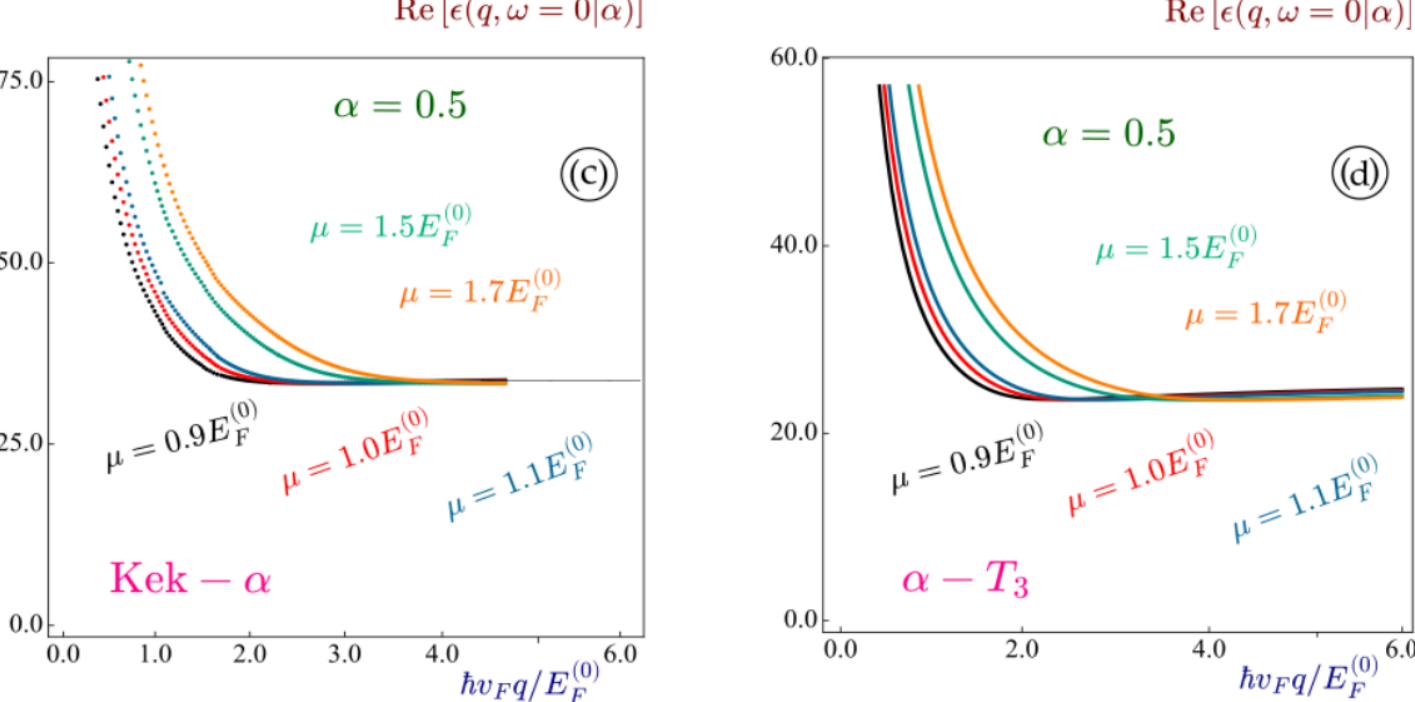}
\caption{
(Color online) Static dielectric function $\epsilon(q,\omega=0 \, \vert \, \Delta_0)$ for the Kek-modulated $\alpha$-$\mathcal{T}_3$ model 
(left panel $(a)$) and the regular unstrained $\alpha$-$\mathcal{T}_3$ model (right panel $(b)$) as a function of wave vector $q$.In both 
panels, each curve corresponds to a fixed Fermi energy (doping level)  $\mu=0.9 E_F^{(0)} $, $1.0E_F^{(0)}$ (flat-band location), 
$1.5 E_F^{(0)}$, $1.7 E_F^{(0)}$, and $2.0 E_F^{(0)}$. In all calculations, the wave vector is fixed at $Q^{(0)}=0.5E_F^{(0)}/(\hbar v_F)$, 
according to our labels.
}
\label{FIG:6}
\end{figure}

\begin{figure} 
\centering
\includegraphics[width=0.6\textwidth]{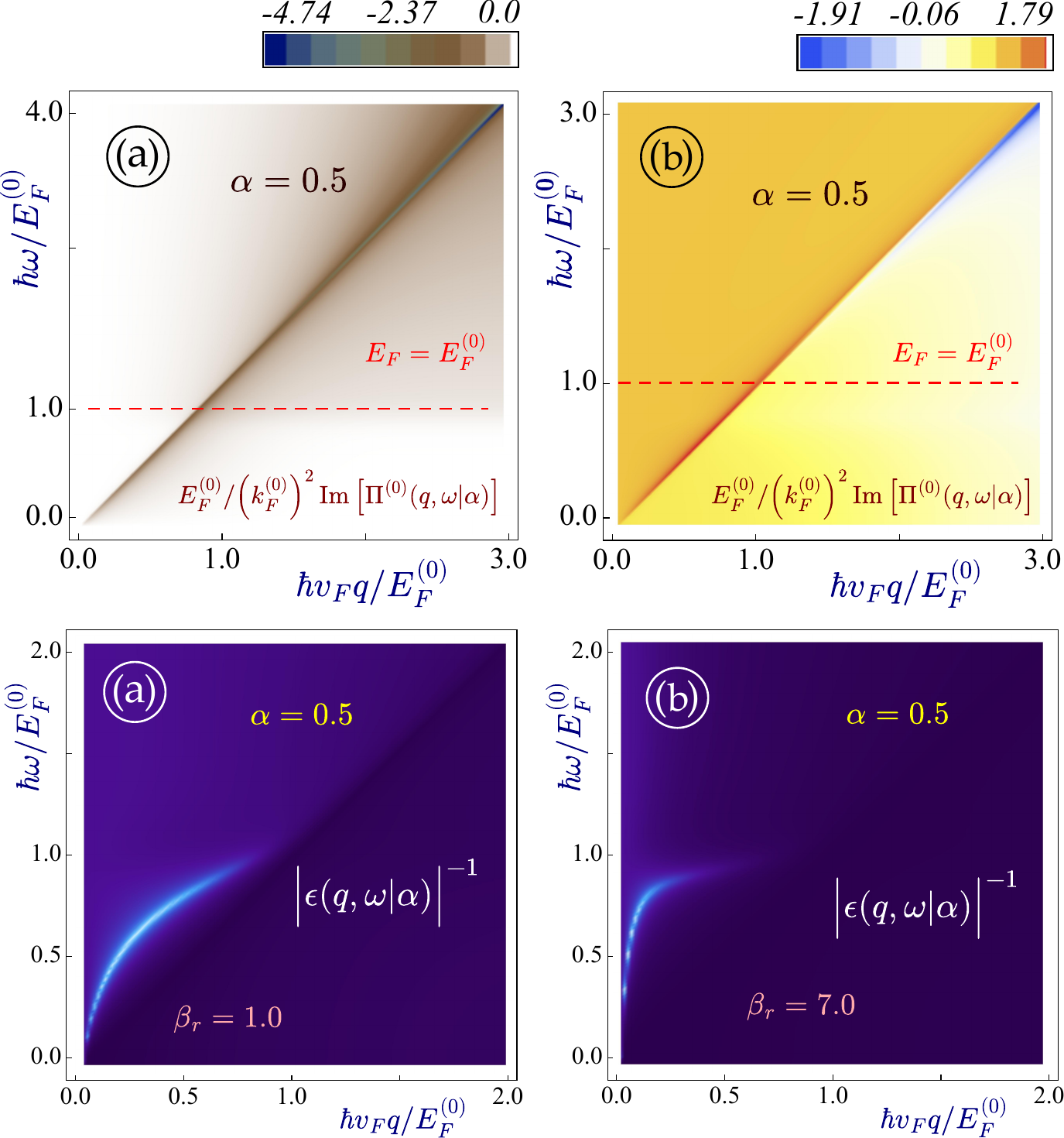}
\caption{(Color online) Numerically calculated dynamical polarization function $\Pi^{(0)}(q,\omega \, \vert \, \alpha)]$ (upper panels  $(a)$ and $(b)$ for its real and imaginary parts, correspondingly) and inverse dielectric function $\Big| \epsilon(q,\omega \vert \alpha)\Big|^{-1}$ (lower panels  $(c)$ and $(d)$ describe the situation for the relative dielectric constant $\beta_r =1.0$ and $\beta_r =7.0$) for a regular (non-strained) $\alpha-\mc{T}_3$ model. 
The relative hopping parameter $\alpha$ was chosen to be $\alpha = 0.5$ for all cases.}
\label{FIG:7}
\end{figure}

To summarize our discussion of the dynamical polarization function and plasmon excitations in various $\alpha-\mc{T}_3$ materials, let us briefly review the known results for the conventional, unstrained $\alpha-\mc{T}_3$ model, shown in Fig.~\ref{FIG:7}. The particle-hole continuum is qualitatively similar to that of graphene. However, an important difference arises from the presence of the flat band: an additional single-particle excitation region emerges due to electronic transitions between the flat band and the states at the Fermi level, and the plasmons experience additional damping. Consequently, undamped plasmons can exist only within the triangular region above the main diagonal $\omega \geq v_F \, q$ and below the Fermi energy $\hbar \omega = E_F^{(0)}$.

\par
The plasmon mode is obtained as the zeros of the real part of the dielectric function $\epsilon(q,\omega \vert \alpha)$. We find that at larger frequencies and wave vectors these plasmon excitations are strongly damped. Another remarkable feature of the conventional $\alpha-\mc{T}_3$ model is the so-called pinching effect: all plasmon branches, corresponding to the different values of the relative dielectric constant $\beta_r$, intersect at a single point $\hbar \omega= ( \hbar v_F) \, q = E_F^{(0)}$, which also belongs to the boundary of the undamped plasmon region.  As the relative dielectric constant increases, the plasmon branches shift to higher frequencies and become damped at smaller wave vectors.

\par
The real part of the dielectric function differs substantially from that of graphene: its shape and location of the plasmon branches remain qualitatively similar; however, their damping is much stronger because of the additional electronic transitions involving the flat band.

\section{Static screening}
\label{sec5}

The static screening and the real-valued Lindhard function (the zero-frequency limit of the polarization function) for the regular $\alpha-\mc{T}_3$  model have been studied well \,\cite{malcolm2016frequency}, and there is a significant difference between the $\alpha-\mc{T}_3$ model and graphene. In both systems, the static polarization function consists of two contributions. One of these terms is relevant for all wave vectors, while the other arises at $2 k_F$. In graphene, the first term is constant, whereas the second term is parabolic and has a finite slope. As a result, the polarization function exhibits a kink, or a discontinuity in its derivative, at $2 k_F$. In contrast, for the $\alpha-\mc{T}_3$  model, the resulting curve is smooth and does not exhibit any breaking point or discontinuity in its derivative. The existence of such a breaking point plays a crucial role in determining the static screening properties and the long-range behavior of the Friedel potential and the screening of a charged impurity.

\medskip 
Let us consider a dilute distribution of charged impurities. The  screened potential due to a dielectric medium in the vicinity of a point charge $Q_0$ is determined by the Fourier transformation of the screened electron-electron interaction, written as \,\cite{sadhukhan2017anisotropic}

\begin{equation}
\Phi(r) = \frac{Q_0}{\alpha_r\epsilon_0} \int_0^\infty d q \,  \frac{J_0(q r)}{\epsilon(q, \omega \, \vert \, \Delta_0)} \ , 
\label{screen}
\end{equation}
where $J_0(q r)$ is the zero-order Bessel function which results from the angular integration of $\tet{exp}[i {\bf q} \cdot {\bf r}]$. As is known for most 2D materials, the $r$-dependence for Eq.\,\eqref{screen} is determined by the Thomas-Fermi decay $\backsim 1/r^3$.  

\par
Therefore, the Friedel oscillations and static screening exhibit substantially different behaviors in graphene and in other $\alpha-\mc{T}_3$ materials. Importantly, this feature remains unchanged in the presence of Kek-$\alpha$ strain. The behavior of the strained system is qualitatively identical to that of the unstrained one. In particular, the asymptotic behavior of the Friedel potential scales as $\backsim 1/r^3$ for graphene and as s $\backsim 1/r^4$ for all other $\alpha-\mc{T}_3$ materials, regardless of the presence of strain.

\par
Our numerical results for the static screening in the Kek-$\alpha$ model are presented in Fig.~\ref{FIG:6}. We observe that the curves are qualitatively very similar to one another and are similar for both strained and unstrained systems. This demonstrates a general feature of static screening in these materials: the presence of strain does not significantly affect its behavior.

\section{Summary and Remarks}
\label{sec6}

In this paper, we have studied the polarization function, plasmon excitations, and their damping in the Kek-$\alpha$ model. This system is formed by introducing atoms with Kekule periodicity on a honeycomb lattice that couple to only one of its sublattices. As a result, it represents a hybrid model of a two-dimensional material that combines the characteristic properties of the $\alpha-\mc{T}_3$ model and Kekule-distorted graphene, and is referred to as the Kek-strained $\alpha-\mc{T}_3$ model.

\medskip 
This atomic structure gives rise to an unusual low-energy band structure which consists of two degenerate flat bands and two inequivalent Dirac cones, commonly referred to as the fast and slow cones. This model combines the features of both Kek-Y graphene, which possesses two Dirac cones, and the $\alpha-\mc{T}_3$  model with its flat bands. However, because the Kek-Y coupling affects only one sublattice, the resulting electronic spectrum is qualitatively different and is determined by a single tunable parameter, $\alpha$. Importantly, the $\alpha\longrightarrow 0$ limit reproduces pristine graphene rather than Kek-Y graphene. The peculiar band structure allows for additional inter-cone transitions between the fast and slow Dirac bands, as well as intervalley transitions between states belonging to different cones and different valleys. Unlike the conventional $\alpha-\mc{T}_3$ model, where $\alpha$ determines only the coupling between the flat and dispersive bands without modifying the energy spectrum,  the electronic dispersions in our model directly depend on the parameter $\alpha$. 

\medskip
We have discovered several unusual features of the particle-hole (Landau damping) modes, plasmon excitations, and plasmon damping in the Kek-$\alpha$ model. First, the particle-hole continuum responsible for plasmon damping consists of two distinct peaks associated with electronic transitions involving the two inequivalent Dirac cones. One peak corresponds to transitions characterized by the smaller Fermi velocity. It follows the main diagonal of the energy-momentum spectrum, while the second appears at higher energies due to the larger Fermi velocity of the fast Dirac cone. Interestingly, as $\alpha$ increases, the higher-energy contribution becomes increasingly dominant.

\par
We have also investigated the real part of the polarization function and demonstrated that this two-peak structure leads to substantial modifications of the plasmon dispersion. However, these plasmon modes predominantly lie within the particle-hole continuum, where they become strongly damped. In general, plasmons are stable and could be observed only within a relatively small region of wave vector and frequency space, partly due to additional transitions involving the two degenerate flat bands, similarly to the conventional $\alpha-\mc{T}_3$  model. Consequently, stable plasmon excitations can exist only within small range of the wave vector, or for a small $\alpha$ close to graphene. Importantly, we find that the existence of the kinks in the static polarizability (Lindhard function) remains unaffected by the presence of Kek-$\alpha$ strain. Consequently, the long-range behavior of the Friedel oscillations is qualitatively the same in both the strained and unstrained $\alpha-\mc{T}_3$ models.

\medskip 
We anticipate that our newly uncovered unusual electronic and collective properties of this model for a wide class of low-dimensional materials will lead to promising device applications in the near future. In particular, the ability to engineer particle-hole modes and suppress plasmons by enhancing the plasmon damping in previously inaccessible regions of the energy-momentum spectrum offers new opportunities for controlling collective excitations, which is crucial for the development of next-generation nanoscale plasmonic and optoelectronic devices.

\acknowledgements
J.M. and A.I. have been supported by the funding received from TRACK1-57-79, PSC-CUNY Award \# GR-00017504. G.G. gratefully acknowledges funding from the U.S. National Aeronautics and Space Administration (NASA) via the NASA-Hunter College Center for Advanced Energy Storage for Space under cooperative agreement 80NSSC24M0177. We also acknowledge support from the Science and Technology Facilities Council (STFC), UK (Reference No. ST/Y005147/1). The views expressed are those of the author and do not necessarily reflect the official policy or position of the Department of
the Air Force, the Department of Defense, or the U.S. government.

\appendix

\section{Derivation of the dynamical polarization function based on the projector operators}
\label{apa}

Although we calculate the polarization function and the wave function overlap factors directly from the wave functions, it is also instructive to demonstrate they could be also calculated equivalently using projection operators. In particular, we derive several of the overlap factors within this formalism. The band projector, or projection operator, is defined as

\begin{equation}
\hat{\mc{P}}_{\lambda,\lambda'}
(\mathbf{k})=\Big| \Psi_\lambda (\mathbf{k}) \Big \rangle  \Big \langle  \Psi_{\lambda'} (\mathbf{k}) \Big| = \Big \vert  \Psi_{\rho}^{\sigma} ({\bf k}) \Big \rangle \Big \langle \Psi_{\rho'}^{\sigma'} ({\bf k}) \Big \vert  \, . 
\end{equation}

By finding such outer product directly, we obtain the following expression for the diagonal elements 

\begin{eqnarray}
\hat{\mc P}_{0,0}(\sigma, {\bf k}) & = & \Big| \Psi_0 (\mathbf{k}) \Big \rangle  \Big \langle  \Psi_{0} (\mathbf{k}) \Big| = \Big \vert  \Psi_{0}^{0} ({\bf k}) \Big \rangle \Big \langle \Psi_{0}^{0} ({\bf k}) \Big \vert  \, = \\
\nonumber 
&& = \frac{1}{2}
\begin{bmatrix}
0&0&0&0&0&0\\
0&0&0&0&0&0\\
0&0&1&0&0&-e^{2i\Theta_{\bf k}}\\
0&0&0&0&0&0\\
0&0&0&0&0&0\\
0&0&-e^{-2i\Theta_{\bf k}}&0&0&1
\end{bmatrix} \, , 
\end{eqnarray}
and

\begin{eqnarray}
\hat{\mc P}_{1,1}(\sigma, {\bf k}) & = & \Big| \Psi_1 (\mathbf{k}) \Big \rangle  \Big \langle  \Psi_{1} (\mathbf{k}) \Big| = \Big \vert  \Psi_{1}^{0} ({\bf k}) \Big \rangle \Big \langle \Psi_{1}^{0} ({\bf k}) \Big \vert  \, = \\
\nonumber 
&& = \frac1{1+2\alpha^2}
\begin{bmatrix}
\alpha^2 &0&-\alpha e^{-2i\Theta_{\bf k}}&\alpha^2&0&0\\
0&0&0&0&0&0\\
-\alpha e^{2 i\Theta_{\bf k}}&0&1&-\alpha&0&0\\
\alpha^2&0&-\alpha&\alpha^2&0&0\\
0&0&0&0&0&0\\
0&0&0&0&0&0
\end{bmatrix} \, . 
\end{eqnarray}

The next set of projectors is obtained as 

\begin{eqnarray}
\hat{\mc P}_{(3,4),(3,4)}(\sigma, \sigma' \, \vert \, {\bf k}) & = & \Big| \Psi_{(3,4)} (\mathbf{k}) \Big \rangle  \Big \langle  \Psi_{(3,4)} (\mathbf{k}) \Big| = \Big \vert  \Psi_{0}^{\sigma = \pm 1} ({\bf k}) \Big \rangle \Big \langle \Psi_{0}^{\sigma' = \pm 1} ({\bf k}) \Big \vert  \, = \\
\nonumber 
&& =\frac14
\begin{bmatrix}
1 & -\sigma\tet{e}^{-i\Theta_{\bf k}}&0&-\tet{e}^{-2i\Theta_{\bf k}}& \sigma\tet{e}^{-i\Theta_{\bf k}}&0\\
-\sigma' \tet{e}^{i\Theta_{\bf k}}&1 & 0 & \sigma \tet{e}^{i\Theta_{\bf k}} & -1 & 0\\
0&0&0&0&0&0\\
-\tet{e}^{2i\Theta_{\bf k}}&\sigma' \tet{e}^{-i\Theta_{\bf k}}&0&1&-\sigma \tet{e}^{i\Theta_{\bf k}}&0\\
\sigma' \tet{e}^{i\Theta_{\bf k}}&-1&0&-\sigma \tet{e}^{-i\Theta_{\bf k}}&1&0\\
0&0&0&0&0&0
\end{bmatrix} \, . 
\end{eqnarray}

Finally, for $\Big| \Psi_{(5,6)} (\mathbf{k}) \Big \rangle = \Big| \Psi_{1}^{\pm 1} (\mathbf{k}) \Big \rangle$ we derive 

\begin{eqnarray}
&& \hat{\mc P}_{(5,6),(5,6)}(\sigma, {\mathbf{k}}) = \hat P_{1,1}^{\pm,\pm}(\sigma, \mathbf{k}) 
=\frac{1}{4 r_{[\alpha]}^2} \, \times \\
\nonumber 
&& \times \, \left\{ \begin{array}{ccc|ccc}
1& \sigma \, r_{[\alpha]} \, \tet{e}^{-i\Theta_{\bf k}}& 2 \alpha \tet{e}^{-2i\Theta_{\bf k}}& e^{-2i\Theta_{\bf k}}&-\sigma r_{[\alpha]} \, \tet{e}^{-i\Theta_{\bf k}}& 2 \alpha \\[2 mm]
\sigma \, r_{[\alpha]} \, \tet{e}^{i\Theta_{\bf k}}& r_{[\alpha]}^2 & 2 \sigma \alpha  \tet{e}^{-i\Theta_{\bf k}}& \sigma r_{[\alpha]} \tet{e}^{-i\Theta_{\bf k}}&-r_{[\alpha]}^2&2 \sigma \alpha r_{[\alpha]} \tet{e}^{i\Theta_{\bf k}}\\[2 mm]
2\alpha \tet{e}^{2i\Theta_{\bf k}}& 2 \sigma \alpha r_{[\alpha]} \tet{e}^{i\Theta_{\bf k}}& 4 \alpha^2 &
2 \alpha & - 2 \sigma \alpha r_{[\alpha]} \tet{e}^{i\Theta_{\bf k}}& 4 \alpha^2 \tet{e}^{2i\Theta_{\bf k}}\\[2 mm] 
\hline
\tet{e}^{2i\Theta_{\bf k}}& \sigma r_{[\alpha]} \tet{e}^{i\Theta_{\bf k}}&2\alpha & 1 &
-\sigma r_{[\alpha]} \tet{e}^{i\Theta_{\bf k}}& 2\alpha \tet{e}^{2i\Theta_{\bf k}}\\[2 mm] 
-\sigma r_{[\alpha]} \tet{e}^{i\Theta_{\bf k}}&-r_{[\alpha]}^2 & -2 \sigma \alpha r_{[\alpha]} \tet{e}^{-i\Theta_{\bf k}}&
-\sigma r_{[\alpha]} \tet{e}^{-i\Theta_{\bf k}} &  r_{[\alpha]}^2 & -2 \sigma \alpha r_{[\alpha]} \tet{e}^{i\Theta_{\bf k}}\\[2 mm] 
2\alpha & 2 \sigma \alpha r_{[\alpha]} \tet{e}^{-i\Theta_{\bf k}}& 4\alpha^2 \tet{e}^{-2i\Theta_{\bf k}}&
2\alpha \tet{e}^{-2 i \Theta_{\bf k}}& -2 \sigma \alpha r_{[\alpha]} \tet{e}^{-i\Theta_{\bf k}} & 4\alpha^2
\end{array}
\right\}
\end{eqnarray}

These four expressions give a set of projector operators associated with our eigenfunctions \eqref{psm1}-\eqref{psm6}. They are Hermitian and satisfy $\hat{\mc P}_n^2=\hat{\mc P}_n$ and $\hat{\mc P}_n \cdot \hat{\mc P}_{n'} = \delta_{n,n'} \hat{\mc P}_n$, which could be verified by a direct calculation.  

\medskip

To evaluate the wave function overlap using the projectors (projector operators), we need not only the diagonal projectors but also the 21 off-diagonal elements (one side of the diagonal, above the main diagonal). It definitely requires many more calculations and falls out of the scope. However, we only present the results for the simplest case of P zero zero. Indeed, we obtain.

\begin{equation}
\mbb{O}_{1,1}({\bf k},{\bf q}) = \mbb{O}_{0,0}^{0,0} 
=
\Big|
\Big \langle
\Psi_1(\mathbf k)
\Big|
\psi_{1}(\mathbf k+\mathbf q)
\Big \rangle
\Big|^2 = \text{Tr} \Big[ \hat{\mc P}_{0,0}({\bf k}) \cdot \hat{\mc P}_{0,0}({\bf k}+{\bf q}) \Big] \, . 
\end{equation}

\begin{equation}
 \hat{\mc P}_{0,0}({\bf k}) \cdot \hat{\mc P}_{0,0}({\bf k}+{\bf q}) = \frac{1}{4} \, 
\left\{
\begin{array}{cccccc}
0 & 0 & 0 & 0 & 0 & 0\\
0 & 0 & 0 & 0 & 0 & 0\\
0 & 0 &
1+e^{2i(\Theta_{\bf k}-\Theta_{{\bf k}+{\bf q}})}
& 0 & 0 &
-\left(e^{2i\theta}+e^{2i\Theta_{{\bf k}+{\bf q}}}\right)\\
0 & 0 & 0 & 0 & 0 & 0\\
0 & 0 & 0 & 0 & 0 & 0\\
0 & 0 &
-\left(e^{-2i\Theta_{\bf k}}+e^{-2i\Theta_{{\bf k}+{\bf q}}}\right)
& 0 & 0 &
1+e^{-2i(\Theta_{\bf k}-\Theta_{{\bf k}+{\bf q}})}
\end{array}
\right\} \, .
\end{equation}

Its trace is immediately calculated as 

\begin{equation}
4 \text{Tr} \Big[  \hat{\mc P}_{0,0}({\bf k}) \cdot \hat{\mc P}_{0,0}({\bf k}+{\bf q}) \Big] = 
2 + e^{-2i(\Theta_{\bf k}-\Theta_{{\bf k}+{\bf q}})} + e^{2i(\Theta_{\bf k}-\Theta_{{\bf k}+{\bf q}})}
= 2+2\cos\left[2(\Theta_{\bf k}-\Theta_{{\bf k}+{\bf q}})\right] = 4 \cos^2 (\Delta_{{\bf k},{\bf k}+{\bf q}}) \, ,
\end{equation}
which confirms our result given by Eq.~\eqref{O00001}. All the remaining 20 overlap factors $\mbb{O}_{[\lambda = 1,\cdots,6,\,\lambda' = 1,\cdots,6]}({\bf k},{\bf q})$ can be obtained in a similar manner.

\section{Graphene limit $\alpha \longrightarrow 0$}
\label{apb}

For graphene with $\alpha = 0$, we obtain

\begin{equation}
\label{grlim0012}
\hat{\mc{H}}_\alpha ({\bf k}) = \left\{
\begin{array}{cc|c|cc|c}
0 & e^{-i\theta}k & 0 & 0 & 0 & 0 \\[6pt]
e^{i\theta}k & 0 & 0 & 0 & 0 & 0 \\[6pt]
\hline
0 & 0 & 0 & 0 & 0 & 0 \\[6pt]
\hline
0 & 0 & 0 & 0 & e^{i\theta}k & 0 \\[6pt]
0 & 0 & 0 & e^{-i\theta}k & 0 & 0 \\[6pt]
\hline
0 & 0 & 0 & 0 & 0 & 0
\end{array}
\right\} 
\end{equation}
The energy spectrum for Hamiltonian \eqref{grlim0012} is given by three double-degenerate bands $\varepsilon_\sigma ({\bf k}) = \sigma \, (\hbar v_F \, k) = \Big\{0,0, - (\hbar v_F) k, - (\hbar v_F) k, (\hbar v_F) k, (\hbar v_F) k \Big\}$,  where $\sigma = 0, \pm 1$, correspond to the flat, conduction, and valence bands. We see that this dispersions represent two degenerate flat bands and two degenerate Dirac cones, identical to the energy band structure of graphene. The flat band persists even in the graphene limit, however, the overlap matrix elements connecting the flat band to the conduction and valence bands vanish in this limit.

\medskip 
 The corresponding wave functions are obtained as.
 
\begin{equation}
\Big| \Psi_1^{(G)} ({\bf k}) \Big \rangle = \Big| \Psi_{\rho = 0}^{\sigma = 0} ({\bf k}) \Big \rangle =
\begin{pmatrix}
0\\
0\\
0\\
0\\
0\\
1
\end{pmatrix}
\end{equation}

\begin{equation}
\Big| \Psi_2^{(G)} ({\bf k}) \Big \rangle = \Big| \Psi_{\rho = 1}^{\sigma = 0} ({\bf k}) \Big \rangle =
\begin{pmatrix}
0\\
0\\
1\\
0\\
0\\
0
\end{pmatrix}
\end{equation}

\begin{equation}
\Big| \Psi_{3,5}^{(G)} ({\bf k}) \Big \rangle = \Big| \Psi_{\rho = 0}^{\sigma = \pm 1} ({\bf k}) \Big \rangle =
\frac{1}{\sqrt{2}}
\begin{pmatrix}
0\\
0\\
0\\
\pm e^{i\theta_{\bf k}}\\
1\\
0
\end{pmatrix} = \frac{1}{\sqrt{2}}
\begin{pmatrix}
0\\
0\\
0\\
\sigma \tet{e}^{i\theta_{\bf k}}\\
1\\
0
\end{pmatrix}
\end{equation}

\begin{equation}
\Big| \Psi_{4,6}^{(G)} ({\bf k}) \Big \rangle = \Big| \Psi_{\rho = 1}^{\sigma = \pm} ({\bf k}) \Big \rangle =
\frac{1}{\sqrt{2}}
\begin{pmatrix}
\pm e^{-i\theta_{\bf k}}\\
1\\
0\\
0\\
0\\
0
\end{pmatrix} = \frac{1}{\sqrt{2}}
\begin{pmatrix}
\sigma e^{-i\theta_{\bf k}}\\
1\\
0\\
0\\
0\\
0
\end{pmatrix}
\end{equation}

\begin{equation}
\Big| \Psi_5^{(G)} ({\bf k}) \Big \rangle = \Big| \Psi_{\rho = 0}^{\sigma = 0} ({\bf k}) \Big \rangle =
\frac{1}{\sqrt{2}}
\begin{pmatrix}
0\\
0\\
0\\
e^{i\theta_{\bf k}}\\
1\\
0
\end{pmatrix}
\end{equation}

\begin{equation}
\Big| \Psi_6^{(G)} ({\bf k}) \Big \rangle = \Big| \Psi_{\rho = 0}^{\sigma = 0} ({\bf k}) \Big \rangle =
\frac{1}{\sqrt{2}}
\begin{pmatrix}
e^{-i\theta_{\bf k}}\\
1\\
0\\
0\\
0\\
0
\end{pmatrix}
\end{equation}

These wave functions are not equivalent to $\alpha = 0$ limit of our general wave functions \eqref{psm1}-\eqref{psm6}

\begin{equation}
\Big| \Psi_1 ({\bf k} \, \vert \, \alpha = 0 ) \Big \rangle = \Big| \Psi_{\rho = 0}^{\sigma = 0} ({\bf k}) \Big \rangle =
\frac{1}{\sqrt2}
\begin{pmatrix}
0\\0\\-e^{i\Theta_{{\bf k}}}\\0\\0\\e^{-i\Theta_{{\bf k}}}
\end{pmatrix}
\end{equation}

\begin{equation}
\Big| \Psi_2 ({\bf k} \, \vert \, \alpha = 0)  \Big \rangle  = \Big| \Psi_{\rho = 1}^{\sigma = 0} ({\bf k})  \Big \rangle =
\begin{pmatrix}
0\\0\\- \tet{e}^{i\Theta_{{\bf k}}}\\0\\0\\0
\end{pmatrix}
\end{equation}
and

\begin{equation}
\Big| \Psi_{\rho = 0}^{\sigma = \pm 1} ({\bf k}  \, \vert \, \alpha = 0)  \Big \rangle  = \frac{1}{2}
\begin{pmatrix}
-\sigma \tet{e}^{-i\Theta_{{\bf k}}}\\-1\\0\\ \sigma \tet{e}^{i\Theta_{{\bf k}}}\\1\\0
\end{pmatrix}
\end{equation}

as well as 

\begin{equation}
\Big| \Psi_{\rho = 1}^{\sigma = \pm 1} ({\bf k}) \Big \rangle = \frac{1}{2}
\begin{pmatrix}
1\\[2mm]
\sigma \tet{e}^{i\Theta_{{\bf k}}}\\[2mm]
0\\[2mm]
\tet{e}^{2i\theta}\\[2mm]
\sigma \tet{e}^{i \Theta_{{\bf k}}}\\[2mm]
0
\end{pmatrix} 
\, .
\end{equation}

However, one can easily find the connection between them:

\begin{eqnarray}
\label{transP001}
&& \Big| \Psi_2^{(G)} ({\bf k}) \Big \rangle = \tet{e}^{\pi - \Theta_{\bf k}} \, \Big| \Psi_2 ({\bf k} \, \vert \, \alpha = 0)  \Big \rangle \, , \\
\nonumber 
&& \Big| \Psi_1^{(G)} ({\bf k}) \Big \rangle = \frac{\tet{e}^{\pi - \Theta_{\bf k}}}{\sqrt{2}} \Big[ \sqrt{2} \, \Big| \Psi_1 ({\bf k} \, \vert \, \alpha = 0) 
\Big \rangle  - \Big| \Psi_2 ({\bf k} \, \vert \, \alpha = 0) 
\Big \rangle \Big] \, , \\
\nonumber 
&& \Big| \Psi_{3,4} ({\bf k})  \, \vert \, \alpha = 0 \Big \rangle = \frac{1}{\sqrt{2}} \, \Big[  
\Big| \Psi_{3,5} ({\bf k}) \, \Big \rangle - \Big| \Psi_{4,6} ({\bf k}) \, \Big \rangle
\Big] \, , \\
\nonumber
&& \Big| \Psi_{5,6} ({\bf k})  \, \vert \, \alpha = 0 \Big \rangle = \frac{\tet{e}^{\Theta_{\bf k}}}{\sqrt{2}} \, \Big[  
\Big| \Psi_{4,6} ({\bf k}) \, \Big \rangle + \Big| \Psi_{3,5} ({\bf k}) \, \Big \rangle
\Big] \, . 
\end{eqnarray}
Importantly, we notice that the eigenstates for Hamiltonian \eqref{grlim0012}, which is the $\alpha \longrightarrow 0$ limit of Hamiltonian \eqref{MainHamalpha}, which corresponds to the general case of Kek-Y $\alpha-\mc{T}_3$ model, is not equivalent to the $\alpha \longrightarrow 0$ limit of eigenvectors \eqref{psm1}-\eqref{psm6}. This is a common situation in condensed matter problems since for a finite $\alpha$  all six eigenvalues are different but two bands become degenerate for  $\alpha \longrightarrow 0$ and the corresponding eigenvectors are no longer unique. Any linear combination of the degenerate eigenvectors is also an eigenvector for Hamiltonian \eqref{grlim0012}, as we just demonstrated by equations \eqref{transP001}. 

\medskip

As expected from the physical considerations discussed above, most of the overlap matrix elements vanish. The only nonzero contributions are the intraband overlap associated with the flat band, which is independent of the scattering angle, and the graphene-like overlap factors between the valence and conduction bands. 

\medskip 
Thus, the overlap functions could be summarized in the following matrix

\begin{equation}
\label{zeroa01}
\overleftrightarrow{\mbb{O}}^{(G)}({\bf k},{\bf q})=
\begin{pmatrix}
1&0&0&0&0&0\\
0&1&0&0&0&0\\
0&0& \mbb{O}_{+}({\bf k},{\bf q}) &0&\mbb{O}_{-}({\bf k},{\bf q})&0\\
0&0&0&\mbb{O}_{+}({\bf k},{\bf q})&0&\mbb{O}_{-}({\bf k},{\bf q})\\
0&0&\mbb{O}_{-}({\bf k},{\bf q})&0&\mbb{O}_{+}({\bf k},{\bf q})&0\\
0&0&0&\mbb{O}_{-}({\bf k},{\bf q})&0&\mbb{O}_{+}({\bf k},{\bf q})
\end{pmatrix} \, , 
\end{equation}

where 

\begin{equation}
\mbb{O}_{\pm}({\bf k},{\bf q}) = \frac{1}{2} \, \Big[ 1 \pm \cos (\delta_{{\bf k},{\bf k}+{\bf q}}) \Big] = \left\{   \begin{array}{c}
\cos^2 (\delta_{{\bf k},{\bf k}+{\bf q}})\\[2mm] 
\sin^2 (\delta_{{\bf k},{\bf k}+{\bf q}})
\end{array} \right. \, . 
\end{equation}

As the last step here, we provide the $\alpha \longrightarrow 0$ limits of earlier obtained wave function overlaps, which of course are not equivalent to the corresponding overlap matrix \eqref{zeroa01}. 

\medskip 
Thus, we derive 

\begin{equation}
\label{O00001g}
\mbb{O}_{1,1}({\bf k},{\bf q}) = \mbb{O}_{0,0}^{0,0}
=\cos^2 (\Delta_{{\bf k},{\bf k}+{\bf q}}) \, , 
\end{equation}
also $\mbb{O}_{2,2}(\mathbf k,\mathbf q \, \vert \, \alpha =0) = \mbb{O}_{0,1}^{0,1} = 1$ and

\begin{equation}
\mbb{O}_{(3,4),(3,4)}(\mathbf k,\mathbf q) = \mbb{O}_{\sigma=\pm 1,0}^{\sigma'\pm 1,0} = 
\frac{1}{4}\,
\Big[
1+\sigma \sigma' \cos (\delta_{{\bf k},{\bf k}+{\bf q}}) 
\Big]^2 \, ,
\end{equation}
as well as  

\begin{eqnarray}
&& \mbb{O}_{(5,6),(5,6)}(\mathbf k,\mathbf q) \equiv \mbb{O}_{\sigma = \pm 1,1}^{\sigma'=\pm 1,1} = \mbb{O}_{1,(3,4)}({\bf k},{\bf q}) \equiv  \mbb{O}_{0,0}^{0,\sigma = \pm1} = \mbb{O}_{1,(5,6)}({\bf k},{\bf q}) \equiv  \mbb{O}_{0,0}^{1,\sigma = \pm1} \, =  \\
\nonumber
&& = \, \mbb{O}_{2,(3,4)}({\bf k},{\bf q}) \equiv  \mbb{O}_{1,\sigma = 0}^{0,\sigma' = \pm1}  = \mbb{O}_{2,(5,6)}({\bf k},{\bf q}) \equiv  \mbb{O}_{1,\sigma = 0}^{1,\sigma' = \pm1} =  0 \, .
\end{eqnarray}
Finally, we obtain $\mbb{O}_{1,2}({\bf k},{\bf q}) = \mbb{O}_{0,0}^{0,1} =\frac{1}{2}$, as well as 

\begin{equation}
\mbb{O}_{(3,4),(5,6)}({\bf k},{\bf q}) = \mbb{O}_{0,\sigma = \pm1}^{1,\sigma' = \pm1} = \frac{1}{4} \, \sin^2(\delta_{{\bf k},{\bf k}+{\bf q}}) \,  .
\end{equation}

\section{Dice lattice limit $\alpha \longrightarrow 1$}
\label{apc}

To finalize the presentation of the model, let us briefly provide the expressions for the wave functions and the wave  overlap functions, corresponding to the other important limit of a dice lattice $\alpha = 1$,.

\medskip 
The wave functions are modified as 

\begin{equation}
\Big| \Psi_1 ({\bf k}) \Big \rangle = \Big| \Psi_{\rho = 0}^{\sigma = 0} ({\bf k}) \Big \rangle =
\frac{1}{\sqrt2}
\begin{pmatrix}
0\\0\\-e^{i\Theta_{{\bf k}}}\\0\\0\\e^{-i\Theta_{{\bf k}}}
\end{pmatrix} \, , 
\end{equation}

\begin{equation}
\Big| \Psi_2 ({\bf k})  \Big \rangle  = \Big| \Psi_{\rho = 1}^{\sigma = 0} ({\bf k})  \Big \rangle =
\frac{1}{\sqrt{3}}
\begin{pmatrix}
\tet{e}^{-i\Theta_{\bf k}}\\0\\- \tet{e}^{i\Theta_{{\bf k}}}\\ \tet{e}^{i\Theta_{\bf k}}\\0\\0
\end{pmatrix}
\end{equation}

and

\begin{equation}
 \Big| \Psi_{3,4}\Big \rangle = \Big| \Psi_{\rho = 0}^{\sigma = \pm 1} ({\bf k})  \Big \rangle  =
\frac{1}{2}
\begin{pmatrix}
\mp \tet{e}^{-i\Theta_{{\bf k}}}\\-1\\0\\ \pm \tet{e}^{i\Theta_{{\bf k}}}\\1\\0
\end{pmatrix} = \frac{1}{2}
\begin{pmatrix}
-\sigma \tet{e}^{-i\Theta_{{\bf k}}}\\-1\\0\\ \sigma \tet{e}^{i\Theta_{{\bf k}}}\\1\\0
\end{pmatrix}
\end{equation}
as well as 

\begin{equation}
\Big| \Psi_{5,6}\Big \rangle  =  \Big| \Psi_{\rho = 1}^{\sigma = \pm 1} ({\bf k}) \Big \rangle = \frac{1}{2 \sqrt{5}}
\begin{pmatrix}
1\\[2mm]
\pm \tet{e}^{i\Theta_{{\bf k}}}\sqrt{5}\\[2mm]
2\tet{e}^{2 i \Theta_{{\bf k}}}\\[2mm]
\tet{e}^{2i\theta}\\[2mm]
\pm \tet{e}^{i \Theta_{{\bf k}}}\sqrt{5}\\[2mm]
2
\end{pmatrix} 
\, .
\end{equation}

The wave function overlap factors obtained in the equations \eqref{O00001}-\eqref{O000013}, for a dice lattice, are reduced to the following expressions 

\begin{equation}
\label{O00001d}
\mbb{O}_{1,1}({\bf k},{\bf q}) = \mbb{O}_{0,0}^{0,0}
=\cos^2 (\Delta_{{\bf k},{\bf k}+{\bf q}}) \, ,
\end{equation}

\begin{equation}
\mbb{O}_{2,2}(\mathbf k,\mathbf q) = \mbb{O}_{0,1}^{0,1} = 
\frac{1}
{9} \,  \Big[  1+8 \cos(\Delta_{{\bf k},{\bf k}+{\bf q}}) \Big] \, , 
\end{equation}
and 

\begin{equation}
\mbb{O}_{(3,4),(3,4)}(\mathbf k,\mathbf q) = \mbb{O}_{\sigma=\pm 1,0}^{\sigma'\pm 1,0} = 
\frac{1}{4}\,
\Big[
1+\sigma \sigma' \cos (\Delta_{{\bf k},{\bf k}+{\bf q}}) 
\Big]^2 \, .
\end{equation}

\begin{eqnarray}
&& \mbb{O}_{(5,6),(5,6)}(\mathbf k,\mathbf q) = \mbb{O}_{\sigma = \pm 1,1}^{\sigma'=\pm 1,1} \, = \\
\nonumber  
&& = \, 2 \left\{
2\Big[\sigma_1 \sigma_2+\cos(\delta_{{\bf k},{\bf k}+{\bf q}})\Big]^2 \, \cos^2(\delta_{{\bf k},{\bf k}+{\bf q}})
+ \frac{1}{2} \, \Big[53+\cos(2\delta_{{\bf k},{\bf k}+{\bf q}})\Big]\sin^2(\delta_{{\bf k},{\bf k}+{\bf q}}) 
\right\}
\, .
\end{eqnarray}
Also, we derive $\mbb{O}_{1,2}({\bf k},{\bf q}) = \mbb{O}_{0,0}^{0,1} =\frac{1}{6}$, $\mbb{O}_{1,(3,4)}({\bf k},{\bf q}) = \mbb{O}_{0,0}^{0,\sigma = \pm1} = 0$, 
$\mbb{O}_{1,(5,6)}({\bf k},{\bf q}) = \mbb{O}_{0,0}^{1,\sigma = \pm1} =  \frac{2}{5} \,  \sin^2(\Delta_{{\bf k},{\bf k}+{\bf q}})$, as well as 

\begin{equation}
\label{On1}
\mbb{O}_{2,(3,4)}({\bf k},{\bf q}) = \mbb{O}_{1,\sigma = 0}^{0,\sigma' = \pm1} = \frac{1}{3}
\sin^2(\delta_{{\bf k},{\bf k}+{\bf q}})
 \, , 
\end{equation}

\begin{equation}
\label{On2}
\mbb{O}_{2,(5,6)}({\bf k},{\bf q}) = \mbb{O}_{1,\sigma = 0}^{1,\sigma' = \pm1} =
\frac{1}{15}
 \sin^2(\delta_{{\bf k},{\bf k}+{\bf q}}) \,  
\end{equation}
and

\begin{equation}
\label{On3}
\mbb{O}_{(3,4),(5,6)}({\bf k},{\bf q}) = \mbb{O}_{0,\sigma = \pm1}^{1,\sigma' = \pm1} = \frac{1}{20} \, \sin^2(\delta_{{\bf k},{\bf k}+{\bf q}}) \, . 
\end{equation}
Three latter overlap functions \eqref{On1}-\eqref{On3} demonstrate a $\backsim  \sin^2(\delta_{{\bf k},{\bf k}+{\bf q}})$ dependence which corresponds to the electron transitions to and from the flat band in a regular $\alpha-\mc{T}_3$ model.

\medskip 
As we can see, the results are less symmetric and exhibit a greater variety of equations and expressions compared to graphene, while remaining considerably simpler than those of the general $\alpha-\mc{T}_3$ model.

\bibliography{KekAlpha}

\begin{thebibliography}{124}
\expandafter\ifx\csname natexlab\endcsname\relax\def\natexlab#1{#1}\fi
\expandafter\ifx\csname bibnamefont\endcsname\relax
  \def\bibnamefont#1{#1}\fi
\expandafter\ifx\csname bibfnamefont\endcsname\relax
  \def\bibfnamefont#1{#1}\fi
\expandafter\ifx\csname citenamefont\endcsname\relax
  \def\citenamefont#1{#1}\fi
\expandafter\ifx\csname url\endcsname\relax
  \def\url#1{\texttt{#1}}\fi
\expandafter\ifx\csname urlprefix\endcsname\relax\def\urlprefix{URL }\fi
\providecommand{\bibinfo}[2]{#2}
\providecommand{\eprint}[2][]{\url{#2}}

\bibitem[{\citenamefont{Geim}(2009)}]{geim2009graphene}
\bibinfo{author}{\bibfnamefont{A.~K.} \bibnamefont{Geim}},
  \bibinfo{journal}{science} \textbf{\bibinfo{volume}{324}},
  \bibinfo{pages}{1530} (\bibinfo{year}{2009}).

\bibitem[{\citenamefont{Geim and Novoselov}(2007)}]{geim2007rise}
\bibinfo{author}{\bibfnamefont{A.~K.} \bibnamefont{Geim}} \bibnamefont{and}
  \bibinfo{author}{\bibfnamefont{K.~S.} \bibnamefont{Novoselov}},
  \bibinfo{journal}{Nature materials} \textbf{\bibinfo{volume}{6}},
  \bibinfo{pages}{183} (\bibinfo{year}{2007}).

\bibitem[{\citenamefont{Castro~Neto et~al.}(2009)\citenamefont{Castro~Neto,
  Guinea, Peres, Novoselov, and Geim}}]{castro2009electronic}
\bibinfo{author}{\bibfnamefont{A.~H.} \bibnamefont{Castro~Neto}},
  \bibinfo{author}{\bibfnamefont{F.}~\bibnamefont{Guinea}},
  \bibinfo{author}{\bibfnamefont{N.~M.} \bibnamefont{Peres}},
  \bibinfo{author}{\bibfnamefont{K.~S.} \bibnamefont{Novoselov}},
  \bibnamefont{and} \bibinfo{author}{\bibfnamefont{A.~K.} \bibnamefont{Geim}},
  \bibinfo{journal}{Reviews of modern physics} \textbf{\bibinfo{volume}{81}},
  \bibinfo{pages}{109} (\bibinfo{year}{2009}).

\bibitem[{\citenamefont{Ando}(2009)}]{ando2009electronic}
\bibinfo{author}{\bibfnamefont{T.}~\bibnamefont{Ando}}, \bibinfo{journal}{NPG
  asia materials} \textbf{\bibinfo{volume}{1}}, \bibinfo{pages}{17}
  (\bibinfo{year}{2009}).

\bibitem[{\citenamefont{Gumbs et~al.}(2014{\natexlab{a}})\citenamefont{Gumbs,
  Iurov, Huang, and Zhemchuzhna}}]{gumbs2014revealing}
\bibinfo{author}{\bibfnamefont{G.}~\bibnamefont{Gumbs}},
  \bibinfo{author}{\bibfnamefont{A.}~\bibnamefont{Iurov}},
  \bibinfo{author}{\bibfnamefont{D.}~\bibnamefont{Huang}}, \bibnamefont{and}
  \bibinfo{author}{\bibfnamefont{L.}~\bibnamefont{Zhemchuzhna}},
  \bibinfo{journal}{Physical Review B} \textbf{\bibinfo{volume}{89}},
  \bibinfo{pages}{241407} (\bibinfo{year}{2014}{\natexlab{a}}).

\bibitem[{\citenamefont{Wehling et~al.}(2014)\citenamefont{Wehling,
  Black-Schaffer, and Balatsky}}]{wehling2014dirac}
\bibinfo{author}{\bibfnamefont{T.~O.} \bibnamefont{Wehling}},
  \bibinfo{author}{\bibfnamefont{A.~M.} \bibnamefont{Black-Schaffer}},
  \bibnamefont{and} \bibinfo{author}{\bibfnamefont{A.~V.}
  \bibnamefont{Balatsky}}, \bibinfo{journal}{Advances in Physics}
  \textbf{\bibinfo{volume}{63}}, \bibinfo{pages}{1} (\bibinfo{year}{2014}).

\bibitem[{\citenamefont{Pedersen et~al.}(2009)\citenamefont{Pedersen, Jauho,
  and Pedersen}}]{pedersen2009optical}
\bibinfo{author}{\bibfnamefont{T.~G.} \bibnamefont{Pedersen}},
  \bibinfo{author}{\bibfnamefont{A.-P.} \bibnamefont{Jauho}}, \bibnamefont{and}
  \bibinfo{author}{\bibfnamefont{K.}~\bibnamefont{Pedersen}},
  \bibinfo{journal}{Physical Review B Condensed Matter And Materials Physics}
  \textbf{\bibinfo{volume}{79}}, \bibinfo{pages}{113406}
  (\bibinfo{year}{2009}).

\bibitem[{\citenamefont{Pereira et~al.}(2008)\citenamefont{Pereira, Kotov, and
  Castro~Neto}}]{pereira2008supercritical}
\bibinfo{author}{\bibfnamefont{V.~M.} \bibnamefont{Pereira}},
  \bibinfo{author}{\bibfnamefont{V.~N.} \bibnamefont{Kotov}}, \bibnamefont{and}
  \bibinfo{author}{\bibfnamefont{A.}~\bibnamefont{Castro~Neto}},
  \bibinfo{journal}{Physical Review B—Condensed Matter and Materials Physics}
  \textbf{\bibinfo{volume}{78}}, \bibinfo{pages}{085101}
  (\bibinfo{year}{2008}).

\bibitem[{\citenamefont{Kibis}(2010)}]{kibis2010metal}
\bibinfo{author}{\bibfnamefont{O.}~\bibnamefont{Kibis}},
  \bibinfo{journal}{Physical Review B} \textbf{\bibinfo{volume}{81}},
  \bibinfo{pages}{165433} (\bibinfo{year}{2010}).

\bibitem[{\citenamefont{Ibarra-Sierra et~al.}(2019)\citenamefont{Ibarra-Sierra,
  Sandoval-Santana, Kunold, and Naumis}}]{ibarra2019dynamical}
\bibinfo{author}{\bibfnamefont{V.}~\bibnamefont{Ibarra-Sierra}},
  \bibinfo{author}{\bibfnamefont{J.}~\bibnamefont{Sandoval-Santana}},
  \bibinfo{author}{\bibfnamefont{A.}~\bibnamefont{Kunold}}, \bibnamefont{and}
  \bibinfo{author}{\bibfnamefont{G.~G.} \bibnamefont{Naumis}},
  \bibinfo{journal}{Physical Review B} \textbf{\bibinfo{volume}{100}},
  \bibinfo{pages}{125302} (\bibinfo{year}{2019}).

\bibitem[{\citenamefont{Iurov et~al.}(2022{\natexlab{a}})\citenamefont{Iurov,
  Zhemchuzhna, Gumbs, Huang, Tse, Blaise, and Ejiogu}}]{iurov2022floquet}
\bibinfo{author}{\bibfnamefont{A.}~\bibnamefont{Iurov}},
  \bibinfo{author}{\bibfnamefont{L.}~\bibnamefont{Zhemchuzhna}},
  \bibinfo{author}{\bibfnamefont{G.}~\bibnamefont{Gumbs}},
  \bibinfo{author}{\bibfnamefont{D.}~\bibnamefont{Huang}},
  \bibinfo{author}{\bibfnamefont{W.-K.} \bibnamefont{Tse}},
  \bibinfo{author}{\bibfnamefont{K.}~\bibnamefont{Blaise}}, \bibnamefont{and}
  \bibinfo{author}{\bibfnamefont{C.}~\bibnamefont{Ejiogu}},
  \bibinfo{journal}{Scientific Reports} \textbf{\bibinfo{volume}{12}},
  \bibinfo{pages}{21348} (\bibinfo{year}{2022}{\natexlab{a}}).

\bibitem[{\citenamefont{Oka and Kitamura}(2019)}]{oka2019floquet}
\bibinfo{author}{\bibfnamefont{T.}~\bibnamefont{Oka}} \bibnamefont{and}
  \bibinfo{author}{\bibfnamefont{S.}~\bibnamefont{Kitamura}},
  \bibinfo{journal}{Annual Review of Condensed Matter Physics}
  \textbf{\bibinfo{volume}{10}}, \bibinfo{pages}{387} (\bibinfo{year}{2019}).

\bibitem[{\citenamefont{Iurov et~al.}(2024)\citenamefont{Iurov, Mattis,
  Zhemchuzhna, Gumbs, and Huang}}]{iurov2024floquet}
\bibinfo{author}{\bibfnamefont{A.}~\bibnamefont{Iurov}},
  \bibinfo{author}{\bibfnamefont{M.}~\bibnamefont{Mattis}},
  \bibinfo{author}{\bibfnamefont{L.}~\bibnamefont{Zhemchuzhna}},
  \bibinfo{author}{\bibfnamefont{G.}~\bibnamefont{Gumbs}}, \bibnamefont{and}
  \bibinfo{author}{\bibfnamefont{D.}~\bibnamefont{Huang}},
  \bibinfo{journal}{Applied Sciences} \textbf{\bibinfo{volume}{14}},
  \bibinfo{pages}{6027} (\bibinfo{year}{2024}).

\bibitem[{\citenamefont{Bukov et~al.}(2015)\citenamefont{Bukov, D'Alessio, and
  Polkovnikov}}]{bukov2015universal}
\bibinfo{author}{\bibfnamefont{M.}~\bibnamefont{Bukov}},
  \bibinfo{author}{\bibfnamefont{L.}~\bibnamefont{D'Alessio}},
  \bibnamefont{and}
  \bibinfo{author}{\bibfnamefont{A.}~\bibnamefont{Polkovnikov}},
  \bibinfo{journal}{Advances in Physics} \textbf{\bibinfo{volume}{64}},
  \bibinfo{pages}{139} (\bibinfo{year}{2015}).

\bibitem[{\citenamefont{Iurov et~al.}(2020{\natexlab{a}})\citenamefont{Iurov,
  Zhemchuzhna, Dahal, Gumbs, and Huang}}]{iurov2020quantum}
\bibinfo{author}{\bibfnamefont{A.}~\bibnamefont{Iurov}},
  \bibinfo{author}{\bibfnamefont{L.}~\bibnamefont{Zhemchuzhna}},
  \bibinfo{author}{\bibfnamefont{D.}~\bibnamefont{Dahal}},
  \bibinfo{author}{\bibfnamefont{G.}~\bibnamefont{Gumbs}}, \bibnamefont{and}
  \bibinfo{author}{\bibfnamefont{D.}~\bibnamefont{Huang}},
  \bibinfo{journal}{Physical Review B} \textbf{\bibinfo{volume}{101}},
  \bibinfo{pages}{035129} (\bibinfo{year}{2020}{\natexlab{a}}).

\bibitem[{\citenamefont{Islam and Saha}(2018)}]{islam2018driven}
\bibinfo{author}{\bibfnamefont{S.~F.} \bibnamefont{Islam}} \bibnamefont{and}
  \bibinfo{author}{\bibfnamefont{A.}~\bibnamefont{Saha}},
  \bibinfo{journal}{Physical Review B} \textbf{\bibinfo{volume}{98}},
  \bibinfo{pages}{235424} (\bibinfo{year}{2018}).

\bibitem[{\citenamefont{Iurov et~al.}(2022{\natexlab{b}})\citenamefont{Iurov,
  Zhemchuzhna, Gumbs, Huang, Dahal, and Abranyos}}]{iurov2022finite}
\bibinfo{author}{\bibfnamefont{A.}~\bibnamefont{Iurov}},
  \bibinfo{author}{\bibfnamefont{L.}~\bibnamefont{Zhemchuzhna}},
  \bibinfo{author}{\bibfnamefont{G.}~\bibnamefont{Gumbs}},
  \bibinfo{author}{\bibfnamefont{D.}~\bibnamefont{Huang}},
  \bibinfo{author}{\bibfnamefont{D.}~\bibnamefont{Dahal}}, \bibnamefont{and}
  \bibinfo{author}{\bibfnamefont{Y.}~\bibnamefont{Abranyos}},
  \bibinfo{journal}{Physical Review B} \textbf{\bibinfo{volume}{105}},
  \bibinfo{pages}{245414} (\bibinfo{year}{2022}{\natexlab{b}}).

\bibitem[{\citenamefont{Aitouni et~al.}(2026)\citenamefont{Aitouni, Azar,
  Cortes, D{\'\i}az, Laroze, and Jellal}}]{aitouni2026laser}
\bibinfo{author}{\bibfnamefont{R.~E.} \bibnamefont{Aitouni}},
  \bibinfo{author}{\bibfnamefont{M.~E.} \bibnamefont{Azar}},
  \bibinfo{author}{\bibfnamefont{C.}~\bibnamefont{Cortes}},
  \bibinfo{author}{\bibfnamefont{P.}~\bibnamefont{D{\'\i}az}},
  \bibinfo{author}{\bibfnamefont{D.}~\bibnamefont{Laroze}}, \bibnamefont{and}
  \bibinfo{author}{\bibfnamefont{A.}~\bibnamefont{Jellal}},
  \bibinfo{journal}{Annalen der Physik} \textbf{\bibinfo{volume}{538}},
  \bibinfo{pages}{e70222} (\bibinfo{year}{2026}).

\bibitem[{\citenamefont{Kristinsson et~al.}(2016)\citenamefont{Kristinsson,
  Kibis, Morina, and Shelykh}}]{kristinsson2016control}
\bibinfo{author}{\bibfnamefont{K.}~\bibnamefont{Kristinsson}},
  \bibinfo{author}{\bibfnamefont{O.~V.} \bibnamefont{Kibis}},
  \bibinfo{author}{\bibfnamefont{S.}~\bibnamefont{Morina}}, \bibnamefont{and}
  \bibinfo{author}{\bibfnamefont{I.~A.} \bibnamefont{Shelykh}},
  \bibinfo{journal}{Scientific reports} \textbf{\bibinfo{volume}{6}},
  \bibinfo{pages}{1} (\bibinfo{year}{2016}).

\bibitem[{\citenamefont{Iurov et~al.}(2017{\natexlab{a}})\citenamefont{Iurov,
  Zhemchuzhna, Gumbs, and Huang}}]{iurov2017exploring}
\bibinfo{author}{\bibfnamefont{A.}~\bibnamefont{Iurov}},
  \bibinfo{author}{\bibfnamefont{L.}~\bibnamefont{Zhemchuzhna}},
  \bibinfo{author}{\bibfnamefont{G.}~\bibnamefont{Gumbs}}, \bibnamefont{and}
  \bibinfo{author}{\bibfnamefont{D.}~\bibnamefont{Huang}},
  \bibinfo{journal}{Journal of Applied Physics} \textbf{\bibinfo{volume}{122}}
  (\bibinfo{year}{2017}{\natexlab{a}}).

\bibitem[{\citenamefont{Liu et~al.}(2014)\citenamefont{Liu, Liu, and
  Wu}}]{liu2014exotic}
\bibinfo{author}{\bibfnamefont{Z.}~\bibnamefont{Liu}},
  \bibinfo{author}{\bibfnamefont{F.}~\bibnamefont{Liu}}, \bibnamefont{and}
  \bibinfo{author}{\bibfnamefont{Y.-S.} \bibnamefont{Wu}},
  \bibinfo{journal}{Chinese Physics B} \textbf{\bibinfo{volume}{23}},
  \bibinfo{pages}{077308} (\bibinfo{year}{2014}).

\bibitem[{\citenamefont{Roy and
  Juri{\v{c}}i{\'c}}(2019)}]{roy2019unconventional}
\bibinfo{author}{\bibfnamefont{B.}~\bibnamefont{Roy}} \bibnamefont{and}
  \bibinfo{author}{\bibfnamefont{V.}~\bibnamefont{Juri{\v{c}}i{\'c}}},
  \bibinfo{journal}{Physical Review B} \textbf{\bibinfo{volume}{99}},
  \bibinfo{pages}{121407} (\bibinfo{year}{2019}).

\bibitem[{\citenamefont{Milicevic et~al.}(2019)\citenamefont{Milicevic,
  Montambaux, Ozawa, Jamadi, Real, Sagnes, Lemaitre, Le~Gratiet, Harouri, Bloch
  et~al.}}]{milicevic2019type}
\bibinfo{author}{\bibfnamefont{M.}~\bibnamefont{Milicevic}},
  \bibinfo{author}{\bibfnamefont{G.}~\bibnamefont{Montambaux}},
  \bibinfo{author}{\bibfnamefont{T.}~\bibnamefont{Ozawa}},
  \bibinfo{author}{\bibfnamefont{O.}~\bibnamefont{Jamadi}},
  \bibinfo{author}{\bibfnamefont{B.}~\bibnamefont{Real}},
  \bibinfo{author}{\bibfnamefont{I.}~\bibnamefont{Sagnes}},
  \bibinfo{author}{\bibfnamefont{A.}~\bibnamefont{Lemaitre}},
  \bibinfo{author}{\bibfnamefont{L.}~\bibnamefont{Le~Gratiet}},
  \bibinfo{author}{\bibfnamefont{A.}~\bibnamefont{Harouri}},
  \bibinfo{author}{\bibfnamefont{J.}~\bibnamefont{Bloch}},
  \bibnamefont{et~al.}, \bibinfo{journal}{Physical Review X}
  \textbf{\bibinfo{volume}{9}}, \bibinfo{pages}{031010} (\bibinfo{year}{2019}).

\bibitem[{\citenamefont{Slot et~al.}(2017)\citenamefont{Slot, Gardenier,
  Jacobse, Van~Miert, Kempkes, Zevenhuizen, Smith, Vanmaekelbergh, and
  Swart}}]{slot2017experimental}
\bibinfo{author}{\bibfnamefont{M.~R.} \bibnamefont{Slot}},
  \bibinfo{author}{\bibfnamefont{T.~S.} \bibnamefont{Gardenier}},
  \bibinfo{author}{\bibfnamefont{P.~H.} \bibnamefont{Jacobse}},
  \bibinfo{author}{\bibfnamefont{G.~C.} \bibnamefont{Van~Miert}},
  \bibinfo{author}{\bibfnamefont{S.~N.} \bibnamefont{Kempkes}},
  \bibinfo{author}{\bibfnamefont{S.~J.} \bibnamefont{Zevenhuizen}},
  \bibinfo{author}{\bibfnamefont{C.~M.} \bibnamefont{Smith}},
  \bibinfo{author}{\bibfnamefont{D.}~\bibnamefont{Vanmaekelbergh}},
  \bibnamefont{and} \bibinfo{author}{\bibfnamefont{I.}~\bibnamefont{Swart}},
  \bibinfo{journal}{Nature physics} \textbf{\bibinfo{volume}{13}},
  \bibinfo{pages}{672} (\bibinfo{year}{2017}).

\bibitem[{\citenamefont{Mukherjee et~al.}(2015)\citenamefont{Mukherjee,
  Spracklen, Choudhury, Goldman, Ohberg, Andersson, and
  Thomson}}]{mukherjee2015observation}
\bibinfo{author}{\bibfnamefont{S.}~\bibnamefont{Mukherjee}},
  \bibinfo{author}{\bibfnamefont{A.}~\bibnamefont{Spracklen}},
  \bibinfo{author}{\bibfnamefont{D.}~\bibnamefont{Choudhury}},
  \bibinfo{author}{\bibfnamefont{N.}~\bibnamefont{Goldman}},
  \bibinfo{author}{\bibfnamefont{P.}~\bibnamefont{Ohberg}},
  \bibinfo{author}{\bibfnamefont{E.}~\bibnamefont{Andersson}},
  \bibnamefont{and} \bibinfo{author}{\bibfnamefont{R.~R.}
  \bibnamefont{Thomson}}, \bibinfo{journal}{Physical review letters}
  \textbf{\bibinfo{volume}{114}}, \bibinfo{pages}{245504}
  (\bibinfo{year}{2015}).

\bibitem[{\citenamefont{Jo et~al.}(2012)\citenamefont{Jo, Guzman, Thomas,
  Hosur, Vishwanath, and Stamper-Kurn}}]{jo2012ultracold}
\bibinfo{author}{\bibfnamefont{G.-B.} \bibnamefont{Jo}},
  \bibinfo{author}{\bibfnamefont{J.}~\bibnamefont{Guzman}},
  \bibinfo{author}{\bibfnamefont{C.~K.} \bibnamefont{Thomas}},
  \bibinfo{author}{\bibfnamefont{P.}~\bibnamefont{Hosur}},
  \bibinfo{author}{\bibfnamefont{A.}~\bibnamefont{Vishwanath}},
  \bibnamefont{and} \bibinfo{author}{\bibfnamefont{D.~M.}
  \bibnamefont{Stamper-Kurn}}, \bibinfo{journal}{Physical review letters}
  \textbf{\bibinfo{volume}{108}}, \bibinfo{pages}{045305}
  (\bibinfo{year}{2012}).

\bibitem[{\citenamefont{Xue et~al.}(2019)\citenamefont{Xue, Yang, Gao, Chong,
  and Zhang}}]{xue2019acoustic}
\bibinfo{author}{\bibfnamefont{H.}~\bibnamefont{Xue}},
  \bibinfo{author}{\bibfnamefont{Y.}~\bibnamefont{Yang}},
  \bibinfo{author}{\bibfnamefont{F.}~\bibnamefont{Gao}},
  \bibinfo{author}{\bibfnamefont{Y.}~\bibnamefont{Chong}}, \bibnamefont{and}
  \bibinfo{author}{\bibfnamefont{B.}~\bibnamefont{Zhang}},
  \bibinfo{journal}{Nature materials} \textbf{\bibinfo{volume}{18}},
  \bibinfo{pages}{108} (\bibinfo{year}{2019}).

\bibitem[{\citenamefont{Leykam et~al.}(2018)\citenamefont{Leykam, Andreanov,
  and Flach}}]{Leykam2018}
\bibinfo{author}{\bibfnamefont{D.}~\bibnamefont{Leykam}},
  \bibinfo{author}{\bibfnamefont{A.}~\bibnamefont{Andreanov}},
  \bibnamefont{and} \bibinfo{author}{\bibfnamefont{S.}~\bibnamefont{Flach}},
  \bibinfo{journal}{Advances in Physics: X} \textbf{\bibinfo{volume}{3}},
  \bibinfo{pages}{1473052} (\bibinfo{year}{2018}).

\bibitem[{\citenamefont{Cao et~al.}(2018{\natexlab{a}})\citenamefont{Cao,
  Fatemi, Fang, Watanabe, Taniguchi, Kaxiras, and Jarillo-Herrero}}]{Cao2018a}
\bibinfo{author}{\bibfnamefont{Y.}~\bibnamefont{Cao}},
  \bibinfo{author}{\bibfnamefont{V.}~\bibnamefont{Fatemi}},
  \bibinfo{author}{\bibfnamefont{S.}~\bibnamefont{Fang}},
  \bibinfo{author}{\bibfnamefont{K.}~\bibnamefont{Watanabe}},
  \bibinfo{author}{\bibfnamefont{T.}~\bibnamefont{Taniguchi}},
  \bibinfo{author}{\bibfnamefont{E.}~\bibnamefont{Kaxiras}}, \bibnamefont{and}
  \bibinfo{author}{\bibfnamefont{P.}~\bibnamefont{Jarillo-Herrero}},
  \bibinfo{journal}{Nature} \textbf{\bibinfo{volume}{556}}, \bibinfo{pages}{43}
  (\bibinfo{year}{2018}{\natexlab{a}}).

\bibitem[{\citenamefont{Cao et~al.}(2018{\natexlab{b}})\citenamefont{Cao,
  Fatemi, Demir, Fang, Tomarken, Luo, Sanchez-Yamagishi, Watanabe, Taniguchi,
  Kaxiras et~al.}}]{Cao2018b}
\bibinfo{author}{\bibfnamefont{Y.}~\bibnamefont{Cao}},
  \bibinfo{author}{\bibfnamefont{V.}~\bibnamefont{Fatemi}},
  \bibinfo{author}{\bibfnamefont{A.}~\bibnamefont{Demir}},
  \bibinfo{author}{\bibfnamefont{S.}~\bibnamefont{Fang}},
  \bibinfo{author}{\bibfnamefont{S.~L.} \bibnamefont{Tomarken}},
  \bibinfo{author}{\bibfnamefont{J.~Y.} \bibnamefont{Luo}},
  \bibinfo{author}{\bibfnamefont{P.}~\bibnamefont{Sanchez-Yamagishi}},
  \bibinfo{author}{\bibfnamefont{K.}~\bibnamefont{Watanabe}},
  \bibinfo{author}{\bibfnamefont{T.}~\bibnamefont{Taniguchi}},
  \bibinfo{author}{\bibfnamefont{E.}~\bibnamefont{Kaxiras}},
  \bibnamefont{et~al.}, \bibinfo{journal}{Nature}
  \textbf{\bibinfo{volume}{556}}, \bibinfo{pages}{80}
  (\bibinfo{year}{2018}{\natexlab{b}}).

\bibitem[{\citenamefont{Ortiz et~al.}(2019)\citenamefont{Ortiz, Gomes, Morey,
  Winiarski, Bordelon, Mangum, Oswald, Rodriguez-Rivera, Neilson, Wilson
  et~al.}}]{Ortiz2019}
\bibinfo{author}{\bibfnamefont{B.~R.} \bibnamefont{Ortiz}},
  \bibinfo{author}{\bibfnamefont{L.~C.} \bibnamefont{Gomes}},
  \bibinfo{author}{\bibfnamefont{J.~R.} \bibnamefont{Morey}},
  \bibinfo{author}{\bibfnamefont{M.}~\bibnamefont{Winiarski}},
  \bibinfo{author}{\bibfnamefont{M.}~\bibnamefont{Bordelon}},
  \bibinfo{author}{\bibfnamefont{J.~S.} \bibnamefont{Mangum}},
  \bibinfo{author}{\bibfnamefont{I.~W.~H.} \bibnamefont{Oswald}},
  \bibinfo{author}{\bibfnamefont{J.~A.} \bibnamefont{Rodriguez-Rivera}},
  \bibinfo{author}{\bibfnamefont{J.~R.} \bibnamefont{Neilson}},
  \bibinfo{author}{\bibfnamefont{S.~D.} \bibnamefont{Wilson}},
  \bibnamefont{et~al.}, \bibinfo{journal}{Physical Review Materials}
  \textbf{\bibinfo{volume}{3}}, \bibinfo{pages}{094407} (\bibinfo{year}{2019}).

\bibitem[{\citenamefont{Yin et~al.}(2022)}]{Yin2022}
\bibinfo{author}{\bibfnamefont{J.-X.} \bibnamefont{Yin}} \bibnamefont{et~al.},
  \bibinfo{journal}{Nature Reviews Physics} \textbf{\bibinfo{volume}{4}},
  \bibinfo{pages}{440} (\bibinfo{year}{2022}).

\bibitem[{\citenamefont{Gorbar et~al.}(2019)\citenamefont{Gorbar, Gusynin, and
  Oriekhov}}]{gorbar2019electron}
\bibinfo{author}{\bibfnamefont{E.}~\bibnamefont{Gorbar}},
  \bibinfo{author}{\bibfnamefont{V.}~\bibnamefont{Gusynin}}, \bibnamefont{and}
  \bibinfo{author}{\bibfnamefont{D.}~\bibnamefont{Oriekhov}},
  \bibinfo{journal}{Physical Review B} \textbf{\bibinfo{volume}{99}},
  \bibinfo{pages}{155124} (\bibinfo{year}{2019}).

\bibitem[{\citenamefont{Illes}(2017)}]{illes2017properties}
\bibinfo{author}{\bibfnamefont{E.}~\bibnamefont{Illes}}, Ph.D. thesis,
  \bibinfo{school}{University of Guelph} (\bibinfo{year}{2017}).

\bibitem[{\citenamefont{Roldan et~al.}(2011)\citenamefont{Roldan, Goerbig, and
  Fuchs}}]{roldan2011theory}
\bibinfo{author}{\bibfnamefont{R.}~\bibnamefont{Roldan}},
  \bibinfo{author}{\bibfnamefont{M.}~\bibnamefont{Goerbig}}, \bibnamefont{and}
  \bibinfo{author}{\bibfnamefont{J.-N.} \bibnamefont{Fuchs}},
  \bibinfo{journal}{Physical Review B} \textbf{\bibinfo{volume}{83}},
  \bibinfo{pages}{205406} (\bibinfo{year}{2011}).

\bibitem[{\citenamefont{Weekes et~al.}(2021)\citenamefont{Weekes, Iurov,
  Zhemchuzhna, Gumbs, and Huang}}]{weekes2021generalized}
\bibinfo{author}{\bibfnamefont{N.}~\bibnamefont{Weekes}},
  \bibinfo{author}{\bibfnamefont{A.}~\bibnamefont{Iurov}},
  \bibinfo{author}{\bibfnamefont{L.}~\bibnamefont{Zhemchuzhna}},
  \bibinfo{author}{\bibfnamefont{G.}~\bibnamefont{Gumbs}}, \bibnamefont{and}
  \bibinfo{author}{\bibfnamefont{D.}~\bibnamefont{Huang}},
  \bibinfo{journal}{Physical Review B} \textbf{\bibinfo{volume}{103}},
  \bibinfo{pages}{165429} (\bibinfo{year}{2021}).

\bibitem[{\citenamefont{Raoux et~al.}(2014)\citenamefont{Raoux, Morigi, Fuchs,
  Pi{\'e}chon, and Montambaux}}]{raoux2014dia}
\bibinfo{author}{\bibfnamefont{A.}~\bibnamefont{Raoux}},
  \bibinfo{author}{\bibfnamefont{M.}~\bibnamefont{Morigi}},
  \bibinfo{author}{\bibfnamefont{J.-N.} \bibnamefont{Fuchs}},
  \bibinfo{author}{\bibfnamefont{F.}~\bibnamefont{Pi{\'e}chon}},
  \bibnamefont{and}
  \bibinfo{author}{\bibfnamefont{G.}~\bibnamefont{Montambaux}},
  \bibinfo{journal}{Physical review letters} \textbf{\bibinfo{volume}{112}},
  \bibinfo{pages}{026402} (\bibinfo{year}{2014}).

\bibitem[{\citenamefont{Islam et~al.}(2023{\natexlab{a}})\citenamefont{Islam,
  Biswas, and Basu}}]{islam2023role}
\bibinfo{author}{\bibfnamefont{M.}~\bibnamefont{Islam}},
  \bibinfo{author}{\bibfnamefont{T.}~\bibnamefont{Biswas}}, \bibnamefont{and}
  \bibinfo{author}{\bibfnamefont{S.}~\bibnamefont{Basu}},
  \bibinfo{journal}{arXiv preprint arXiv:2304.08830}
  (\bibinfo{year}{2023}{\natexlab{a}}).

\bibitem[{\citenamefont{Islam et~al.}(2023{\natexlab{b}})\citenamefont{Islam,
  Biswas, and Basu}}]{islam2023effect}
\bibinfo{author}{\bibfnamefont{M.}~\bibnamefont{Islam}},
  \bibinfo{author}{\bibfnamefont{T.}~\bibnamefont{Biswas}}, \bibnamefont{and}
  \bibinfo{author}{\bibfnamefont{S.}~\bibnamefont{Basu}},
  \bibinfo{journal}{Physical Review B} \textbf{\bibinfo{volume}{108}},
  \bibinfo{pages}{085423} (\bibinfo{year}{2023}{\natexlab{b}}).

\bibitem[{\citenamefont{Malcolm and Nicol}(2016)}]{malcolm2016frequency}
\bibinfo{author}{\bibfnamefont{J.}~\bibnamefont{Malcolm}} \bibnamefont{and}
  \bibinfo{author}{\bibfnamefont{E.}~\bibnamefont{Nicol}},
  \bibinfo{journal}{Physical Review B} \textbf{\bibinfo{volume}{93}},
  \bibinfo{pages}{165433} (\bibinfo{year}{2016}).

\bibitem[{\citenamefont{Oriekho}(2023)}]{oriekho2023quantum}
\bibinfo{author}{\bibfnamefont{D.}~\bibnamefont{Oriekho}}, Ph.D. thesis,
  \bibinfo{school}{Leiden University} (\bibinfo{year}{2023}).

\bibitem[{\citenamefont{Oriekhov and Gusynin}(2020)}]{oriekhov2020rkky}
\bibinfo{author}{\bibfnamefont{D.}~\bibnamefont{Oriekhov}} \bibnamefont{and}
  \bibinfo{author}{\bibfnamefont{V.}~\bibnamefont{Gusynin}},
  \bibinfo{journal}{Physical Review B} \textbf{\bibinfo{volume}{101}},
  \bibinfo{pages}{235162} (\bibinfo{year}{2020}).

\bibitem[{\citenamefont{Huang et~al.}(2019)\citenamefont{Huang, Iurov, Xu, Lai,
  and Gumbs}}]{huang2019interplay}
\bibinfo{author}{\bibfnamefont{D.}~\bibnamefont{Huang}},
  \bibinfo{author}{\bibfnamefont{A.}~\bibnamefont{Iurov}},
  \bibinfo{author}{\bibfnamefont{H.-Y.} \bibnamefont{Xu}},
  \bibinfo{author}{\bibfnamefont{Y.-C.} \bibnamefont{Lai}}, \bibnamefont{and}
  \bibinfo{author}{\bibfnamefont{G.}~\bibnamefont{Gumbs}},
  \bibinfo{journal}{Physical Review B} \textbf{\bibinfo{volume}{99}},
  \bibinfo{pages}{245412} (\bibinfo{year}{2019}).

\bibitem[{\citenamefont{Illes and Nicol}(2017)}]{illes2017klein}
\bibinfo{author}{\bibfnamefont{E.}~\bibnamefont{Illes}} \bibnamefont{and}
  \bibinfo{author}{\bibfnamefont{E.}~\bibnamefont{Nicol}},
  \bibinfo{journal}{Physical Review B} \textbf{\bibinfo{volume}{95}},
  \bibinfo{pages}{235432} (\bibinfo{year}{2017}).

\bibitem[{\citenamefont{Iurov et~al.}(2020{\natexlab{b}})\citenamefont{Iurov,
  Zhemchuzhna, Fekete, Gumbs, and Huang}}]{iurov2020klein}
\bibinfo{author}{\bibfnamefont{A.}~\bibnamefont{Iurov}},
  \bibinfo{author}{\bibfnamefont{L.}~\bibnamefont{Zhemchuzhna}},
  \bibinfo{author}{\bibfnamefont{P.}~\bibnamefont{Fekete}},
  \bibinfo{author}{\bibfnamefont{G.}~\bibnamefont{Gumbs}}, \bibnamefont{and}
  \bibinfo{author}{\bibfnamefont{D.}~\bibnamefont{Huang}},
  \bibinfo{journal}{Physical Review Research} \textbf{\bibinfo{volume}{2}},
  \bibinfo{pages}{043245} (\bibinfo{year}{2020}{\natexlab{b}}).

\bibitem[{\citenamefont{Roslyak et~al.}(2010)\citenamefont{Roslyak, Iurov,
  Gumbs, and Huang}}]{roslyak2010unimpeded}
\bibinfo{author}{\bibfnamefont{O.}~\bibnamefont{Roslyak}},
  \bibinfo{author}{\bibfnamefont{A.}~\bibnamefont{Iurov}},
  \bibinfo{author}{\bibfnamefont{G.}~\bibnamefont{Gumbs}}, \bibnamefont{and}
  \bibinfo{author}{\bibfnamefont{D.}~\bibnamefont{Huang}},
  \bibinfo{journal}{Journal of Physics: Condensed Matter}
  \textbf{\bibinfo{volume}{22}}, \bibinfo{pages}{165301}
  (\bibinfo{year}{2010}).

\bibitem[{\citenamefont{Tamang and Biswas}(2023)}]{tamang2023probing}
\bibinfo{author}{\bibfnamefont{L.}~\bibnamefont{Tamang}} \bibnamefont{and}
  \bibinfo{author}{\bibfnamefont{T.}~\bibnamefont{Biswas}},
  \bibinfo{journal}{Physical Review B} \textbf{\bibinfo{volume}{107}},
  \bibinfo{pages}{085408} (\bibinfo{year}{2023}).

\bibitem[{\citenamefont{Iurov et~al.}(2019)\citenamefont{Iurov, Gumbs, and
  Huang}}]{iurov2019peculiar}
\bibinfo{author}{\bibfnamefont{A.}~\bibnamefont{Iurov}},
  \bibinfo{author}{\bibfnamefont{G.}~\bibnamefont{Gumbs}}, \bibnamefont{and}
  \bibinfo{author}{\bibfnamefont{D.}~\bibnamefont{Huang}},
  \bibinfo{journal}{Physical Review B} \textbf{\bibinfo{volume}{99}},
  \bibinfo{pages}{205135} (\bibinfo{year}{2019}).

\bibitem[{\citenamefont{Dey and Ghosh}(2018)}]{dey2018photoinduced}
\bibinfo{author}{\bibfnamefont{B.}~\bibnamefont{Dey}} \bibnamefont{and}
  \bibinfo{author}{\bibfnamefont{T.~K.} \bibnamefont{Ghosh}},
  \bibinfo{journal}{Physical Review B} \textbf{\bibinfo{volume}{98}},
  \bibinfo{pages}{075422} (\bibinfo{year}{2018}).

\bibitem[{\citenamefont{Hou et~al.}(2007)\citenamefont{Hou, Chamon, and
  Mudry}}]{Hou2007}
\bibinfo{author}{\bibfnamefont{C.-Y.} \bibnamefont{Hou}},
  \bibinfo{author}{\bibfnamefont{C.}~\bibnamefont{Chamon}}, \bibnamefont{and}
  \bibinfo{author}{\bibfnamefont{C.}~\bibnamefont{Mudry}},
  \bibinfo{journal}{Physical Review Letters} \textbf{\bibinfo{volume}{98}},
  \bibinfo{pages}{186809} (\bibinfo{year}{2007}).

\bibitem[{\citenamefont{Jackiw and Pi}(2007)}]{Jackiw2007}
\bibinfo{author}{\bibfnamefont{R.}~\bibnamefont{Jackiw}} \bibnamefont{and}
  \bibinfo{author}{\bibfnamefont{S.-Y.} \bibnamefont{Pi}},
  \bibinfo{journal}{Physical Review Letters} \textbf{\bibinfo{volume}{98}},
  \bibinfo{pages}{266402} (\bibinfo{year}{2007}).

\bibitem[{\citenamefont{Herrera and
  Naumis}(2020{\natexlab{a}})}]{herrera2020dynamic}
\bibinfo{author}{\bibfnamefont{S.~A.} \bibnamefont{Herrera}} \bibnamefont{and}
  \bibinfo{author}{\bibfnamefont{G.~G.} \bibnamefont{Naumis}},
  \bibinfo{journal}{Physical Review B} \textbf{\bibinfo{volume}{102}},
  \bibinfo{pages}{205429} (\bibinfo{year}{2020}{\natexlab{a}}).

\bibitem[{\citenamefont{Naumis et~al.}(2017)\citenamefont{Naumis,
  Barraza-Lopez, Oliva-Leyva, and Terrones}}]{Naumis2017}
\bibinfo{author}{\bibfnamefont{G.~G.} \bibnamefont{Naumis}},
  \bibinfo{author}{\bibfnamefont{S.}~\bibnamefont{Barraza-Lopez}},
  \bibinfo{author}{\bibfnamefont{M.}~\bibnamefont{Oliva-Leyva}},
  \bibnamefont{and} \bibinfo{author}{\bibfnamefont{H.}~\bibnamefont{Terrones}},
  \bibinfo{journal}{Reports on Progress in Physics}
  \textbf{\bibinfo{volume}{80}}, \bibinfo{pages}{096501}
  (\bibinfo{year}{2017}).

\bibitem[{\citenamefont{Oliva-Leyva and Naumis}(2013)}]{OlivaLeyva2013}
\bibinfo{author}{\bibfnamefont{M.}~\bibnamefont{Oliva-Leyva}} \bibnamefont{and}
  \bibinfo{author}{\bibfnamefont{G.~G.} \bibnamefont{Naumis}},
  \bibinfo{journal}{Physical Review B} \textbf{\bibinfo{volume}{88}},
  \bibinfo{pages}{085430} (\bibinfo{year}{2013}).

\bibitem[{\citenamefont{Pereira et~al.}(2009)\citenamefont{Pereira,
  Castro~Neto, and Peres}}]{Pereira2009}
\bibinfo{author}{\bibfnamefont{V.~M.} \bibnamefont{Pereira}},
  \bibinfo{author}{\bibfnamefont{A.~H.} \bibnamefont{Castro~Neto}},
  \bibnamefont{and} \bibinfo{author}{\bibfnamefont{N.~M.~R.}
  \bibnamefont{Peres}}, \bibinfo{journal}{Physical Review B}
  \textbf{\bibinfo{volume}{80}}, \bibinfo{pages}{045401}
  (\bibinfo{year}{2009}).

\bibitem[{\citenamefont{Mojarro et~al.}(2020)\citenamefont{Mojarro,
  Ibarra-Sierra, Sandoval-Santana, Carrillo-Bastos, and
  Naumis}}]{mojarro2020dynamical}
\bibinfo{author}{\bibfnamefont{M.}~\bibnamefont{Mojarro}},
  \bibinfo{author}{\bibfnamefont{V.}~\bibnamefont{Ibarra-Sierra}},
  \bibinfo{author}{\bibfnamefont{J.}~\bibnamefont{Sandoval-Santana}},
  \bibinfo{author}{\bibfnamefont{R.}~\bibnamefont{Carrillo-Bastos}},
  \bibnamefont{and} \bibinfo{author}{\bibfnamefont{G.~G.}
  \bibnamefont{Naumis}}, \bibinfo{journal}{Physical Review B}
  \textbf{\bibinfo{volume}{102}}, \bibinfo{pages}{165301}
  (\bibinfo{year}{2020}).

\bibitem[{\citenamefont{Herrera and
  Naumis}(2020{\natexlab{b}})}]{herrera2020electronic}
\bibinfo{author}{\bibfnamefont{S.~A.} \bibnamefont{Herrera}} \bibnamefont{and}
  \bibinfo{author}{\bibfnamefont{G.~G.} \bibnamefont{Naumis}},
  \bibinfo{journal}{Physical Review B} \textbf{\bibinfo{volume}{101}},
  \bibinfo{pages}{205413} (\bibinfo{year}{2020}{\natexlab{b}}).

\bibitem[{\citenamefont{Iurov et~al.}(2023{\natexlab{a}})\citenamefont{Iurov,
  Zhemchuzhna, Gumbs, and Huang}}]{iurov2023application}
\bibinfo{author}{\bibfnamefont{A.}~\bibnamefont{Iurov}},
  \bibinfo{author}{\bibfnamefont{L.}~\bibnamefont{Zhemchuzhna}},
  \bibinfo{author}{\bibfnamefont{G.}~\bibnamefont{Gumbs}}, \bibnamefont{and}
  \bibinfo{author}{\bibfnamefont{D.}~\bibnamefont{Huang}},
  \bibinfo{journal}{Applied Sciences} \textbf{\bibinfo{volume}{13}},
  \bibinfo{pages}{6095} (\bibinfo{year}{2023}{\natexlab{a}}).

\bibitem[{\citenamefont{de~Juan et~al.}(2012)\citenamefont{de~Juan, Sturla, and
  Vozmediano}}]{deJuan2012}
\bibinfo{author}{\bibfnamefont{F.}~\bibnamefont{de~Juan}},
  \bibinfo{author}{\bibfnamefont{M.}~\bibnamefont{Sturla}}, \bibnamefont{and}
  \bibinfo{author}{\bibfnamefont{M.~A.~H.} \bibnamefont{Vozmediano}},
  \bibinfo{journal}{Physical Review Letters} \textbf{\bibinfo{volume}{108}},
  \bibinfo{pages}{227205} (\bibinfo{year}{2012}).

\bibitem[{\citenamefont{Andrade et~al.}(2020)\citenamefont{Andrade,
  Carrillo-Bastos, Pantale{\'o}n, and Mireles}}]{Andrade2019Nanoribbons}
\bibinfo{author}{\bibfnamefont{E.}~\bibnamefont{Andrade}},
  \bibinfo{author}{\bibfnamefont{R.}~\bibnamefont{Carrillo-Bastos}},
  \bibinfo{author}{\bibfnamefont{P.~A.} \bibnamefont{Pantale{\'o}n}},
  \bibnamefont{and} \bibinfo{author}{\bibfnamefont{F.}~\bibnamefont{Mireles}},
  \bibinfo{journal}{Physical Review B} \textbf{\bibinfo{volume}{101}},
  \bibinfo{pages}{035416} (\bibinfo{year}{2020}).

\bibitem[{\citenamefont{Ruiz-Tijerina et~al.}(2019)\citenamefont{Ruiz-Tijerina,
  Andrade, Carrillo-Bastos, Mireles, and Naumis}}]{RuizTijerina2019}
\bibinfo{author}{\bibfnamefont{D.~A.} \bibnamefont{Ruiz-Tijerina}},
  \bibinfo{author}{\bibfnamefont{E.}~\bibnamefont{Andrade}},
  \bibinfo{author}{\bibfnamefont{R.}~\bibnamefont{Carrillo-Bastos}},
  \bibinfo{author}{\bibfnamefont{F.}~\bibnamefont{Mireles}}, \bibnamefont{and}
  \bibinfo{author}{\bibfnamefont{G.~G.} \bibnamefont{Naumis}},
  \bibinfo{journal}{Physical Review B} \textbf{\bibinfo{volume}{100}},
  \bibinfo{pages}{075431} (\bibinfo{year}{2019}).

\bibitem[{\citenamefont{Andrade et~al.}(2019)\citenamefont{Andrade,
  Carrillo-Bastos, and Naumis}}]{andrade2019valley}
\bibinfo{author}{\bibfnamefont{E.}~\bibnamefont{Andrade}},
  \bibinfo{author}{\bibfnamefont{R.}~\bibnamefont{Carrillo-Bastos}},
  \bibnamefont{and} \bibinfo{author}{\bibfnamefont{G.~G.}
  \bibnamefont{Naumis}}, \bibinfo{journal}{Physical Review B}
  \textbf{\bibinfo{volume}{99}}, \bibinfo{pages}{035411}
  (\bibinfo{year}{2019}).

\bibitem[{\citenamefont{Andrade et~al.}(2025)\citenamefont{Andrade,
  Carrillo-Bastos, and Naumis}}]{andrade2025topical}
\bibinfo{author}{\bibfnamefont{E.}~\bibnamefont{Andrade}},
  \bibinfo{author}{\bibfnamefont{R.}~\bibnamefont{Carrillo-Bastos}},
  \bibnamefont{and} \bibinfo{author}{\bibfnamefont{G.~G.}
  \bibnamefont{Naumis}}, \bibinfo{journal}{Journal of Physics: Condensed
  Matter} \textbf{\bibinfo{volume}{37}}, \bibinfo{pages}{193003}
  (\bibinfo{year}{2025}).

\bibitem[{\citenamefont{Eom and Koo}(2020)}]{Eom2020}
\bibinfo{author}{\bibfnamefont{D.}~\bibnamefont{Eom}} \bibnamefont{and}
  \bibinfo{author}{\bibfnamefont{J.-Y.} \bibnamefont{Koo}},
  \bibinfo{journal}{Nanoscale} \textbf{\bibinfo{volume}{12}},
  \bibinfo{pages}{19604} (\bibinfo{year}{2020}).

\bibitem[{\citenamefont{S{\'a}nchez-Gonz{\'a}lez
  et~al.}(2025)\citenamefont{S{\'a}nchez-Gonz{\'a}lez, Mojarro, Maytorena, and
  Carrillo-Bastos}}]{sanchez2025band}
\bibinfo{author}{\bibfnamefont{L.~E.} \bibnamefont{S{\'a}nchez-Gonz{\'a}lez}},
  \bibinfo{author}{\bibfnamefont{M.}~\bibnamefont{Mojarro}},
  \bibinfo{author}{\bibfnamefont{J.~A.} \bibnamefont{Maytorena}},
  \bibnamefont{and}
  \bibinfo{author}{\bibfnamefont{R.}~\bibnamefont{Carrillo-Bastos}},
  \bibinfo{journal}{Physical Review B} \textbf{\bibinfo{volume}{111}},
  \bibinfo{pages}{115417} (\bibinfo{year}{2025}).

\bibitem[{\citenamefont{Politano and Chiarello}(2014)}]{politano2014plasmon}
\bibinfo{author}{\bibfnamefont{A.}~\bibnamefont{Politano}} \bibnamefont{and}
  \bibinfo{author}{\bibfnamefont{G.}~\bibnamefont{Chiarello}},
  \bibinfo{journal}{Nanoscale} \textbf{\bibinfo{volume}{6}},
  \bibinfo{pages}{10927} (\bibinfo{year}{2014}).

\bibitem[{\citenamefont{Yan et~al.}(2012)\citenamefont{Yan, Li, Li, Zhu,
  Avouris, and Xia}}]{yan2012infrared}
\bibinfo{author}{\bibfnamefont{H.}~\bibnamefont{Yan}},
  \bibinfo{author}{\bibfnamefont{Z.}~\bibnamefont{Li}},
  \bibinfo{author}{\bibfnamefont{X.}~\bibnamefont{Li}},
  \bibinfo{author}{\bibfnamefont{W.}~\bibnamefont{Zhu}},
  \bibinfo{author}{\bibfnamefont{P.}~\bibnamefont{Avouris}}, \bibnamefont{and}
  \bibinfo{author}{\bibfnamefont{F.}~\bibnamefont{Xia}}, \bibinfo{journal}{Nano
  letters} \textbf{\bibinfo{volume}{12}}, \bibinfo{pages}{3766}
  (\bibinfo{year}{2012}).

\bibitem[{\citenamefont{Hwang and Sarma}(2007)}]{hwang2007dielectric}
\bibinfo{author}{\bibfnamefont{E.}~\bibnamefont{Hwang}} \bibnamefont{and}
  \bibinfo{author}{\bibfnamefont{S.~D.} \bibnamefont{Sarma}},
  \bibinfo{journal}{Physical Review B} \textbf{\bibinfo{volume}{75}},
  \bibinfo{pages}{205418} (\bibinfo{year}{2007}).

\bibitem[{\citenamefont{Polini et~al.}(2008)\citenamefont{Polini, Asgari,
  Borghi, Barlas, Pereg-Barnea, and MacDonald}}]{polini2008plasmons}
\bibinfo{author}{\bibfnamefont{M.}~\bibnamefont{Polini}},
  \bibinfo{author}{\bibfnamefont{R.}~\bibnamefont{Asgari}},
  \bibinfo{author}{\bibfnamefont{G.}~\bibnamefont{Borghi}},
  \bibinfo{author}{\bibfnamefont{Y.}~\bibnamefont{Barlas}},
  \bibinfo{author}{\bibfnamefont{T.}~\bibnamefont{Pereg-Barnea}},
  \bibnamefont{and}
  \bibinfo{author}{\bibfnamefont{A.}~\bibnamefont{MacDonald}},
  \bibinfo{journal}{Physical Review B} \textbf{\bibinfo{volume}{77}},
  \bibinfo{pages}{081411} (\bibinfo{year}{2008}).

\bibitem[{\citenamefont{Jablan et~al.}(2013)\citenamefont{Jablan,
  Solja{\v{c}}i{\'c}, and Buljan}}]{jablan2013plasmons}
\bibinfo{author}{\bibfnamefont{M.}~\bibnamefont{Jablan}},
  \bibinfo{author}{\bibfnamefont{M.}~\bibnamefont{Solja{\v{c}}i{\'c}}},
  \bibnamefont{and} \bibinfo{author}{\bibfnamefont{H.}~\bibnamefont{Buljan}},
  \bibinfo{journal}{Proceedings of the IEEE} \textbf{\bibinfo{volume}{101}},
  \bibinfo{pages}{1689} (\bibinfo{year}{2013}).

\bibitem[{\citenamefont{Zhang et~al.}(2012)\citenamefont{Zhang, Zhang, and
  Xu}}]{zhang2012surface}
\bibinfo{author}{\bibfnamefont{J.}~\bibnamefont{Zhang}},
  \bibinfo{author}{\bibfnamefont{L.}~\bibnamefont{Zhang}}, \bibnamefont{and}
  \bibinfo{author}{\bibfnamefont{W.}~\bibnamefont{Xu}},
  \bibinfo{journal}{Journal of Physics D: Applied Physics}
  \textbf{\bibinfo{volume}{45}}, \bibinfo{pages}{113001}
  (\bibinfo{year}{2012}).

\bibitem[{\citenamefont{Wunsch et~al.}(2006)\citenamefont{Wunsch, Stauber,
  Sols, and Guinea}}]{wunsch2006dynamical}
\bibinfo{author}{\bibfnamefont{B.}~\bibnamefont{Wunsch}},
  \bibinfo{author}{\bibfnamefont{T.}~\bibnamefont{Stauber}},
  \bibinfo{author}{\bibfnamefont{F.}~\bibnamefont{Sols}}, \bibnamefont{and}
  \bibinfo{author}{\bibfnamefont{F.}~\bibnamefont{Guinea}},
  \bibinfo{journal}{New Journal of Physics} \textbf{\bibinfo{volume}{8}},
  \bibinfo{pages}{318} (\bibinfo{year}{2006}).

\bibitem[{\citenamefont{Constant et~al.}(2016)\citenamefont{Constant, Hornett,
  Chang, and Hendry}}]{constant2016all}
\bibinfo{author}{\bibfnamefont{T.~J.} \bibnamefont{Constant}},
  \bibinfo{author}{\bibfnamefont{S.~M.} \bibnamefont{Hornett}},
  \bibinfo{author}{\bibfnamefont{D.~E.} \bibnamefont{Chang}}, \bibnamefont{and}
  \bibinfo{author}{\bibfnamefont{E.}~\bibnamefont{Hendry}},
  \bibinfo{journal}{Nature Physics} \textbf{\bibinfo{volume}{12}},
  \bibinfo{pages}{124} (\bibinfo{year}{2016}).

\bibitem[{\citenamefont{Gamayun}(2011)}]{gamayun2011dynamical}
\bibinfo{author}{\bibfnamefont{O.}~\bibnamefont{Gamayun}},
  \bibinfo{journal}{Physical Review B—Condensed Matter and Materials Physics}
  \textbf{\bibinfo{volume}{84}}, \bibinfo{pages}{085112}
  (\bibinfo{year}{2011}).

\bibitem[{\citenamefont{Pyatkovskiy}(2008)}]{pyatkovskiy2008dynamical}
\bibinfo{author}{\bibfnamefont{P.}~\bibnamefont{Pyatkovskiy}},
  \bibinfo{journal}{Journal of Physics: Condensed Matter}
  \textbf{\bibinfo{volume}{21}}, \bibinfo{pages}{025506}
  (\bibinfo{year}{2008}).

\bibitem[{\citenamefont{Tabert and Nicol}(2014)}]{tabert2014dynamical}
\bibinfo{author}{\bibfnamefont{C.~J.} \bibnamefont{Tabert}} \bibnamefont{and}
  \bibinfo{author}{\bibfnamefont{E.~J.} \bibnamefont{Nicol}},
  \bibinfo{journal}{Physical Review B} \textbf{\bibinfo{volume}{89}},
  \bibinfo{pages}{195410} (\bibinfo{year}{2014}).

\bibitem[{\citenamefont{Torbatian et~al.}(2021)\citenamefont{Torbatian, Novko,
  and Asgari}}]{torbatian2021hyperbolic}
\bibinfo{author}{\bibfnamefont{Z.}~\bibnamefont{Torbatian}},
  \bibinfo{author}{\bibfnamefont{D.}~\bibnamefont{Novko}}, \bibnamefont{and}
  \bibinfo{author}{\bibfnamefont{R.}~\bibnamefont{Asgari}},
  \bibinfo{journal}{Physical Review B} \textbf{\bibinfo{volume}{104}},
  \bibinfo{pages}{075432} (\bibinfo{year}{2021}).

\bibitem[{\citenamefont{Hayn et~al.}(2021)\citenamefont{Hayn, Wei, Silkin, and
  van~den Brink}}]{hayn2021plasmons}
\bibinfo{author}{\bibfnamefont{R.}~\bibnamefont{Hayn}},
  \bibinfo{author}{\bibfnamefont{T.}~\bibnamefont{Wei}},
  \bibinfo{author}{\bibfnamefont{V.~M.} \bibnamefont{Silkin}},
  \bibnamefont{and} \bibinfo{author}{\bibfnamefont{J.}~\bibnamefont{van~den
  Brink}}, \bibinfo{journal}{Physical Review Materials}
  \textbf{\bibinfo{volume}{5}}, \bibinfo{pages}{024201} (\bibinfo{year}{2021}).

\bibitem[{\citenamefont{Gomes and Ramos}(2021)}]{gomes2021tilted}
\bibinfo{author}{\bibfnamefont{Y.}~\bibnamefont{Gomes}} \bibnamefont{and}
  \bibinfo{author}{\bibfnamefont{R.~O.} \bibnamefont{Ramos}},
  \bibinfo{journal}{Physical Review B} \textbf{\bibinfo{volume}{104}},
  \bibinfo{pages}{245111} (\bibinfo{year}{2021}).

\bibitem[{\citenamefont{Ross-Harvey et~al.}(2025)\citenamefont{Ross-Harvey,
  Iurov, Zhemchuzhna, Gumbs, Huang, and Fekete}}]{ross2025dynamical}
\bibinfo{author}{\bibfnamefont{G.}~\bibnamefont{Ross-Harvey}},
  \bibinfo{author}{\bibfnamefont{A.}~\bibnamefont{Iurov}},
  \bibinfo{author}{\bibfnamefont{L.}~\bibnamefont{Zhemchuzhna}},
  \bibinfo{author}{\bibfnamefont{G.}~\bibnamefont{Gumbs}},
  \bibinfo{author}{\bibfnamefont{D.}~\bibnamefont{Huang}}, \bibnamefont{and}
  \bibinfo{author}{\bibfnamefont{P.}~\bibnamefont{Fekete}},
  \bibinfo{journal}{Physical Review B} \textbf{\bibinfo{volume}{111}},
  \bibinfo{pages}{045413} (\bibinfo{year}{2025}).

\bibitem[{\citenamefont{Badalyan et~al.}(2009)\citenamefont{Badalyan,
  Matos-Abiague, Vignale, and Fabian}}]{badalyan2009anisotropic}
\bibinfo{author}{\bibfnamefont{S.~M.} \bibnamefont{Badalyan}},
  \bibinfo{author}{\bibfnamefont{A.}~\bibnamefont{Matos-Abiague}},
  \bibinfo{author}{\bibfnamefont{G.}~\bibnamefont{Vignale}}, \bibnamefont{and}
  \bibinfo{author}{\bibfnamefont{J.}~\bibnamefont{Fabian}},
  \bibinfo{journal}{Physical Review B—Condensed Matter and Materials Physics}
  \textbf{\bibinfo{volume}{79}}, \bibinfo{pages}{205305}
  (\bibinfo{year}{2009}).

\bibitem[{\citenamefont{Trescher et~al.}(2015)\citenamefont{Trescher, Sbierski,
  Brouwer, and Bergholtz}}]{trescher2015quantum}
\bibinfo{author}{\bibfnamefont{M.}~\bibnamefont{Trescher}},
  \bibinfo{author}{\bibfnamefont{B.}~\bibnamefont{Sbierski}},
  \bibinfo{author}{\bibfnamefont{P.~W.} \bibnamefont{Brouwer}},
  \bibnamefont{and} \bibinfo{author}{\bibfnamefont{E.~J.}
  \bibnamefont{Bergholtz}}, \bibinfo{journal}{Physical Review B}
  \textbf{\bibinfo{volume}{91}}, \bibinfo{pages}{115135}
  (\bibinfo{year}{2015}).

\bibitem[{\citenamefont{Sriram et~al.}(2020)\citenamefont{Sriram, Manikandan,
  Chuang, and Chueh}}]{sriram2020hybridizing}
\bibinfo{author}{\bibfnamefont{P.}~\bibnamefont{Sriram}},
  \bibinfo{author}{\bibfnamefont{A.}~\bibnamefont{Manikandan}},
  \bibinfo{author}{\bibfnamefont{F.-C.} \bibnamefont{Chuang}},
  \bibnamefont{and} \bibinfo{author}{\bibfnamefont{Y.-L.} \bibnamefont{Chueh}},
  \bibinfo{journal}{Small} \textbf{\bibinfo{volume}{16}},
  \bibinfo{pages}{1904271} (\bibinfo{year}{2020}).

\bibitem[{\citenamefont{Andersen and Thygesen}(2013)}]{andersen2013plasmons}
\bibinfo{author}{\bibfnamefont{K.}~\bibnamefont{Andersen}} \bibnamefont{and}
  \bibinfo{author}{\bibfnamefont{K.~S.} \bibnamefont{Thygesen}},
  \bibinfo{journal}{arXiv preprint arXiv:1311.0158}  (\bibinfo{year}{2013}).

\bibitem[{\citenamefont{Scholz et~al.}(2013)\citenamefont{Scholz, Stauber, and
  Schliemann}}]{scholz2013plasmons}
\bibinfo{author}{\bibfnamefont{A.}~\bibnamefont{Scholz}},
  \bibinfo{author}{\bibfnamefont{T.}~\bibnamefont{Stauber}}, \bibnamefont{and}
  \bibinfo{author}{\bibfnamefont{J.}~\bibnamefont{Schliemann}},
  \bibinfo{journal}{arXiv preprint arXiv:1306.1666}  (\bibinfo{year}{2013}).

\bibitem[{\citenamefont{Sadhukhan et~al.}(2020)\citenamefont{Sadhukhan,
  Politano, and Agarwal}}]{sadhukhan2020novel}
\bibinfo{author}{\bibfnamefont{K.}~\bibnamefont{Sadhukhan}},
  \bibinfo{author}{\bibfnamefont{A.}~\bibnamefont{Politano}}, \bibnamefont{and}
  \bibinfo{author}{\bibfnamefont{A.}~\bibnamefont{Agarwal}},
  \bibinfo{journal}{Physical Review Letters} \textbf{\bibinfo{volume}{124}},
  \bibinfo{pages}{046803} (\bibinfo{year}{2020}).

\bibitem[{\citenamefont{Dutta et~al.}(2023)\citenamefont{Dutta, Chakraborty,
  and Agarwal}}]{dutta2023intrinsic}
\bibinfo{author}{\bibfnamefont{D.}~\bibnamefont{Dutta}},
  \bibinfo{author}{\bibfnamefont{A.}~\bibnamefont{Chakraborty}},
  \bibnamefont{and} \bibinfo{author}{\bibfnamefont{A.}~\bibnamefont{Agarwal}},
  \bibinfo{journal}{Physical Review B} \textbf{\bibinfo{volume}{107}},
  \bibinfo{pages}{165404} (\bibinfo{year}{2023}).

\bibitem[{\citenamefont{Dey and Ghosh}(2022)}]{dey2022dynamical}
\bibinfo{author}{\bibfnamefont{B.}~\bibnamefont{Dey}} \bibnamefont{and}
  \bibinfo{author}{\bibfnamefont{T.~K.} \bibnamefont{Ghosh}},
  \bibinfo{journal}{Journal of Physics: Condensed Matter}
  \textbf{\bibinfo{volume}{34}}, \bibinfo{pages}{255701}
  (\bibinfo{year}{2022}).

\bibitem[{\citenamefont{Shitrit et~al.}(2013)\citenamefont{Shitrit, Yulevich,
  Kleiner, and Hasman}}]{shitrit2013spin}
\bibinfo{author}{\bibfnamefont{N.}~\bibnamefont{Shitrit}},
  \bibinfo{author}{\bibfnamefont{I.}~\bibnamefont{Yulevich}},
  \bibinfo{author}{\bibfnamefont{V.}~\bibnamefont{Kleiner}}, \bibnamefont{and}
  \bibinfo{author}{\bibfnamefont{E.}~\bibnamefont{Hasman}},
  \bibinfo{journal}{Applied Physics Letters} \textbf{\bibinfo{volume}{103}}
  (\bibinfo{year}{2013}).

\bibitem[{\citenamefont{Sarma and Li}(2013)}]{sarma2013intrinsic}
\bibinfo{author}{\bibfnamefont{S.~D.} \bibnamefont{Sarma}} \bibnamefont{and}
  \bibinfo{author}{\bibfnamefont{Q.}~\bibnamefont{Li}},
  \bibinfo{journal}{Physical Review B} \textbf{\bibinfo{volume}{87}},
  \bibinfo{pages}{235418} (\bibinfo{year}{2013}).

\bibitem[{\citenamefont{Karimi and Knezevic}(2017)}]{karimi2017plasmons}
\bibinfo{author}{\bibfnamefont{F.}~\bibnamefont{Karimi}} \bibnamefont{and}
  \bibinfo{author}{\bibfnamefont{I.}~\bibnamefont{Knezevic}},
  \bibinfo{journal}{Physical Review B} \textbf{\bibinfo{volume}{96}},
  \bibinfo{pages}{125417} (\bibinfo{year}{2017}).

\bibitem[{\citenamefont{Brey and Fertig}(2007)}]{brey2007elementary}
\bibinfo{author}{\bibfnamefont{L.}~\bibnamefont{Brey}} \bibnamefont{and}
  \bibinfo{author}{\bibfnamefont{H.}~\bibnamefont{Fertig}},
  \bibinfo{journal}{Physical Review B} \textbf{\bibinfo{volume}{75}},
  \bibinfo{pages}{125434} (\bibinfo{year}{2007}).

\bibitem[{\citenamefont{Fei et~al.}(2015)\citenamefont{Fei, Goldflam, Wu, Dai,
  Wagner, McLeod, Liu, Post, Zhu, Janssen et~al.}}]{fei2015edge}
\bibinfo{author}{\bibfnamefont{Z.}~\bibnamefont{Fei}},
  \bibinfo{author}{\bibfnamefont{M.}~\bibnamefont{Goldflam}},
  \bibinfo{author}{\bibfnamefont{J.-S.} \bibnamefont{Wu}},
  \bibinfo{author}{\bibfnamefont{S.}~\bibnamefont{Dai}},
  \bibinfo{author}{\bibfnamefont{M.}~\bibnamefont{Wagner}},
  \bibinfo{author}{\bibfnamefont{A.}~\bibnamefont{McLeod}},
  \bibinfo{author}{\bibfnamefont{M.}~\bibnamefont{Liu}},
  \bibinfo{author}{\bibfnamefont{K.}~\bibnamefont{Post}},
  \bibinfo{author}{\bibfnamefont{S.}~\bibnamefont{Zhu}},
  \bibinfo{author}{\bibfnamefont{G.}~\bibnamefont{Janssen}},
  \bibnamefont{et~al.}, \bibinfo{journal}{Nano letters}
  \textbf{\bibinfo{volume}{15}}, \bibinfo{pages}{8271} (\bibinfo{year}{2015}).

\bibitem[{\citenamefont{Wang}(2005)}]{wang2005plasmon}
\bibinfo{author}{\bibfnamefont{X.-F.} \bibnamefont{Wang}},
  \bibinfo{journal}{Physical Review B} \textbf{\bibinfo{volume}{72}},
  \bibinfo{pages}{085317} (\bibinfo{year}{2005}).

\bibitem[{\citenamefont{Kajiwara et~al.}(2016)\citenamefont{Kajiwara, Urade,
  Nakata, Nakanishi, and Kitano}}]{kajiwara2016observation}
\bibinfo{author}{\bibfnamefont{S.}~\bibnamefont{Kajiwara}},
  \bibinfo{author}{\bibfnamefont{Y.}~\bibnamefont{Urade}},
  \bibinfo{author}{\bibfnamefont{Y.}~\bibnamefont{Nakata}},
  \bibinfo{author}{\bibfnamefont{T.}~\bibnamefont{Nakanishi}},
  \bibnamefont{and} \bibinfo{author}{\bibfnamefont{M.}~\bibnamefont{Kitano}},
  \bibinfo{journal}{Physical Review B} \textbf{\bibinfo{volume}{93}},
  \bibinfo{pages}{075126} (\bibinfo{year}{2016}).

\bibitem[{\citenamefont{Iurov et~al.}(2021)\citenamefont{Iurov, Zhemchuzhna,
  Gumbs, Huang, Fekete, Anwar, Dahal, and Weekes}}]{iurov2021tailoring}
\bibinfo{author}{\bibfnamefont{A.}~\bibnamefont{Iurov}},
  \bibinfo{author}{\bibfnamefont{L.}~\bibnamefont{Zhemchuzhna}},
  \bibinfo{author}{\bibfnamefont{G.}~\bibnamefont{Gumbs}},
  \bibinfo{author}{\bibfnamefont{D.}~\bibnamefont{Huang}},
  \bibinfo{author}{\bibfnamefont{P.}~\bibnamefont{Fekete}},
  \bibinfo{author}{\bibfnamefont{F.}~\bibnamefont{Anwar}},
  \bibinfo{author}{\bibfnamefont{D.}~\bibnamefont{Dahal}}, \bibnamefont{and}
  \bibinfo{author}{\bibfnamefont{N.}~\bibnamefont{Weekes}},
  \bibinfo{journal}{Scientific reports} \textbf{\bibinfo{volume}{11}},
  \bibinfo{pages}{20577} (\bibinfo{year}{2021}).

\bibitem[{\citenamefont{Gumbs et~al.}(2015{\natexlab{a}})\citenamefont{Gumbs,
  Iurov, and Horing}}]{gumbs2015nonlocal}
\bibinfo{author}{\bibfnamefont{G.}~\bibnamefont{Gumbs}},
  \bibinfo{author}{\bibfnamefont{A.}~\bibnamefont{Iurov}}, \bibnamefont{and}
  \bibinfo{author}{\bibfnamefont{N.}~\bibnamefont{Horing}},
  \bibinfo{journal}{Physical Review B} \textbf{\bibinfo{volume}{91}},
  \bibinfo{pages}{235416} (\bibinfo{year}{2015}{\natexlab{a}}).

\bibitem[{\citenamefont{Yao et~al.}(2018)\citenamefont{Yao, Liu, Huang, Choi,
  Xie, Flor~Flores, Wu, Yu, Kwong, Huang et~al.}}]{yao2018broadband}
\bibinfo{author}{\bibfnamefont{B.}~\bibnamefont{Yao}},
  \bibinfo{author}{\bibfnamefont{Y.}~\bibnamefont{Liu}},
  \bibinfo{author}{\bibfnamefont{S.-W.} \bibnamefont{Huang}},
  \bibinfo{author}{\bibfnamefont{C.}~\bibnamefont{Choi}},
  \bibinfo{author}{\bibfnamefont{Z.}~\bibnamefont{Xie}},
  \bibinfo{author}{\bibfnamefont{J.}~\bibnamefont{Flor~Flores}},
  \bibinfo{author}{\bibfnamefont{Y.}~\bibnamefont{Wu}},
  \bibinfo{author}{\bibfnamefont{M.}~\bibnamefont{Yu}},
  \bibinfo{author}{\bibfnamefont{D.-L.} \bibnamefont{Kwong}},
  \bibinfo{author}{\bibfnamefont{Y.}~\bibnamefont{Huang}},
  \bibnamefont{et~al.}, \bibinfo{journal}{Nature Photonics}
  \textbf{\bibinfo{volume}{12}}, \bibinfo{pages}{22} (\bibinfo{year}{2018}).

\bibitem[{\citenamefont{Li et~al.}(2017)\citenamefont{Li, Ren, and
  He}}]{li2017first}
\bibinfo{author}{\bibfnamefont{P.}~\bibnamefont{Li}},
  \bibinfo{author}{\bibfnamefont{X.}~\bibnamefont{Ren}}, \bibnamefont{and}
  \bibinfo{author}{\bibfnamefont{L.}~\bibnamefont{He}},
  \bibinfo{journal}{Physical Review B} \textbf{\bibinfo{volume}{96}},
  \bibinfo{pages}{165417} (\bibinfo{year}{2017}).

\bibitem[{\citenamefont{Iurov et~al.}(2017{\natexlab{b}})\citenamefont{Iurov,
  Huang, and Zhemchuzhna}}]{iurov2017controlling}
\bibinfo{author}{\bibfnamefont{G.}~\bibnamefont{Iurov},
  \bibfnamefont{Andrii~and}},
  \bibinfo{author}{\bibfnamefont{D.}~\bibnamefont{Huang}}, \bibnamefont{and}
  \bibinfo{author}{\bibfnamefont{L.}~\bibnamefont{Zhemchuzhna}},
  \bibinfo{journal}{Journal of Applied Physics} \textbf{\bibinfo{volume}{121}}
  (\bibinfo{year}{2017}{\natexlab{b}}).

\bibitem[{\citenamefont{Sarma and Madhukar}(1981)}]{sarma1981collective}
\bibinfo{author}{\bibfnamefont{S.~D.} \bibnamefont{Sarma}} \bibnamefont{and}
  \bibinfo{author}{\bibfnamefont{A.}~\bibnamefont{Madhukar}},
  \bibinfo{journal}{Physical Review B} \textbf{\bibinfo{volume}{23}},
  \bibinfo{pages}{805} (\bibinfo{year}{1981}).

\bibitem[{\citenamefont{Yerin et~al.}(2023)\citenamefont{Yerin, Varlamov,
  Felici, and Di~Carlo}}]{yerin2023dielectric}
\bibinfo{author}{\bibfnamefont{Y.}~\bibnamefont{Yerin}},
  \bibinfo{author}{\bibfnamefont{A.}~\bibnamefont{Varlamov}},
  \bibinfo{author}{\bibfnamefont{R.}~\bibnamefont{Felici}}, \bibnamefont{and}
  \bibinfo{author}{\bibfnamefont{A.}~\bibnamefont{Di~Carlo}},
  \bibinfo{journal}{Physical Review B} \textbf{\bibinfo{volume}{108}},
  \bibinfo{pages}{075134} (\bibinfo{year}{2023}).

\bibitem[{\citenamefont{Ju et~al.}(1993)\citenamefont{Ju, Bulgac, and
  Keller}}]{ju1993excitation}
\bibinfo{author}{\bibfnamefont{N.}~\bibnamefont{Ju}},
  \bibinfo{author}{\bibfnamefont{A.}~\bibnamefont{Bulgac}}, \bibnamefont{and}
  \bibinfo{author}{\bibfnamefont{J.~W.} \bibnamefont{Keller}},
  \bibinfo{journal}{Physical Review B} \textbf{\bibinfo{volume}{48}},
  \bibinfo{pages}{9071} (\bibinfo{year}{1993}).

\bibitem[{\citenamefont{Gumbs et~al.}(2014{\natexlab{b}})\citenamefont{Gumbs,
  Balassis, Iurov, and Fekete}}]{gumbs2014strongly}
\bibinfo{author}{\bibfnamefont{G.}~\bibnamefont{Gumbs}},
  \bibinfo{author}{\bibfnamefont{A.}~\bibnamefont{Balassis}},
  \bibinfo{author}{\bibfnamefont{A.}~\bibnamefont{Iurov}}, \bibnamefont{and}
  \bibinfo{author}{\bibfnamefont{P.}~\bibnamefont{Fekete}},
  \bibinfo{journal}{The Scientific World Journal}
  \textbf{\bibinfo{volume}{2014}}, \bibinfo{pages}{726303}
  (\bibinfo{year}{2014}{\natexlab{b}}).

\bibitem[{\citenamefont{Berini and De~Leon}(2012)}]{berini2012surface}
\bibinfo{author}{\bibfnamefont{P.}~\bibnamefont{Berini}} \bibnamefont{and}
  \bibinfo{author}{\bibfnamefont{I.}~\bibnamefont{De~Leon}},
  \bibinfo{journal}{Nature photonics} \textbf{\bibinfo{volume}{6}},
  \bibinfo{pages}{16} (\bibinfo{year}{2012}).

\bibitem[{\citenamefont{Barnes}(2006)}]{barnes2006surface}
\bibinfo{author}{\bibfnamefont{W.~L.} \bibnamefont{Barnes}},
  \bibinfo{journal}{Journal of optics A: pure and applied optics}
  \textbf{\bibinfo{volume}{8}}, \bibinfo{pages}{S87} (\bibinfo{year}{2006}).

\bibitem[{\citenamefont{Balassis et~al.}(2020)\citenamefont{Balassis, Dahal,
  Gumbs, Iurov, Huang, and Roslyak}}]{balassis2020magnetoplasmons}
\bibinfo{author}{\bibfnamefont{A.}~\bibnamefont{Balassis}},
  \bibinfo{author}{\bibfnamefont{D.}~\bibnamefont{Dahal}},
  \bibinfo{author}{\bibfnamefont{G.}~\bibnamefont{Gumbs}},
  \bibinfo{author}{\bibfnamefont{A.}~\bibnamefont{Iurov}},
  \bibinfo{author}{\bibfnamefont{D.}~\bibnamefont{Huang}}, \bibnamefont{and}
  \bibinfo{author}{\bibfnamefont{O.}~\bibnamefont{Roslyak}},
  \bibinfo{journal}{Journal of Physics: Condensed Matter}
  \textbf{\bibinfo{volume}{32}}, \bibinfo{pages}{485301}
  (\bibinfo{year}{2020}).

\bibitem[{\citenamefont{Mawrie et~al.}(2014)\citenamefont{Mawrie, Biswas, and
  Ghosh}}]{mawrie2014magnetotransport}
\bibinfo{author}{\bibfnamefont{A.}~\bibnamefont{Mawrie}},
  \bibinfo{author}{\bibfnamefont{T.}~\bibnamefont{Biswas}}, \bibnamefont{and}
  \bibinfo{author}{\bibfnamefont{T.~K.} \bibnamefont{Ghosh}},
  \bibinfo{journal}{Journal of Physics: Condensed Matter}
  \textbf{\bibinfo{volume}{26}}, \bibinfo{pages}{405301}
  (\bibinfo{year}{2014}).

\bibitem[{\citenamefont{Tamang et~al.}(2023)\citenamefont{Tamang, Verma, and
  Biswas}}]{tamang2023orbital}
\bibinfo{author}{\bibfnamefont{L.}~\bibnamefont{Tamang}},
  \bibinfo{author}{\bibfnamefont{S.}~\bibnamefont{Verma}}, \bibnamefont{and}
  \bibinfo{author}{\bibfnamefont{T.}~\bibnamefont{Biswas}},
  \bibinfo{journal}{arXiv preprint arXiv:2309.07074}  (\bibinfo{year}{2023}).

\bibitem[{\citenamefont{Dutta et~al.}(2022)\citenamefont{Dutta, Ghosh, Singh,
  Lin, Politano, Bansil, and Agarwal}}]{dutta2022collective}
\bibinfo{author}{\bibfnamefont{D.}~\bibnamefont{Dutta}},
  \bibinfo{author}{\bibfnamefont{B.}~\bibnamefont{Ghosh}},
  \bibinfo{author}{\bibfnamefont{B.}~\bibnamefont{Singh}},
  \bibinfo{author}{\bibfnamefont{H.}~\bibnamefont{Lin}},
  \bibinfo{author}{\bibfnamefont{A.}~\bibnamefont{Politano}},
  \bibinfo{author}{\bibfnamefont{A.}~\bibnamefont{Bansil}}, \bibnamefont{and}
  \bibinfo{author}{\bibfnamefont{A.}~\bibnamefont{Agarwal}},
  \bibinfo{journal}{Physical Review B} \textbf{\bibinfo{volume}{105}},
  \bibinfo{pages}{165104} (\bibinfo{year}{2022}).

\bibitem[{\citenamefont{Gumbs et~al.}(2015{\natexlab{b}})\citenamefont{Gumbs,
  Iurov, Huang, and Pan}}]{gumbs2015tunable}
\bibinfo{author}{\bibfnamefont{G.}~\bibnamefont{Gumbs}},
  \bibinfo{author}{\bibfnamefont{A.}~\bibnamefont{Iurov}},
  \bibinfo{author}{\bibfnamefont{D.}~\bibnamefont{Huang}}, \bibnamefont{and}
  \bibinfo{author}{\bibfnamefont{W.}~\bibnamefont{Pan}},
  \bibinfo{journal}{Journal of Applied Physics} \textbf{\bibinfo{volume}{118}}
  (\bibinfo{year}{2015}{\natexlab{b}}).

\bibitem[{\citenamefont{Koseki et~al.}(2016)\citenamefont{Koseki, Ryzhii,
  Otsuji, Popov, and Satou}}]{koseki2016giant}
\bibinfo{author}{\bibfnamefont{Y.}~\bibnamefont{Koseki}},
  \bibinfo{author}{\bibfnamefont{V.}~\bibnamefont{Ryzhii}},
  \bibinfo{author}{\bibfnamefont{T.}~\bibnamefont{Otsuji}},
  \bibinfo{author}{\bibfnamefont{V.}~\bibnamefont{Popov}}, \bibnamefont{and}
  \bibinfo{author}{\bibfnamefont{A.}~\bibnamefont{Satou}},
  \bibinfo{journal}{Physical Review B} \textbf{\bibinfo{volume}{93}},
  \bibinfo{pages}{245408} (\bibinfo{year}{2016}).

\bibitem[{\citenamefont{Petrov et~al.}(2017)\citenamefont{Petrov, Svintsov,
  Ryzhii, and Shur}}]{petrov2017amplified}
\bibinfo{author}{\bibfnamefont{A.~S.} \bibnamefont{Petrov}},
  \bibinfo{author}{\bibfnamefont{D.}~\bibnamefont{Svintsov}},
  \bibinfo{author}{\bibfnamefont{V.}~\bibnamefont{Ryzhii}}, \bibnamefont{and}
  \bibinfo{author}{\bibfnamefont{M.~S.} \bibnamefont{Shur}},
  \bibinfo{journal}{Physical Review B} \textbf{\bibinfo{volume}{95}},
  \bibinfo{pages}{045405} (\bibinfo{year}{2017}).

\bibitem[{\citenamefont{Mojarro et~al.}(2026)\citenamefont{Mojarro,
  Carrillo-Bastos, and Maytorena}}]{mojarro2026topical}
\bibinfo{author}{\bibfnamefont{M.}~\bibnamefont{Mojarro}},
  \bibinfo{author}{\bibfnamefont{R.}~\bibnamefont{Carrillo-Bastos}},
  \bibnamefont{and} \bibinfo{author}{\bibfnamefont{J.~A.}
  \bibnamefont{Maytorena}}, \bibinfo{journal}{Journal of Physics: Condensed
  Matter} \textbf{\bibinfo{volume}{38}}, \bibinfo{pages}{243001}
  (\bibinfo{year}{2026}).

\bibitem[{\citenamefont{Xiong et~al.}(2023)\citenamefont{Xiong, Ba, Duan, Deng,
  Wang, and Wang}}]{xiong2023optical}
\bibinfo{author}{\bibfnamefont{Q.-Y.} \bibnamefont{Xiong}},
  \bibinfo{author}{\bibfnamefont{J.-Y.} \bibnamefont{Ba}},
  \bibinfo{author}{\bibfnamefont{H.-J.} \bibnamefont{Duan}},
  \bibinfo{author}{\bibfnamefont{M.-X.} \bibnamefont{Deng}},
  \bibinfo{author}{\bibfnamefont{Y.-M.} \bibnamefont{Wang}}, \bibnamefont{and}
  \bibinfo{author}{\bibfnamefont{R.-Q.} \bibnamefont{Wang}},
  \bibinfo{journal}{Physical Review B} \textbf{\bibinfo{volume}{107}},
  \bibinfo{pages}{155150} (\bibinfo{year}{2023}).

\bibitem[{\citenamefont{Mojarro et~al.}(2021)\citenamefont{Mojarro,
  Carrillo-Bastos, and Maytorena}}]{mojarro2021optical}
\bibinfo{author}{\bibfnamefont{M.}~\bibnamefont{Mojarro}},
  \bibinfo{author}{\bibfnamefont{R.}~\bibnamefont{Carrillo-Bastos}},
  \bibnamefont{and} \bibinfo{author}{\bibfnamefont{J.~A.}
  \bibnamefont{Maytorena}}, \bibinfo{journal}{Physical Review B}
  \textbf{\bibinfo{volume}{103}}, \bibinfo{pages}{165415}
  (\bibinfo{year}{2021}).

\bibitem[{\citenamefont{Stauber et~al.}(2013)\citenamefont{Stauber, San-Jose,
  and Brey}}]{stauber2013optical}
\bibinfo{author}{\bibfnamefont{T.}~\bibnamefont{Stauber}},
  \bibinfo{author}{\bibfnamefont{P.}~\bibnamefont{San-Jose}}, \bibnamefont{and}
  \bibinfo{author}{\bibfnamefont{L.}~\bibnamefont{Brey}}, \bibinfo{journal}{New
  Journal of Physics} \textbf{\bibinfo{volume}{15}}, \bibinfo{pages}{113050}
  (\bibinfo{year}{2013}).

\bibitem[{\citenamefont{Tan et~al.}(2021)\citenamefont{Tan, Yan, Zhao, Guo,
  Chang et~al.}}]{tan2021anisotropic}
\bibinfo{author}{\bibfnamefont{C.-Y.} \bibnamefont{Tan}},
  \bibinfo{author}{\bibfnamefont{C.-X.} \bibnamefont{Yan}},
  \bibinfo{author}{\bibfnamefont{Y.-H.} \bibnamefont{Zhao}},
  \bibinfo{author}{\bibfnamefont{H.}~\bibnamefont{Guo}},
  \bibinfo{author}{\bibfnamefont{H.-R.} \bibnamefont{Chang}},
  \bibnamefont{et~al.}, \bibinfo{journal}{Physical Review B}
  \textbf{\bibinfo{volume}{103}}, \bibinfo{pages}{125425}
  (\bibinfo{year}{2021}).

\bibitem[{\citenamefont{Wareham and Nicol}(2023)}]{wareham2023optical}
\bibinfo{author}{\bibfnamefont{W.~C.} \bibnamefont{Wareham}} \bibnamefont{and}
  \bibinfo{author}{\bibfnamefont{E.}~\bibnamefont{Nicol}},
  \bibinfo{journal}{Physical Review B} \textbf{\bibinfo{volume}{108}},
  \bibinfo{pages}{085424} (\bibinfo{year}{2023}).

\bibitem[{\citenamefont{Iurov et~al.}(2023{\natexlab{b}})\citenamefont{Iurov,
  Zhemchuzhna, Gumbs, and Huang}}]{iurov2023optical}
\bibinfo{author}{\bibfnamefont{A.}~\bibnamefont{Iurov}},
  \bibinfo{author}{\bibfnamefont{L.}~\bibnamefont{Zhemchuzhna}},
  \bibinfo{author}{\bibfnamefont{G.}~\bibnamefont{Gumbs}}, \bibnamefont{and}
  \bibinfo{author}{\bibfnamefont{D.}~\bibnamefont{Huang}},
  \bibinfo{journal}{Physical Review B} \textbf{\bibinfo{volume}{107}},
  \bibinfo{pages}{195137} (\bibinfo{year}{2023}{\natexlab{b}}).

\bibitem[{\citenamefont{Wild et~al.}(2023)\citenamefont{Wild, Mariani, and
  Portnoi}}]{wild2023optical}
\bibinfo{author}{\bibfnamefont{A.}~\bibnamefont{Wild}},
  \bibinfo{author}{\bibfnamefont{E.}~\bibnamefont{Mariani}}, \bibnamefont{and}
  \bibinfo{author}{\bibfnamefont{M.}~\bibnamefont{Portnoi}},
  \bibinfo{journal}{Scientific Reports} \textbf{\bibinfo{volume}{13}},
  \bibinfo{pages}{19211} (\bibinfo{year}{2023}).

\bibitem[{\citenamefont{Oriekhov and Gusynin}(2022)}]{oriekhov2022optical}
\bibinfo{author}{\bibfnamefont{D.}~\bibnamefont{Oriekhov}} \bibnamefont{and}
  \bibinfo{author}{\bibfnamefont{V.}~\bibnamefont{Gusynin}},
  \bibinfo{journal}{Physical Review B} \textbf{\bibinfo{volume}{106}},
  \bibinfo{pages}{115143} (\bibinfo{year}{2022}).

\bibitem[{\citenamefont{Malik et~al.}(2026)\citenamefont{Malik, Ahsan, and
  Islam}}]{malik2026probing}
\bibinfo{author}{\bibfnamefont{S.~A.} \bibnamefont{Malik}},
  \bibinfo{author}{\bibfnamefont{M.}~\bibnamefont{Ahsan}}, \bibnamefont{and}
  \bibinfo{author}{\bibfnamefont{S.~F.} \bibnamefont{Islam}},
  \bibinfo{journal}{Physical Review B} \textbf{\bibinfo{volume}{114}},
  \bibinfo{pages}{055406} (\bibinfo{year}{2026}).

\bibitem[{\citenamefont{Sadhukhan and
  Agarwal}(2017)}]{sadhukhan2017anisotropic}
\bibinfo{author}{\bibfnamefont{K.}~\bibnamefont{Sadhukhan}} \bibnamefont{and}
  \bibinfo{author}{\bibfnamefont{A.}~\bibnamefont{Agarwal}},
  \bibinfo{journal}{Physical Review B} \textbf{\bibinfo{volume}{96}},
  \bibinfo{pages}{035410} (\bibinfo{year}{2017}).

\end{thebibliography}
\end{document}